\documentclass{article}

\usepackage{iclr2025_conference,times}

\iclrfinalcopy

\usepackage[utf8]{inputenc} % allow utf-8 input
\usepackage[T1]{fontenc}    % use 8-bit T1 fonts
\usepackage{hyperref}       % hyperlinks
\usepackage{url}            % simple URL typesetting
\usepackage{booktabs}       % professional-quality tables
\usepackage{multirow}       % table cells spanning multiple rows
\usepackage{amsfonts}       % blackboard math symbols
\usepackage{nicefrac}       % compact symbols for 1/2, etc.
\usepackage{microtype}      % microtypography
\usepackage{xcolor}         % colors

\usepackage{graphicx}
\usepackage{amsmath}
\usepackage{amssymb}
\usepackage{mathtools}
\usepackage{amsthm}
\usepackage{orcidlink}
\usepackage{physics} % Quantum notation macros
\usepackage{bm}
\usepackage{caption}
\usepackage{subcaption}
\usepackage[inline]{enumitem}
\usepackage[ruled,lined,linesnumbered]{algorithm2e}
\SetKwInput{KwInput}{Input}
\SetKwInput{KwOutput}{Output}
\SetKwInput{KwRequire}{Require}
\SetKwComment{Comment}{$\triangleright$~}{}
\SetNlSty{}{}{:} % Set algorithm line numbers
\usepackage[nomain,acronym,shortcuts,toc]{glossaries}
\glsdisablehyper % Remove hyperlinks for glossary entries.
\makeglossaries
\makeatletter
\newcommand{\acx}{\protect\@acx}%
\newcommand{\@acx}[1]{%
  \ifAC@dua
   \acl{#1}%
  \else
   \expandafter\ifx\csname ac@#1\endcsname\AC@used
      \acs{#1}%
   \else
      \acl{#1}%
   \fi
  \fi
}
\makeatother

\newacronym{ours}{eQMARL}{entangled QMARL}
\newacronym{arc}{ARC}{Advanced Research Computing}

\newacronym{ctde}{CTDE}{centralized training with decentralized execution}
\newacronym{fctde}{fCTDE}{fully centralized CTDE}
\newacronym{sctde}{sCTDE}{split CTDE}
\newacronym{qfctde}{qfCTDE}{quantum fully centralized CTDE}

\newacronym{eqmarl}{eQMARL}{entangled quantum multi-agent reinforcement learning}

\newacronym{fl}{FL}{federated learning}

\newacronym{hpc}{HPC}{high-performance computing}

\newacronym{marl}{MARL}{multi-agent reinforcement learning}
\newacronym{mdp}{MDP}{Markov decision process}
\newacronym{ml}{ML}{machine learning}
\newacronym{maa2c}{MAA2C}{multi-agent advantage actor-critic}

\newacronym{nn}{NN}{neural network}
\newacronym{nisq}{NISQ}{noisy intermediate-scale quantum}

\newacronym{pqc}{PQC}{parameterized quantum circuit}
\newacronym{pomdp}{POMDP}{partially observable MDP}% {partially observable Markov decision process}

\newacronym{qrl}{QRL}{quantum reinforcement learning}
\newacronym{qmarl}{QMARL}{quantum multi-agent reinforcement learning}
\newacronym{ql}{QL}{Q-learning}
\newacronym{qql}{QQL}{quantum Q-learning}
\newacronym{qm2arl}{QM2ARL}{quantum meta MARL}
\newacronym{qnn}{QNN}{quantum neural network}
\newacronym{qml}{QML}{quantum machine learning}
\newacronym{qc}{QC}{quantum computing}
\newacronym{qa3c}{QA3C}{quantum asynchronous advantage actor-critic}
\newacronym{qcnn}{QCNN}{quantum convolutional neural network}

\newacronym{rl}{RL}{reinforcement learning}

\newacronym{sl}{SL}{split learning}
\newacronym{qsl}{QSL}{quantum split learning}
\newacronym{snn}{SNN}{split neural network}
\newacronym{qsnn}{QSNN}{quantum split neural network}

\newacronym{vqc}{VQC}{variational quantum circuit}

\glslocalreset{ml}
\glslocalreset{rl}
\glslocalreset{marl}
\usepackage{natbib}
\usepackage[capitalize]{cleveref} % Load last, especially AFTER algorithm2e.
\Crefname{figure}{Fig.}{Figs.}
\crefrangeformat{line}{lines~#3#1#4--#5#2#6}
\crefformat{equation}{(#2#1#3)}
\crefrangeformat{equation}{(#3#1#4--#5#2#6)}

\usepackage{titlesec}
\titlespacing*{\section}
{0pt}{0.25em}{0.25em}
\titlespacing*{\subsection}
{0pt}{0.25em}{0.25em}
\titlespacing*{\subsubsection}
{0pt}{0.25em}{0.25em}
\titlespacing*{\paragraph}
{0pt}{0.25em}{1em} % NeurIPS instructions say `\paragraph` must be followed by `1em` of space.

\newcommand{\bufferspacefromheader}{\vspace*{-1em}}

\title{eQMARL: Entangled Quantum Multi-Agent Reinforcement Learning for Distributed\\Cooperation over Quantum Channels}

\author{%
  Alexander DeRieux~\mbox{\orcidlink{0000-0003-1606-0668}} \quad Walid Saad~\mbox{\orcidlink{0000-0003-2247-2458}}\\
  Bradley Department of Electrical and Computer Engineering, Virginia Tech, USA \\
  \texttt{\{derieux,walids\}@vt.edu}
}

\begin{document}

% Set vertical spacing before/after `align` environment.
\setlength{\abovedisplayskip}{2pt}
\setlength{\belowdisplayskip}{2pt}
\setlength{\abovedisplayshortskip}{-8pt}
\setlength{\belowdisplayshortskip}{0pt}

% Set spacing of floats.
% https://tex.stackexchange.com/a/23316
% \setlength{\textfloatsep}{1\baselineskip} % space between last top float or first bottom float and the text
\setlength{\textfloatsep}{0.5em} % No extra spacing after all floats (figure, table, algorithm, footnote, etc.).
% \setlength{\floatsep}{0pt} % No extra spacing after all floats (figure, table, algorithm, footnote, etc.).

% Gives the width of the current document in pts (shown in log).
% \showthe\textwidth

\maketitle

%TC:break Abstract
\begin{abstract}
    Collaboration is a key challenge in distributed multi-agent reinforcement learning (MARL) environments. Learning frameworks for these decentralized systems must weigh the benefits of explicit player coordination against the communication overhead and computational cost of sharing local observations and environmental data. Quantum computing has sparked a potential synergy between quantum entanglement and cooperation in multi-agent environments, which could enable more efficient distributed collaboration with minimal information sharing. This relationship is largely unexplored, however, as current state-of-the-art quantum MARL (QMARL) implementations rely on classical information sharing rather than entanglement over a quantum channel as a coordination medium. In contrast, in this paper, a novel framework dubbed entangled QMARL (eQMARL) is proposed. The proposed eQMARL is a distributed actor-critic framework that facilitates cooperation over a quantum channel and eliminates local observation sharing via a quantum entangled split critic. Introducing a quantum critic uniquely spread across the agents allows coupling of local observation encoders through entangled input qubits over a quantum channel, which requires no explicit sharing of local observations and reduces classical communication overhead. Further, agent policies are tuned through joint observation-value function estimation via joint quantum measurements, thereby reducing the centralized computational burden. Experimental results show that eQMARL with $\Psi^{+}$ entanglement converges to a cooperative strategy up to $17.8\%$ faster and with a higher overall score compared to split classical and fully centralized classical and quantum baselines. The results also show that eQMARL achieves this performance with a constant factor of $25$-times fewer centralized parameters compared to the split classical baseline.
\end{abstract}

\section{Introduction}

\Ac{qrl} is emerging as a relatively new class of \ac{qml} for decision making. Exploiting the performance and data encoding enhancements of \acl{qc}, \ac{qrl} has many promising applications across diverse areas such as finance \citep{Herman2022SurveyQuantumComputing}, healthcare \citep{Flother2023StateQuantumComputing}, and even wireless networks \citep{Narottama2023QuantumMachineLearning}. Its multi-agent variant, \ac{qmarl}, is of specific interest because of the potential synergies between decentralized agent cooperation and quantum entanglement.
Indeed, in quantum mechanics \citep{einstein1971born}, entanglement is a distinctly quantum property that intrinsically links the behavior of one particle with another regardless of their physical proximity.
The use of entanglement in the broader field of \ac{qml} is a recent notion. Few core works like \citet{Mitarai2018QuantumCircuitLearning} and \citet{Du2020ExpressivePowerParameterizeda} use entangled layers within \ac{vqc} designs to link the behavior of \emph{quantum bits} (or \emph{qubits}) within a single hybrid quantum model. 
Even in the recently proposed \ac{qsl} framework \citep{Yun2023QuantumSplitNeural}, entanglement is only used locally within each \ac{vqc} branch of the \ac{qsnn} model.
What has not yet been explored, however, is using entanglement to \emph{couple} the behavior of multiple \ac{qml} models.
In \ac{qmarl}, the use of entanglement can be further extended to the implicit coordination amongst agents during training time.
Historically, in both purely classical and quantum \ac{marl}, classical communication, shared replay buffers, centralized global networks, and fully-observable environment assumptions have all proven to be viable methods for coordinating a group policy \citep{Yun2022QuantumMultiAgentReinforcement,Yun2023QuantumMultiAgentActorCritic,Yun2022QuantumMultiAgentMeta,Chen2023AsynchronousTrainingQuantum,Park2023QuantumMultiAgentReinforcement,Kolle2024MultiAgentQuantumReinforcement}.
Even \ac{qsl}, which is not exclusive to \acs{marl}, relies fully on classical communication between branches of the \ac{qsnn} \citep{Yun2023QuantumSplitNeural}.
None of these approaches, however, take advantage of the available quantum channel and quantum entanglement as coupling mediums across decentralized agents or model branches, and opt instead for more classical methods of coordination.
In short, entanglement is one such phenomenon of quantum mechanics that has not yet been fully explored in the context of cooperation in \ac{qmarl} settings.

In contrast to prior art, we propose a novel framework dubbed \emph{\ac{ours}}.
The proposed \ac{ours} is a distributed actor-critic framework, intersecting canonical \ac{ctde} and fully decentralized learning, that facilitates collaboration over a quantum channel using a quantum entangled split critic. 
Our design uniquely allows agents to coordinate their policies by, for the first time, splitting the quantum critic architecture over a quantum channel and coupling their localized observation encoders using entangled input qubits. This uniquely allows agents to cooperate over a quantum channel, which eliminates the need for observation sharing, and further reduces classical communication overhead.
Also, agent policies are tuned via joint observation-value function estimation using joint quantum measurements across all qubits in the system, which minimizes the computational burden of a central server.
As will be evident from our analysis, \ac{ours} will be shown to converge to a cooperative strategy faster, with higher overall score on average, and with fewer centralized parameters compared to baselines.
All of our source code and experiments are publicly available on GitHub\footnote{\url{https://github.com/news-vt/eqmarl}}.

\subsection{Related works and their limitations}

\Ac{qmarl} is a nascent field, with few works applying the quantum advantage to scenarios with multiple agents \citep{Yun2022QuantumMultiAgentReinforcement,Yun2023QuantumMultiAgentActorCritic,Yun2022QuantumMultiAgentMeta,Chen2023AsynchronousTrainingQuantum,Park2023QuantumMultiAgentReinforcement,Kolle2024MultiAgentQuantumReinforcement}. Further, the application of quantum to \ac{sl} is even newer, with \citet{Yun2023QuantumSplitNeural} being the only prior work.
A complete overview of prior works is provided as supplementary material in \cref{app:related}.
The resounding theme in these prior works is the use of independent agents or branches that communicate and learn through centralized classical means.
No prior work, however, makes use of the quantum channel as a medium for system coupling or for multi-agent collaboration. 
Indeed, the quantum elements serve as drop-in replacements for classical \ac{nn} counterparts, and, importantly, the quantum channel between agents and the potential for sharing entangled qubit states go largely under-utilized. 
Simply put, entanglement and the quantum channel are potentially useful untapped cooperative resources intrinsic to \ac{qmarl} that have largely unknown benefits.

\subsection{Contributions}

The contributions of this work are summarized as follows:
\vspace{-0.5em}
% \vspace{-0.1em}
% \begin{itemize}[itemsep=0.1em]
\begin{itemize}[itemsep=-0.07em]
    \item We propose a novel \ac{ours} framework that trains decentralized policies via a split quantum critic over a quantum channel with entangled input qubits and joint measurement. 
    \item We propose a new \acs{qmarl} algorithm for training distributed agents via optimizing a split critic \emph{without sharing local environment observations amongst agents or a central server}. 
    \item We show that the split nature of \ac{ours} reduces the computational burden of  a central quantum server by distributing and tuning parameterized quantum gates across agents in the system, and requiring a small number of parameters for joint measurement.
    \item We empirically demonstrate that \ac{ours} with $\Psi^{+}$ entanglement exhibits a faster convergence time that can reach up to $17.8\%$ faster, and with higher overall score, compared to split classical and fully centralized classical and quantum baselines in environments with full and partial information. Further, the results also show that \ac{ours} achieves this level of performance and cooperation with a constant factor of $25$-times fewer centralized parameters compared to the split classical baseline.
\end{itemize}
% \vspace*{-1.2em}
% \vspace*{-0.8em}
\vspace*{-0.5em}
To the best of our knowledge, this is the first application of \ac{qmarl} that exploits the quantum channel and entanglement between agents to learn without sharing local observations, while also reducing the classical communication overhead and central computation burden of leading approaches.
\section{Proposed \ac{ours} framework} \label{sec:solution}

\begin{figure}[t]
    % \vspace*{-0.8em}
    % \vspace*{-1em}
    \bufferspacefromheader
    \centering
    \includegraphics[width=\linewidth]{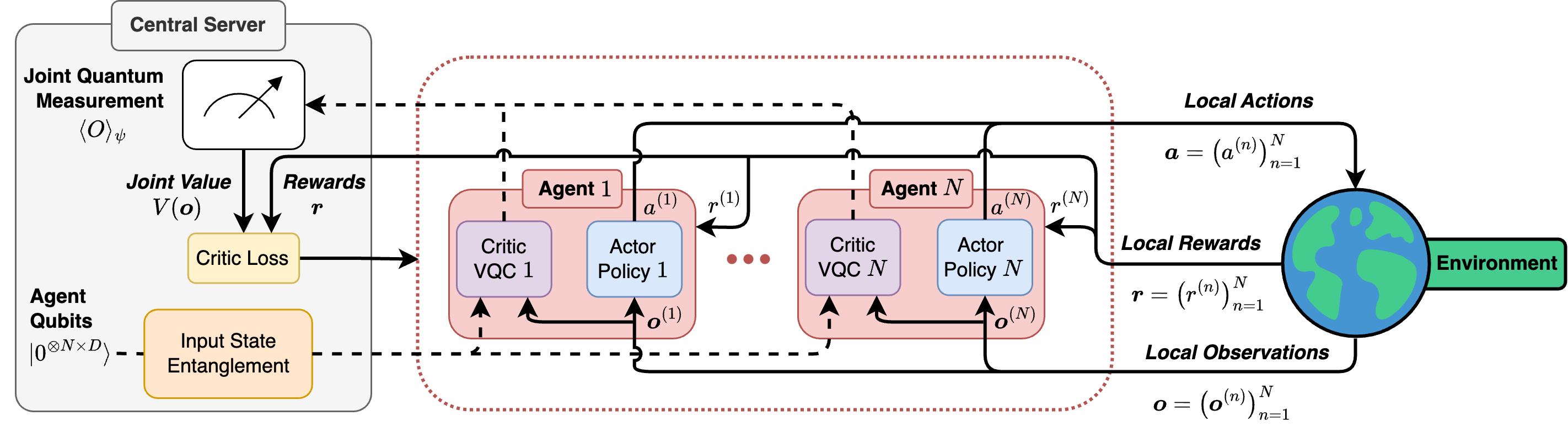}
    \vspace*{-1.7em}
    \caption{General design of our \ac{ours} framework. Dashed (solid) arrows represent quantum (classical) communication. A split quantum critic is deployed across the agents via local \acp{vqc} (purple) coupled via input entanglement (orange) at a trusted central server (gray). Joint quantum measurements across all qubits (white) estimate the joint value from the locally-encoded observations.}
    % \vspace*{-0.8em}
    \vspace*{-0.6em}
    \label{fig:system_design}
\end{figure}

\begin{figure}[t]
    % \vspace*{-1em}
    \bufferspacefromheader
    \centering
    \includegraphics[width=\linewidth]{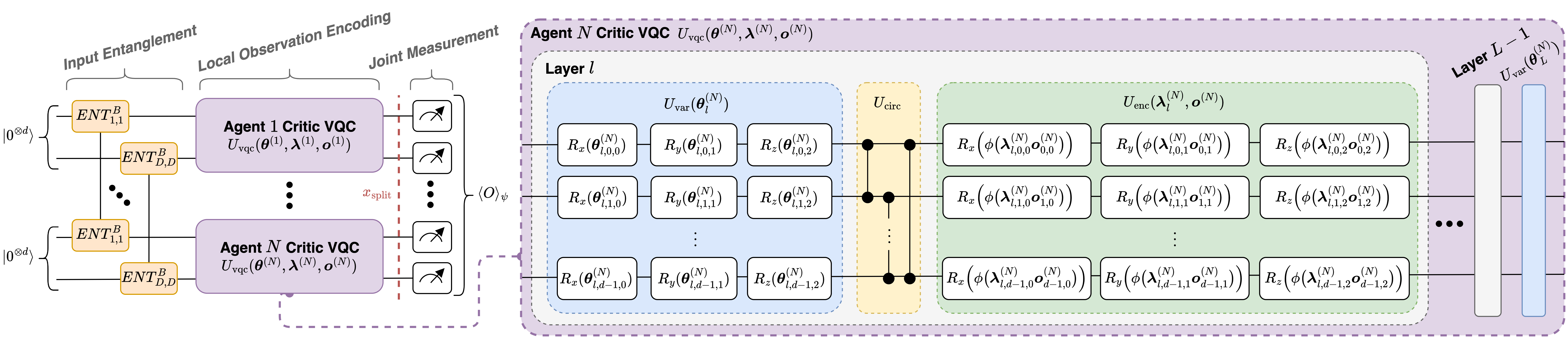}
    \caption{Quantum system design with $N$ agents and $D$ qubits per agent. Input entanglement operators (orange) couple the split critic \acp{vqc} (purple, with split point marked in red), which has cascaded layers of variational (blue), circular entanglement (yellow), and encoding (green) operators.}
    % \vspace*{-1.8em}
    % \vspace*{-0.8em}
    \vspace*{-0.6em}
    \label{fig:system_design_quantum}
\end{figure}

Our proposed \ac{ours} framework is a new approach for training multi-agent actor-critic architectures which lies at the intersection between \ac{ctde} and fully decentralized learning. Inspired from \ac{ctde}, we deploy decentralized agent policies which learn using a joint value function at training time. The key to our approach, however, is that we use quantum entanglement to deploy the joint value function estimator as a critic network \emph{which is spread across the agents to operate in mostly decentralized fashion}. 
An overview of our framework design is shown in \cref{fig:system_design}, and the design of the system architecture from a purely quantum perspective is shown in \cref{fig:system_design_quantum}. 
From \cref{fig:system_design}, the two main elements of \ac{ours} are a \emph{central quantum server} and a set of $N$ \emph{decentralized quantum agents} $\mathcal{N}=\{n\}_{n=1}^{N}$. 
The decentralized agents do not communicate with each other; only communication with the server is necessary during training. 
During execution, the agents interact with the environment independently and are fully decentralized. During training, our \ac{ours} framework is divided into core stages: 1) Centralized quantum input state entanglement preparation, 2) Decentralized agent environment observation encoding and variational rotations, and 3) Joint value estimation through joint quantum measurement. 
\Cref{fig:system_design_quantum} shows how input states are prepared using custom pairwise entanglement operators, followed by agent \acp{vqc}, and then joint measurements. Physically, these operations occur at different locations, 
however, it is equivalent to consider these as a single quantum system, from input state preparation to final measurement. For purposes of quantum state transmission, we assume an ideal quantum channel environment with no losses.

\subsection{Joint input entanglement}

The first stage of \ac{ours} creates an entangled input state for the split quantum critic network, which couples the critic \acp{vqc} spread across the agents. 
To understand how this works, we first provide a brief primer on quantum entanglement. More comprehensive preliminaries are provided as supplementary material in \cref{app:prelim}.
Consider two independent agents $A$ and $B$, with quantum systems described by Hilbert spaces $\mathcal{H}_{A}$ and $\mathcal{H}_{B}$, and arbitrary quantum states $|\psi\rangle_{A}$ and $|\psi\rangle_{B}$. The combined Hilbert space of the two systems can be represented using the Kronecker (i.e., tensor) product $\mathcal{H}_{AB} = \mathcal{H}_{A} \otimes \mathcal{H}_{B}$. The combined quantum system is said to be \emph{separable} if the agent states can be cleanly separated as a tensor product of the two systems, i.e., $|\psi\rangle_{AB} = |\psi\rangle_{A} \otimes |\psi\rangle_{B}$. If, however, the states cannot be represented in this form, then the combined system is said to be inseparable, or \emph{entangled}. For example, if each agent has one qubit in the $\{|0\rangle, |1\rangle\}$ basis, then a separable system could have the state $|\psi_{\textrm{separable}}\rangle_{AB} = |0\rangle_{A} \otimes |1\rangle_{A}$, whereas an entangled system could have the state $|\psi_{\textrm{entangled}}\rangle_{AB} = (|00\rangle_{AB} + |11\rangle_{AB})/\sqrt{2}$. 
In \ac{ours}, each agent $n \in \mathcal{N}$ is assigned a set of $D$ qubits $\mathcal{Q}^{(n)} = \{q_{d}^{(n)}\}_{d=1}^{D}$ chosen based upon the environment state dimension and desired quantum state encoding method. The total number of system qubits is $N \times D$, and is represented by the union of agent qubit sets $\mathcal{Q} = \bigcup_{n=1}^{N} \mathcal{Q}^{(n)}$.
We couple the agents by preparing an input state that pairwise entangles their qubits using a variation of Bell state entanglement \citep{Nielsen2012QuantumComputationQuantum} such that
\vspace*{-0.8em}
\begin{equation}
    \textrm{ENT}^{B}_{\delta(1,d),\dots,\delta(N,d)} = \begin{cases}
        \bigl( \bigotimes_{n=2}^{N} \textrm{CNOT}_{\delta(1,d), \delta(n,d)} \bigr) H_{\delta(1,d)}, & \textrm{if $B=\Phi^{+}$}, \\[0.5em]
        \bigl( \bigotimes_{n=2}^{N} \textrm{CNOT}_{\delta(1,d), \delta(n,d)} \bigr) H_{\delta(1,d)} X_{\delta(1,d)}, & \textrm{if $B=\Phi^{-}$}, \\[0.5em]
        \bigl( \bigotimes_{n=2}^{N} \textrm{CNOT}_{\delta(1,d), \delta(n,d)} \bigr) H_{\delta(1,d)} \bigl( \bigotimes_{k=2}^{N} X_{\delta(k,d)} \bigr), & \textrm{if $B=\Psi^{+}$}, \\[0.5em]
        \bigl( \bigotimes_{n=2}^{N} \textrm{CNOT}_{\delta(1,d), \delta(n,d)} \bigr) H_{\delta(1,d)} \bigl( \bigotimes_{k=1}^{N} X_{\delta(k,d)} \bigr), & \textrm{if $B=\Psi^{-}$},
    \end{cases} \label{eqn:ent}
\end{equation}
is a coupling operator across qubits $\{q^{(n)}_d\}_{n=1}^{N} \subseteq \mathcal{Q}$, where $\mathcal{B} \in \{\Phi^{+}, \Phi^{-}, \Psi^{+}, \Psi^{-}\}$ is the set of Bell states, $B \in \mathcal{B}$ is the selected entanglement scheme, and $\delta(n,d) = (n-1)D + d$ is an index mapping within $\mathcal{Q}$ for agent $n \in \mathcal{N}$ at qubit index $d \in [1,D]$, i.e., $\delta(n,d) \equiv q^{(n)}_d \in \mathcal{Q}^{(n)}$. Importantly, this operator can be applied to entangle any arbitrary set of qubits within the circuit.
Note that in this work we assume the agents receive their entangled qubits in real-time via a trusted central source, i.e., a central server, however, they could be pre-generated and stored in quantum memory at the agent if desired.
The quantum circuits that generate each $B \in \mathcal{B}$ are given in \cref{app:ent}, \cref{app:fig:joint_entanglement_circuit}.

\subsection{Decentralized split critic \ac{vqc} design}

At its core, our joint quantum critic is a \ac{snn} \citep{Vepakomma2018SplitLearningHealth}, with each agent's local \ac{vqc} serving as a branch.
After the input qubits are entangled, they are partitioned back into $N$ sets of $D$ qubits, 
i.e., $\{\mathcal{Q}^{(n)}\}_{n\in\mathcal{N}}$, 
and transmitted to each agent respectively. 
The agents collect and encode local observations from the environment into their assigned qubits using a \ac{vqc}.
We use the \ac{vqc} architectures of \citet{Jerbi2021ParametrizedQuantumPolicies,Skolik2021QuantumAgentsGym} for our hybrid quantum network design.
Each agent in our \ac{qmarl} setting uses the same \ac{vqc} architecture for their branch of the critic, but tunes their own unique set of parameters. The same architecture is a reasonable assumption since all agents are learning in the same environment, and the uniqueness of parameters tailors each branch to local observations.
From \cref{fig:system_design_quantum}, the \ac{vqc} design consists of $L$ cascaded layers of \emph{variational}, \emph{circular entanglement}, and \emph{encoding} operators, with an additional variational layer at the end of the circuit before measurement.
The trainable variational layer performs sequential parameterized Pauli X, Y, and Z-axis rotations, and it can be expressed as the unitary operator
%
% \begin{equation}
$
    U_{\textrm{var}}(\bm{\theta}^{(n)}_l) = \bigotimes_{d=0}^{D-1} R_z(\bm{\theta}^{(n)}_{l,d,2}) R_y(\bm{\theta}^{(n)}_{l,d,1}) R_x(\bm{\theta}^{(n)}_{l,d,0}),
    \label{eqn:U_var}
$
% \end{equation}
%
where $\bm{\theta}^{(n)} \in [0, 2\pi]^{(L+1) \times D \times 3}$ is a matrix of rotation angle parameters for agent $n$.
The non-trainable circular entanglement layer binds neighboring qubits within a single agent using the operator
% 
% \begin{equation}
$
    U_{\textrm{circ}} = CZ_{0,D-1} \bigl( \prod_{d=0}^{D-2} CZ_{d,d+1} \bigr).
    \label{eqn:U_circ}
$
% \end{equation}
% 
% 
The trainable encoding layer maps a matrix of classical features $\bm{o}^{(n)} \in \mathbb{R}^{D \times 3}$, i.e., an agent's environment observation, into a quantum state via the operator:
%
% \begin{equation}
$
    U_{\textrm{enc}}(\bm{\lambda}^{(n)}_l, \bm{o}^{(n)}) = \bigotimes_{d=0}^{D-1} R_z\bigl(\phi\bigl(\bm{\lambda}^{(n)}_{l,d,2} \bm{o}^{(n)}_{d,2}\bigr)\bigr) R_y\bigl(\phi\bigl(\bm{\lambda}^{(n)}_{l,d,1} \bm{o}^{(n)}_{d,1}\bigr)\bigr) R_x\bigl(\phi\bigl(\bm{\lambda}^{(n)}_{l,d,0} \bm{o}^{(n)}_{d,0}\bigr)\bigr),
    \label{eqn:U_enc}
$
% \end{equation}
%
where $\bm{\lambda}^{(n)} \in \mathbb{R}^{L \times D \times 3}$ is a matrix of trainable scaling parameters, and $\phi \colon \mathbb{R} \mapsto \mathbb{R}$ is an optional squash activation function.
% \footnote{We employ $\phi(x) = \textrm{arctan}(x)$ to squash the encoder inputs to range $[-\frac{\pi}{2}, \frac{\pi}{2}]$.}
% 
The entire \ac{vqc} can be expressed as a single operator, as follows:
\begin{equation}
    U_{\textrm{vqc}}(\bm{\theta}^{(n)},\bm{\lambda}^{(n)},\bm{o}^{(n)}) = U_{\textrm{var}}(\bm{\theta}^{(n)}_{L}) \prod_{l=0}^{L-1} U_{\textrm{enc}}(\bm{\lambda}^{(n)}_{L-l-1}, \bm{o}^{(n)}) ~ U_{\textrm{circ}} ~ U_{\textrm{var}}(\bm{\theta}^{(n)}_{L-l-1}),
    \label{eqn:U_vqc}
\end{equation}
which is parameterized by variational angles $\bm{\theta}^{(n)}$ and encoding weights $\bm{\lambda}^{(n)}$.

\subsection{Centralized joint measurement}
The locally encoded qubits for each agent are subsequently forwarded to a central quantum server, which could either be the entanglement source or a different location, for joint measurement.
A joint measurement across all qubits in the system is made in the Pauli $Z$ basis using the observable $O= \bigotimes_{d=1}^{N \times D} Z_d$.
The joint value for the locally-encoded observations
is then estimated as follows:
\begin{equation}
    V(\bm{o}) \simeq w \left(\frac{1 + \langle O \rangle _{\psi}}{2}\right),
    \label{eqn:value-estimate}
\end{equation}
where $w \in \mathbb{R}$ is a learned scaling parameter, $\langle O \rangle _{\psi}$ is the expected value of the joint observable~w.r.t.~an arbitrary system state $\ket{\psi}$ across all qubits, and $\bm{o} = (\bm{o}^{(n)})_{n=1}^{N}$ is a vector of joint observations. This rescaling is necessary because the range of the measured observable is $\langle O \rangle _{\psi} \in [-1, 1]$ (i.e., proportional to the eigenvalues of the operator $O$), whereas $V(\bm{o})\in\mathbb{R}$.
The critic loss with respect to the joint value and local agent rewards is then disseminated amongst the agents for tuning of their localized portion of the split critic network and local policy networks.

\subsection{Split critic loss}

The loss of the split critic is derived in a way similar to \citet{Vepakomma2018SplitLearningHealth}.
Since the input entanglement stage of \ac{ours} has no trainable parameters, it does not exist for the purposes of \acs{sl} backpropagation.
We denote the point of joint quantum measurement as the \emph{split point}, which is preceded by local agent \ac{vqc} \emph{branches}.
Each branch can be individually tuned using the partial gradient of the loss at the split point via partial gradient~w.r.t.~its own local parameters. 
If we define $x_{\textrm{split}}$ as the split point, then the partial gradient of each branch's parameters can be estimated using the central loss, as follows:
\begin{equation}
    \nabla_{\bm{\theta}_{\textrm{critic}}^{(n)}} \mathcal{L}_{\textrm{critic}} = \frac{\partial \mathcal{L}_{\textrm{critic}}}{\partial \bm{\theta}_{\textrm{critic}}^{(n)}} = \frac{\partial \mathcal{L}_{\textrm{critic}}}{\partial x_{\textrm{split}}} \frac{\partial x_{\textrm{split}}}{\partial \bm{\theta}_{\textrm{critic}}^{(n)}} = \underbrace{\left(\nabla_{x_{\textrm{split}}} \mathcal{L}_{\textrm{critic}} \vphantom{\left(\nabla_{\bm{\theta}_{\textrm{critic}}^{(n)}} x_{\textrm{split}}\right)}\right)}_{\textrm{Central server}} \underbrace{\left(\nabla_{\bm{\theta}_{\textrm{critic}}^{(n)}} x_{\textrm{split}}\right)}_{\textrm{Local agent}},
    \label{eqn:joint-critic-local-gradient}
\end{equation}
where $\nabla_{x_{\textrm{split}}} \mathcal{L}_{\textrm{critic}}$ is the gradient of the loss at the split point, and $\nabla_{\bm{\theta}_{\textrm{critic}}^{(n)}} x_{\textrm{split}}$ is the gradient from the split point back to the start of branch $n\in\mathcal{N}$. The value of $\nabla_{x_{\textrm{split}}} \mathcal{L}_{\textrm{critic}}$ is sent classically to the agents, and since \cref{eqn:value-estimate} only uses a single trainable parameter, $w$, the classical communication overhead needed for split backpropagation is minimal.
Here, we use the \texttt{Huber} loss for the critic (see \cref{app:loss}).

\subsection{Coupled agent learning algorithm}

% \vspace*{-1em}
\bufferspacefromheader
\begin{algorithm}[t]
    \caption{Summary of \ac{ours} training using \acrshort{maa2c} for a quantum entangled split critic. The full algorithm is provided in \cref{app:algorithm}, \cref{alg:maa2c}}\label{alg:maa2c-summary}
    \scriptsize
    \KwRequire{Set of $N$ agents $\mathcal{N}$, 1 quantum entanglement source, 1 quantum measurement server}
    Initialize $N$ critic branches $U_{\textrm{vqc}}$ with $D$ qubits and parameters $\bm{\theta}_{\textrm{critic}}^{(n)}$, $\bm{\lambda}^{(n)}$, $\forall n \in \mathcal{N}$\;
    \For{all episodes}{
        \For(\Comment*[h]{\ac{ours} training, notation $\bm{o}_{\tau}=\bigl(\bm{o}_{\tau}^{(n)}\bigr)_{n=1}^{N}$, $\bm{o}_{\tau+1}=\bigl(\bm{o}_{\tau+1}^{(n)}\bigr)_{n=1}^{N}$, $\bm{a}_{\tau}=\bigl(a_{\tau}^{(n)}\bigr)_{n=1}^{N}$.}){all time steps $\tau$}{\label{alg:maa2c-summary:quantum_start}
            Central quantum server generates $2N$ sets of $D$ entangled qubits and sends to agents via \textbf{quantum channel}\;
            \For{each agent $n \in \mathcal{N}$}{
                Apply $U_{\textrm{vqc}}(\bm{\theta}^{(n)}_{\textrm{critic}},\bm{\lambda}^{(n)},\bm{o}_{\tau}^{(n)})$ and $U_{\textrm{vqc}}(\bm{\theta}^{(n)}_{\textrm{critic}},\bm{\lambda}^{(n)},\bm{o}_{\tau+1}^{(n)})$ from \cref{eqn:U_vqc} locally using assigned entangled input qubits\;
                Transmit resulting qubits via \textbf{quantum channel}, and reward $r_{\tau}^{(n)}$ via \textbf{classical channel} to central quantum server\;
            }
            Perform joint measurements on qubits across all agents to estimate $V(\bm{o}_{\tau})$ and $V(\bm{o}_{\tau+1})$ using \cref{eqn:value-estimate}\;
            Estimate $Q(\bm{o}_{\tau}, \bm{a}_{\tau}) = \sum_{n=1}^{N}{r_{\tau}^{(n)}} + \gamma V(\bm{o}_{\tau+1})$ using discount factor $\gamma$\;
            Compute $\nabla_{x_{\textrm{split}}} \mathcal{L}_{\textrm{critic}}$ and transmit via \textbf{classical channel} to each agent to update $\bm{\theta}_{\textrm{critic}}^{(n)}$ locally using partial gradient from \cref{eqn:joint-critic-local-gradient}\;
        }\label{alg:maa2c-summary:quantum_end}
    }
\end{algorithm}

Our \ac{ours} uses a variation of the \ac{maa2c} algorithm \citep{Papoudakis2021BenchmarkingMultiAgentDeep} to train local agent policies with a split quantum joint critic. 
Here, we summarize the algorithm in \cref{alg:maa2c-summary}, which focuses on the elements for necessary for tuning the critic.
In \ac{ours}, there are $N$ quantum agents that are physically separated from each other (no cross-agent communication is assumed) and one central quantum server. Each agent $n\in\mathcal{N}$ employs a \ac{vqc}, given by \cref{eqn:U_vqc}, with unique parameters $\bm{\theta}_{\textrm{critic}}^{(n)}$ and $\bm{\lambda}^{(n)}$, that serves as one branch in the split critic network. 
All agents interact with the environment independently and each has its own local data buffer -- local observations are neither shared amongst agents nor with the server.
The first stage of \ac{ours} is fundamentally similar to traditional \ac{maa2c}.
The second stage is where the uniqueness of \ac{ours} comes into play. The central quantum server prepares $2N$ sets of $D$ entangled qubits using \cref{eqn:ent} for each time step $\tau$, which are then transmitted to the agents via a quantum channel. Each agent then encodes their local observations $\bm{o}^{(n)}_{\tau}$ and $\bm{o}^{(n)}_{\tau+1}$ using \cref{eqn:U_vqc} and their assigned entangled input qubits, and transmits the resulting qubits via quantum channel back to the server. The agents also share their corresponding reward $r^{(n)}_{\tau}$ via a classical channel with the server, which will be used for downstream loss calculations.
Access to the reward is necessary for the critic to evaluate agent policy performance. This is a reasonable assumption in \ac{ours} as the reward value contains no localized environment information, and is also used in \citet{Yun2022QuantumMultiAgentReinforcement,Yun2023QuantumMultiAgentActorCritic,Park2023QuantumMultiAgentReinforcement}; regardless, the classical channel will also be used to transmit partial gradients of the critic loss.
The server then performs a joint measurement on all the qubits associated with $\bm{o}^{(n)}_{\tau}$ and $\bm{o}^{(n)}_{\tau+t}$ to obtain estimates for $V(\bm{o}_{\tau})$ and $V(\bm{o}_{\tau+1})$ using \cref{eqn:value-estimate}.
Subsequently, the server computes the expected cumulative reward $Q(\bm{o}_{\tau}, \bm{a}_{\tau})$
for the joint observations and actions at $\tau$ using $V(\bm{o}_{\tau+1})$, discount factor $\gamma$, and the respective rewards.
The joint critic loss $\mathcal{L}_{\textrm{critic}}$ is then computed, its partial gradient~w.r.t.~the split point $\nabla_{x_{\textrm{split}}} \mathcal{L}_{\textrm{critic}}$ is estimated, and then sent via a classical channel to each agent to update their local weights $\bm{\theta}^{(n)}_{\textrm{critic}}$ using \cref{eqn:joint-critic-local-gradient}.
\section{Experiments and demonstrations} \label{sec:exp}

\subsection{Environments}\label{sec:exp:env}

We use the \texttt{CoinGame} environment first proposed in \citet{Lerer2018MaintainingCooperationComplex}, and as implemented in \citet{Phan2022EmergentCooperationMutual}, which has been widely used \citep{Foerster2018LearningOpponentLearningAwareness,Phan2022EmergentCooperationMutual,Kolle2024MultiAgentQuantumReinforcement}, a multi-agent variant of the canonical \texttt{CartPole} environment \citep{Barto1983NeuronlikeAdaptiveElements}, and a multi-agent variant of the \texttt{MiniGrid} environment \citep{MinigridMiniworld23} as benchmarks for \ac{marl} scenarios.
In particular, \texttt{CoinGame}'s nature as a zero-sum game and the independent natures of both multi-agent \texttt{CartPole} and \texttt{MiniGrid} serve as intriguing case studies for learning cooperative strategies using full, i.e., described by a \ac{mdp}, and partial, i.e., described by a \ac{pomdp}, information.
In \texttt{CoinGame}, we evaluate agents using the \emph{Score} metric, which aggregates all agent undiscounted rewards over a single episode.
In both \texttt{CartPole} and \texttt{MiniGrid} we evaluate agents using the \emph{total reward} metric, which aggregates the number of time steps an agent maintains pole balance and the reward received for maze navigation over a single episode respectively.
See \cref{app:env} for environment details.

\subsection{Experiment setup}\label{sec:exp:setup}

We compare \ac{ours} against three baselines that are considered the current state-of-the-art configurations in actor-critic \ac{ctde}:
\begin{enumerate*}[label=\arabic*)]
    \item \textbf{\Ac{fctde}}, a classical configuration where the critic is a simple fully-connected \acs{nn} located at a central server, like in \citet{sukthankar_cooperative_2017,foerster_counterfactual_2018}, and requires agents to transmit their local observations to the server via a classical channel;
    \item \textbf{\Ac{sctde}}, a classical configuration where the critic is a branching \acs{nn} encoder spread across the agents, which is combined using a centralized \acs{nn} based on \citet{Rashid2018QMIXMonotonicValue} located at a central server, and requires agents to transmit intermediate \ac{nn} activations via a classical channel; and
    \item \textbf{\Ac{qfctde}}, a quantum variant of \ac{fctde} where the critic is located at a central server, as in \citet{Yun2022QuantumMultiAgentReinforcement,Yun2023QuantumMultiAgentActorCritic,Park2023QuantumMultiAgentReinforcement}, and agents transmit their local observations via a classical channel.
\end{enumerate*}
These baselines were carefully chosen to convey how a quantum entangled split critic eliminates local environment observation sharing, while reducing classical communication overhead by leveraging the quantum channel, and minimizing centralized computational complexity.
In our experiments we simplify the setup by using policy sharing across the agents, as done in \citet{Yun2023QuantumMultiAgentActorCritic} and \citet{Chen2023AsynchronousTrainingQuantum}.
All classical models were built using \texttt{tensorflow}, the quantum models using \texttt{tensorflow-quantum}, and \texttt{cirq} for quantum simulations.
For \texttt{CoinGame}, all models were trained for 3000 epochs, with $T=50$ steps, $\gamma=0.99$. 
For \texttt{CartPole}, all models were trained for 1000 epochs, with a maximum of $500$ steps per episode.
For \texttt{MiniGrid}, all models were trained for 1000 epochs, with a maximum of $T=50$ steps per episode.
All models use the \texttt{Adam} optimizer with varying learning rates.
The quantum models use $D=4$ qubits, $L=5$ layers, and $\phi=\texttt{arctan}$ activation.
The classical models use $h=12$ hidden units for \texttt{CoinGame} and \texttt{CartPole}, and $h=100$ for \texttt{MiniGrid}.
See \cref{app:hyp,app:qenctran,app:exp} for further details.
We conduct all experiments on a high-performance computing cluster with 128 CPU cores and 256 GB of memory per node. The training time of \texttt{sCTDE} for \texttt{CoinGame} \ac{mdp} is $\approx 5.5~\textrm{minutes}$, and for \ac{pomdp} is $\approx 7.5~\textrm{minutes}$. In contrast, the training time of \ac{ours} is $\approx 3.5~\textrm{hours}$ and $\approx 8.5~\textrm{hours}$ for \ac{mdp} and \ac{pomdp} respectively; this is in line with many current QMARL works \citep{Yun2022QuantumMultiAgentReinforcement,Yun2023QuantumMultiAgentActorCritic,Yun2022QuantumMultiAgentMeta,Chen2023AsynchronousTrainingQuantum,Park2023QuantumMultiAgentReinforcement,Kolle2024MultiAgentQuantumReinforcement}, and is indicative of the known computational complexities of running quantum simulations on classical hardware.

\subsection{Comparing quantum input entanglement styles}\label{sec:exp:results:ent}

\begin{figure}[t]
    % \vspace*{-0.8em}
    % \vspace*{-1em}
    \bufferspacefromheader
     \centering
     \begin{subfigure}{0.40\linewidth}
         \centering
         \includegraphics[width=\linewidth]{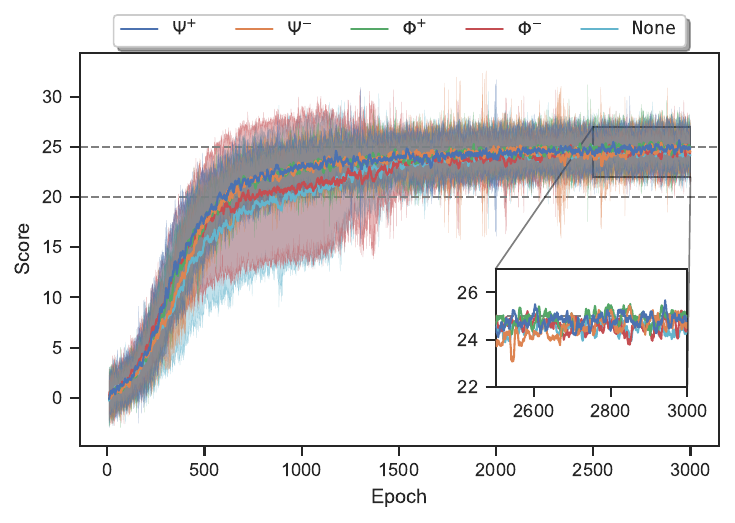}
         
         \vspace*{-0.1em}
         \caption{Score - \ac{mdp}}
         \vspace*{-0.25em}
         % \label{fig:fig_maa2c_mdp_entanglement_compare:undiscounted_reward}
         \label{fig:fig_maa2c_entanglement_compare:mdp}
     \end{subfigure}%
     % \hfill
     \begin{subfigure}{0.40\linewidth}
         \centering
         \includegraphics[width=\linewidth]{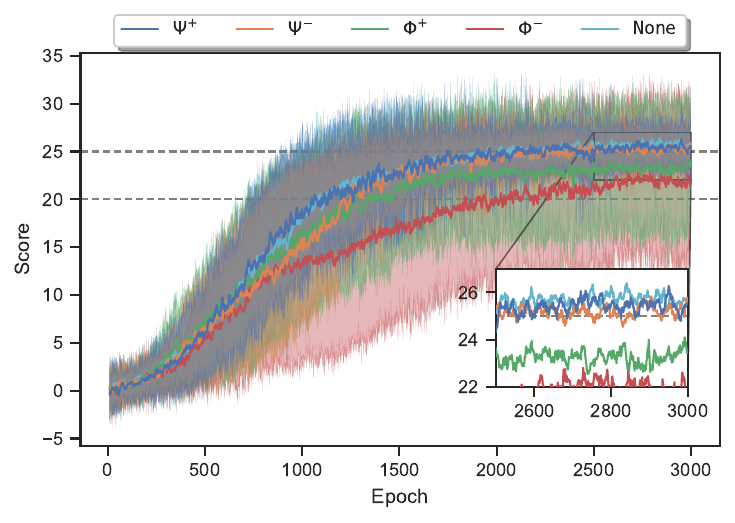}
         
         \vspace*{-0.1em}
         \caption{Score - \ac{pomdp}}
         \vspace*{-0.25em}
         % \label{fig:fig_maa2c_pomdp_entanglement_compare:undiscounted_reward}
         \label{fig:fig_maa2c_entanglement_compare:pomdp}
     \end{subfigure}
    \caption{Comparison of \texttt{CoinGame-2} score performance with (a) \ac{mdp}, and (b) \ac{pomdp} dynamics for \texttt{eQMARL} using $\Psi^{+}$ (blue), $\Psi^{-}$ (orange), $\Phi^{+}$ (green), $\Phi^{-}$ (red), and $\texttt{None}$ (cyan) entanglement averaged over 10 runs of 3,000 epochs, with $\pm 1$ std.\ dev.\ shown as shaded regions. These figures generally show that $\Psi^{+}$ outperforms other entanglement styles across both dynamics.}
    \label{fig:fig_maa2c_entanglement_compare}
    % \vspace*{-0.8em}
    \vspace*{-0.6em}
\end{figure}

The first set of experiments demonstrate the effectiveness of various input entanglement styles used in \ac{ours} approach. We run two separate experiments using the \texttt{CoinGame-2} environment using both \ac{mdp} and \ac{pomdp} dynamics. The score metric results for both dynamics are shown in \cref{fig:fig_maa2c_entanglement_compare}. 
We consider score threshold markers of 20 and 25 to aid our discussion.
In the \ac{mdp} setting of \cref{fig:fig_maa2c_entanglement_compare:mdp}, we see see that $\Psi^{+}$ entanglement converges $4.5\%$ faster to a score threshold of 20 compared to the next closest $\Psi^{-}$.
Similarly, $\Phi^{+}$ converges $5.2\%$ faster to a score of 25 compared to the next closest $\Psi^{-}$.
At the end of training, all peak scores hover slightly above 25.
Looking at the shaded standard deviation regions, we get a sense for the stability of each entanglement style. Specifically, we see that $\Psi^{+}$, $\Psi^{-}$, and $\Phi^{+}$ have similar tight ranges until epoch 1500, whereas both $\Phi^{-}$ and $\texttt{None}$ have far lower minimum values until around epoch 1300. Moreover, $\Psi^{-}$ appears to have large downward spikes toward the end of training.
\Cref{fig:fig_maa2c_entanglement_compare:mdp} shows that there is a gap in convergence between $\Phi^{-}$ and $\texttt{None}$, and the other styles. 
Looking closer, we observe that $\Phi^{+}$ plateaus at earlier epochs, and $\Psi^{-}$ is more unstable (dropping in score) at later epochs.
Hence, \emph{we see a clear advantage for applying $\Psi^{+}$}.
In the \ac{pomdp} setting of \cref{fig:fig_maa2c_entanglement_compare:pomdp}, we see that $\Psi^{+}$ converges $2\%$ faster to a score of 20 compared to the next closest $\mathtt{None}$.
Interestingly, there is a much larger gap in convergence between $\Phi^{-}$ and the others.
A score of 25 is achieved $10.7\%$ faster by $\Psi^{+}$ compared to $\mathtt{None}$, whereas both $\Phi^{+}$ and $\Phi^{-}$ never reach this threshold. 
The final peak scores for $\Psi^{+}$ and $\mathtt{None}$ hover slightly above 26.
The shaded standard deviation regions exhibit a cascade effect between the styles, and, in particular, we observe that $\Phi^{-}$ has the lowest min, followed by $\Phi^{+}$ which has a slightly higher floor.
These groupings are interesting as both $\Phi^{+}$ and $\Phi^{-}$ are similar in composition, only differing by a phase.
Hence, \emph{we again see a clear convergence and score advantage for using $\Psi^{+}$}.

% Cross-referencing
Comparing the performance of the entanglement styles with both dynamics paints a picture of the generalizability of the system as a whole.
Interestingly, the worse performance of $\Phi^{+}$ and $\Phi^{-}$ suggests that same-state entanglement, $\ket{00}$ and $\ket{11}$, regardless of phase, results in less coupling of agents compared to opposite-state entanglement, $\ket{01}$ and $\ket{10}$. 
One explanation for why the performance of $\mathtt{None}$ is similar to $\Psi^{+}$ in certain cases could be the degradation of fidelity (i.e., entanglement strength) within the agent \acp{vqc}. The circular entanglement unitary within an agent's \ac{vqc}, defined as $U_{\textrm{circ}}$ from \cref{eqn:U_circ}, binds the behavior of qubits within an agent by creating a ``weakly'' entangled state (i.e., low fidelity) from the preceding $U_{\textrm{var}}$. Introducing additional input entanglement could, in some cases, lower the fidelity of that entangled state further, resulting in poor model performance. It also could increase the fidelity, however, as similar to the process of entanglement distillation. We believe this decrease in fidelity is the reason why states like $\Phi^{+}$ and $\Phi^{-}$ perform poorly in most cases. Based on this, we also believe $\Psi^{+}$ performs better than $\Psi^{-}$ because of the decoherence associated with the difference in phase of the $|10\rangle$ term. Importantly, this polarization-dependent phase shift changes the structure of the entangled state entirely. Indeed, this phase is affected by the downstream $U_{\textrm{circ}}$, and thus results in entirely different measurement outcomes.
The consistently high performance of $\Psi^{+}$ across both dynamics suggests that it enhances the generalizability of the system, and, since input entanglement does not increase classical computational overhead, we see that $\Psi^{+}$ entanglement can be used to couple the agents in both dynamics while achieving comparable or higher performance. Thus, \emph{we select $\Psi^{+}$ as the entanglement scheme to be used in all subsequent experiments}.

%%%%%%%%%%%%%%%%

\subsection{\texttt{CoinGame} experiments}\label{sec:exp:results:coingame}

\begin{figure*}[t]
    % \vspace*{-0.8em}
    % \vspace*{-1em}
    \bufferspacefromheader
    \centering
    \begin{subfigure}{0.40\linewidth}
        \centering
        \includegraphics[width=\linewidth]{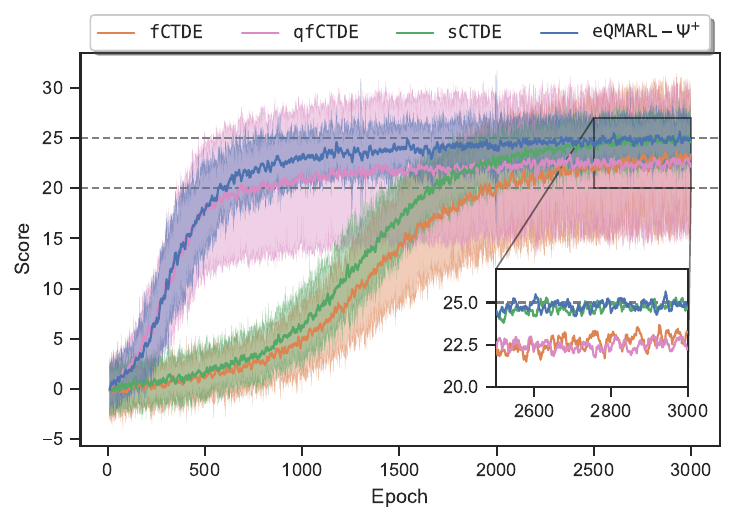}
        
        \vspace*{-0.1em}
        \caption{Score - \ac{mdp}}
        \vspace*{-0.25em}
        \label{fig:fig_maa2c_mdp:undiscounted_reward}
    \end{subfigure}%
    % \hfill%
    \begin{subfigure}{0.40\linewidth}
        \centering
        \includegraphics[width=\linewidth]{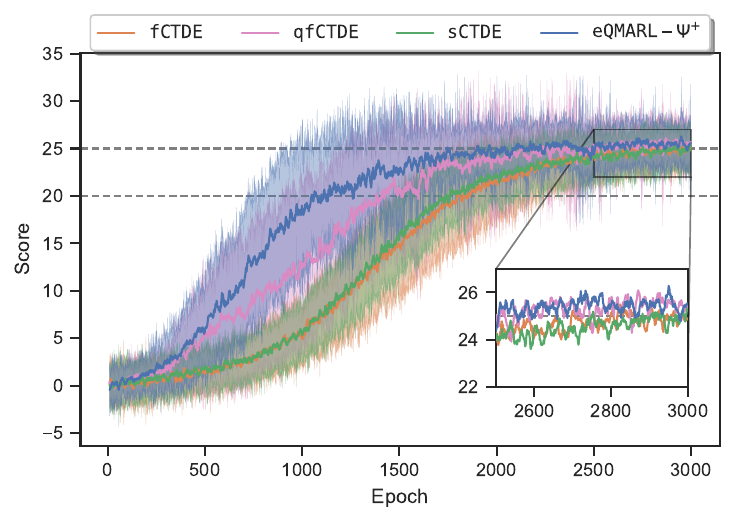}
        
        \vspace*{-0.1em}
        \caption{Score - \ac{pomdp}}
        \vspace*{-0.25em}
        \label{fig:fig_maa2c_pomdp:undiscounted_reward}
    \end{subfigure}
    
   \caption{Comparison of \texttt{CoinGame-2} score performance with (a) \ac{mdp}, and (b) \ac{pomdp} dynamics for \texttt{fCTDE} (orange), \texttt{qfCTDE} (magenta), \texttt{sCTDE} (green), and \texttt{eQMARL-$\Psi^{+}$} (blue) averaged over 10 runs of 3,000 epochs, with $\pm 1$ std.\ dev.\ shown as shaded regions. These figures generally show that \ac{ours} outperforms baselines across both environment dynamics.}
   \label{fig:fig_maa2c_mdp_pomdp}
   % \vspace*{-0.8em}
   \vspace*{-0.6em}
\end{figure*}

We next compare the performance of \texttt{eQMARL-$\Psi^{+}$} with baselines \texttt{fCTDE}, \texttt{sCTDE}, and \texttt{qfCTDE} using the \texttt{CoinGame-2} environment with \ac{mdp} and \ac{pomdp} state dynamics, as shown in \cref{fig:fig_maa2c_mdp:undiscounted_reward,fig:fig_maa2c_pomdp:undiscounted_reward}.
Looking at the \ac{mdp} score metric in \cref{fig:fig_maa2c_mdp:undiscounted_reward}, we see that \texttt{eQMARL-$\Psi^{+}$} converges \emph{$16.2\%$ faster to a score threshold of 20} than the next-closest \texttt{qfCTDE}, and \emph{$10.8\%$ faster to a score threshold of 25} compared to \texttt{sCTDE}.
Overall, we observe a $1.4\%$ increase in max score for \texttt{eQMARL-$\Psi^{+}$} compared to the next highest \texttt{sCTDE}.
Additionally, \texttt{eQMARL-$\Psi^{+}$} is smoother than \texttt{qfCTDE} at later epochs; suggesting that input entanglement stabilizes convergence.
\Cref{fig:fig_maa2c_pomdp:undiscounted_reward} shows that \texttt{eQMARL-$\Psi^{+}$} converges $24\%$ faster to a score of 20 considering the noticeable gap between it and \texttt{qfCTDE}.
This demonstrates that the branching quantum network with input entanglement shortens convergence time compared to the fully centralized variant.
\Cref{fig:fig_maa2c_pomdp:undiscounted_reward} also shows that all models achieve a score of 25, however, in this case, \texttt{eQMARL-$\Psi^{+}$} converges $17.8\%$ faster and with slightly higher score than \texttt{qfCTDE}.
Examining smoothness, we see much greater fluctuation between all curves compared to the \ac{mdp} case. 
The difference in performance between the baselines in \cref{fig:fig_maa2c_mdp:undiscounted_reward,fig:fig_maa2c_pomdp:undiscounted_reward} demonstrates a clear quantum advantage for learning in the presence of partial information. Specifically, the faster convergence to the peak score threshold in \texttt{eQMARL-$\Psi^{+}$} shows that splitting the quantum critic across the quantum channel with entangled input qubits allows the agents to learn a more cooperative strategy without direct access to local observations.
This is interesting because \texttt{qfCTDE} is centralized and has all local observations at its disposal.
This additional information would initially suggest better performance compared to approaches with only local information. However, from \cref{fig:fig_maa2c_pomdp:undiscounted_reward}, we observe a clear benefit for not only splitting the quantum critic as branches across the agents, but also introducing an entangled input state that couples their encoding behavior.
Hence, from \cref{fig:fig_maa2c_mdp:undiscounted_reward,fig:fig_maa2c_pomdp:undiscounted_reward}, we conclude that \emph{our proposed \texttt{eQMARL-$\Psi^{+}$} configuration learns to play the game significantly faster than the classical variants} without sharing local observations, transmitting intermediate activations, nor tuning large \acsp{nn} at the central server. The higher performance and shorter convergence time, compared to both the quantum and classical baselines, shows that splitting the critic amongst the agents results in no apparent loss in capability. In fact, the smoothness of the curves suggests that the input entanglement stabilizes the network over time.

%%%%%%%%%%%%%%%%%%%%%%%%%%%%%%%%%%%%%%%%%%%%%%%%%%%%%%%
\subsection{\texttt{CartPole} Experiments}\label{sec:exp:results:cartpole}

\begin{figure*}[t]
    % \vspace*{-0.8em}
    % \vspace*{-1em}
    \bufferspacefromheader
    \centering
    \begin{subfigure}{0.40\linewidth}
        \centering
        \includegraphics[width=\linewidth]{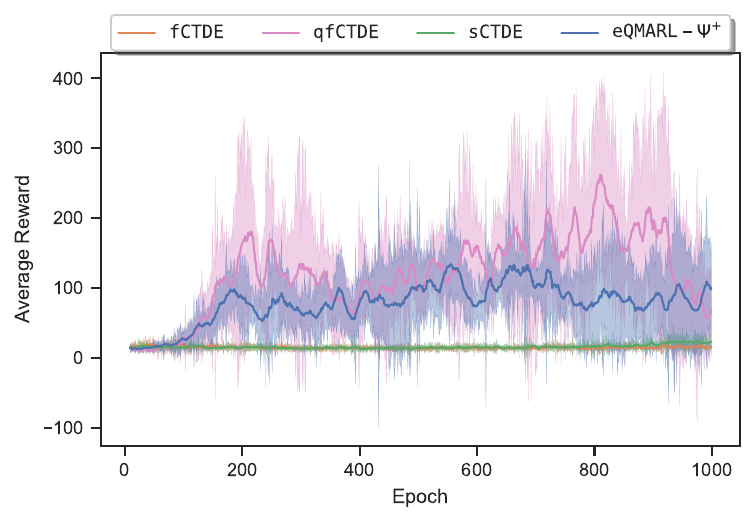}
        
        \vspace*{-0.1em}
        \caption{Average Reward - \ac{mdp}}
        \vspace*{-0.25em}
        \label{fig:fig_cartpole_maa2c_mdp:reward_mean}
    \end{subfigure}%
    % \hfill%
    \begin{subfigure}{0.40\linewidth}
        \centering
        \includegraphics[width=\linewidth]{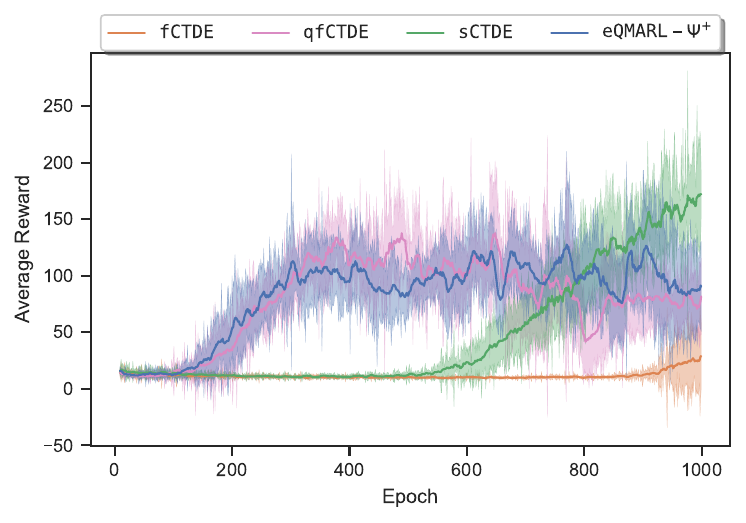}
        
        \vspace*{-0.1em}
        \caption{Average Reward - \ac{pomdp}}
        \vspace*{-0.25em}
        \label{fig:fig_cartpole_maa2c_pomdp:reward_mean}
    \end{subfigure}
   \caption{Comparison of \texttt{CartPole} \ac{mdp} and \ac{pomdp} environment average reward performance for \texttt{fCTDE} (orange), \texttt{qfCTDE} (magenta), \texttt{sCTDE} (green), and \texttt{eQMARL-$\Psi^{+}$} (blue) averaged over 5 runs of 1,000 epochs, with $\pm 1$ std.\ dev.\ shown as shaded regions. These figures generally show that \ac{ours} outperforms classical baselines and is more stable than \texttt{qfCTDE} across both dynamics.}
   \label{fig:fig_cartpole_maa2c_mdp_pomdp}
   \vspace*{-0.8em}
\end{figure*}

The next set of experiments compare the performance of \ac{ours} with baselines using a multi-agent variant of the \texttt{CartPole} environment with both \ac{mdp} and \ac{pomdp} state dynamics. The average reward metric across all agents in the environment for both dynamics is shown in \cref{fig:fig_cartpole_maa2c_mdp_pomdp}. We use reward thresholds of the \emph{mean} and \emph{max} values to draw comparisons.
From \cref{fig:fig_cartpole_maa2c_mdp_pomdp} we see that the classical models do not perform well overall in either setting, and \texttt{qfCTDE} experiences high variance in the \ac{mdp} case. 
For \ac{mdp}, \texttt{qfCTDE} achieves the highest mean and max rewards overall, but with an extremely high standard deviation. In contrast, \texttt{eQMARL-$\Psi^{+}$} reaches its mean and max rewards $12.2\%$ and $31.5\%$ faster than \texttt{qfCTDE} respectively.
For \ac{pomdp}, we see that \texttt{sCTDE} achieves the highest overall max reward at the end of training, but with a very low mean. In contrast, both \texttt{qfCTDE} and \texttt{eQMARL-$\Psi^{+}$} achieve a similar max value significantly earlier, about two-thirds of the way through training. \texttt{eQMARL-$\Psi^{+}$} achieves the highest mean reward overall very early in training, which is $9.1\%$ faster than \texttt{qfCTDE} with a similarly low standard deviation to the \ac{mdp} setting.
The key observation from the \texttt{CartPole} experiments is that \texttt{eQMARL-$\Psi^{+}$} more rapidly learns a strategy with higher average reward than the classical variants in both \ac{mdp} and \ac{pomdp} settings. Further, \texttt{eQMARL-$\Psi^{+}$} slightly outperforms \texttt{qfCTDE} in the \ac{pomdp} setting with a higher average reward, and is much more stable with a significantly lower variance in the \ac{mdp} setting -- all achieved via implicit collaboration though entanglement. These results show that, without observation sharing, \texttt{eQMARL-$\Psi^{+}$} can learn a similarly performant, and dramatically more stable, strategy compared to a fully centralized quantum approach that has access to all agent observations.

%%%%%%%%%%%%%%%%%%%%%%%%%%%%%%%%%%%%%%%%%%%%%%%%%%%%%%%

%%%%%%%%%%%%%%%%%%%%%%%%%%%%%%%%%%%%%%%%%%%%%%%%%%%%%%%
%%%
% MiniGrid experiment.
%%%
%%%%%%%%%%%%%%%%%%%%%%%%%%%%%%%%%%%%%%%%%%%%%%%%%%%%%%%
\subsection{\texttt{MiniGrid} Experiment}\label{sec:exp:results:minigrid}

\begin{figure*}[t]
    \vspace*{-0.8em}
    % \bufferspacefromheader
    \centering
    \includegraphics[width=0.40\linewidth]{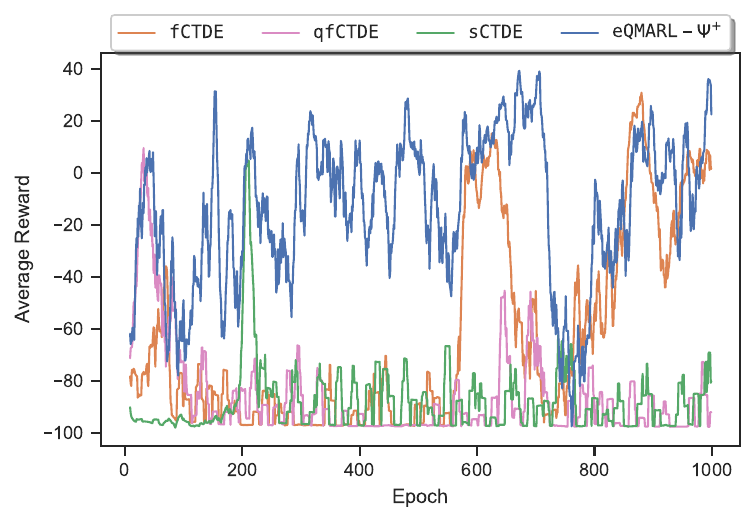}
    \vspace*{-0.5em}
    \caption{Comparison of \texttt{MiniGrid} \ac{pomdp} environment reward performance for \texttt{fCTDE} (orange), \texttt{qfCTDE} (magenta), \texttt{sCTDE} (green), and \texttt{eQMARL-$\Psi^{+}$} (blue) over 1000 epochs. This figure shows that \ac{ours} outperforms baselines by learning direct goal navigation instead of spinning in place.}
    \label{fig:fig_minigrid_maa2c}
    % \vspace*{-0.8em}
    \vspace*{-0.6em}
\end{figure*}

The next experiment compares the performance of \ac{ours} with baselines using a multi-agent variant of the \texttt{MiniGrid} environment with \ac{pomdp} state dynamics, in which agents have a limited field of view. The average reward metric across all agents in the environment is shown in \cref{fig:fig_minigrid_maa2c}. From \cref{fig:fig_minigrid_maa2c} we can see that, for the vast majority of training, the \texttt{fCTDE}, \texttt{qfCTDE}, and \texttt{sCTDE} baselines have an average reward that is clustered near $-100$; meaning that they exhaust many steps by simply spinning in place (since the maximum step size is 50, and $-2$ is the same-position reward). In contrast, we see that the average reward of \texttt{eQMARL-$\Psi^{+}$} is spread out higher over the training regime with a mean of about $-13$, which is nearly 4.5-times higher than the other baselines. Indeed, this negative reward means that \texttt{eQMARL-$\Psi^{+}$} also expends actions turning in place, but the fact the reward is so close to zero implies these events occur at a vastly reduced frequency than the baselines. In testing, \texttt{eQMARL-$\Psi^{+}$} was able to traverse to the goal in as little as 9 steps, whereas \texttt{fCTDE} required 17 steps, and both \texttt{qfCTDE} and \texttt{sCTDE} were unable to find the goal within the 50 step limit. This is a marked 50\% improvement in the exploration and navigation speed of \texttt{eQMARL-$\Psi^{+}$} over \texttt{fCTDE}, with the bonus of no observation sharing. Hence, we have shown that \texttt{eQMARL-$\Psi^{+}$} can indeed be applied to more complex environments, such as grid-world navigation with limited visibility, and provides learning benefits over baselines all without the need for observation sharing.

%%%%%%%%%%%%%%%%%%%%%%%%%%%%%%%%%%%%%%%%%%%%%%%%%%%%%%%

%%%%%%%%%%%%%%%%%%%%%%%%%%%%%%%%%%%%%%%%%%%%%%%%%%%%%%%
%%%
% Ablation study.
%%%
%%%%%%%%%%%%%%%%%%%%%%%%%%%%%%%%%%%%%%%%%%%%%%%%%%%%%%%
\subsection{Ablation Study}\label{sec:exp:results:ablation}

\begin{figure*}[t]
    % \vspace*{-0.8em}
    % \vspace*{-1em}
    \bufferspacefromheader
    \centering
    \begin{subfigure}{0.40\linewidth}
        \centering
        \includegraphics[width=\linewidth]{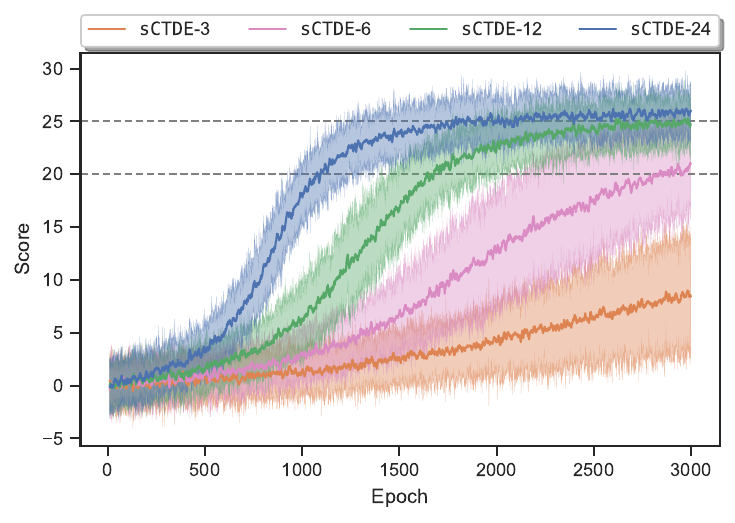}
        \caption{\texttt{sCTDE}}
        \vspace*{-0.25em}
    \end{subfigure}%
    \begin{subfigure}{0.40\linewidth}
        \centering
        \includegraphics[width=\linewidth]{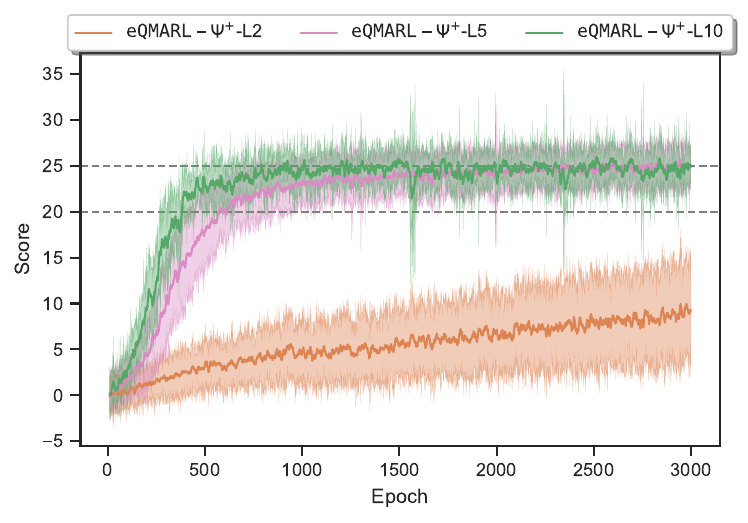}
        \caption{\texttt{eQMARL-$\Psi^{+}$}}
        \vspace*{-0.25em}
    \end{subfigure}
    \caption{Ablation study score performance on \ac{mdp} \texttt{CoinGame-2} for (a) sCTDE, and (b) eQMARL-$\Psi^{+}$ with hidden layer units $h \in \{3,6,12,24\}$ and VQC layers $L \in \{2,5,10\}$, averaged over 10 runs of 3000 epochs, with $\pm 1$ std.\ dev.\ shown as shaded regions. These figures generally show that selecting parameters $h=12$ and $L=5$ results in optimal performance.}
    \label{fig:fig_maa2c_mdp_ablation:score}
    % \vspace*{-0.8em}
    \vspace*{-1em}
\end{figure*}

The last set of experiments we consider is an ablation study to examine the relationship between \ac{nn} layer depth and performance, and to facilitate fair model size comparisons. In particular, we trained \texttt{fCTDE} and \texttt{sCTDE} with hidden layer units $h \in \{3,6,12,24\}$, and \texttt{qfCTDE} and \texttt{eQMARL} with VQC layers $L \in \{2,5,10\}$ on the \texttt{CoinGame-2} environment with \ac{mdp} and \ac{pomdp} state dynamics for 10 experiments of 3000 epochs each. An excerpt of the score metrics results for \texttt{eQMARL} and \texttt{sCTDE} in the \ac{mdp} setting are shown in \cref{fig:fig_maa2c_mdp_ablation:score}, and a comparison of the critic model sizes used in each framework is shown in \cref{tab:model_size_comparison}. The full results are provided in \cref{app:exp:ablation}.
In \cref{fig:fig_maa2c_mdp_ablation:score} we see that \texttt{eQMARL-$\Psi^{+}$} with $L=5$ achieves a mean score 3-times higher than $L=2$, and nearly identical to $L=10$. This trend is similar for \texttt{qfCTDE}. Both \texttt{sCTDE} and \texttt{fCTDE} also exhibit similar behavior for hidden units; that is the performance of $h=12$ is nearly 2-times that of $h=6$, and only marginally less than $h=24$.
Considering the significant performance drops and increased variation incurred by reducing, and the limited gains by increasing, both $h$ and $L$, the selection of $h=12$ and $L=5$ results in the most comparable performance across all baselines.
Looking at \cref{tab:model_size_comparison}, the critic sizes reported are for the entire system. This distinction is important since both \texttt{eQMARL} and \texttt{sCTDE} split the critic network across the agents the total size of the agent-specific network is a fraction of the total size.
For \ac{mdp} dynamics, we observe that the quantum models are 4 times smaller than the classical variants.
For \ac{pomdp}, we observe that the quantum models are slightly larger than their classical counterparts. This is because, in \ac{pomdp}, the quantum models require a classical \acs{nn} for dimensionality reduction at the input of each encoder. While the overall system size is larger, however, the number of central parameters is significantly reduced in the quantum cases -- requiring only 1 parameter instead of 25. 
This is important for scaleability because the baselines implement a full \acs{nn} at the central server and its size scales with $N$. In contrast, \ac{ours} has only a single trainable parameter tied to the measurement observable, which will remain fixed regardless of $N$.

\begin{table}[t]
    \vspace*{-1.8em}
    \caption{Comparison of the best critic model size in number of trainable parameters for each framework used on the \texttt{CoinGame-2} environment with \ac{mdp} and \ac{pomdp} dynamics.}
    \scriptsize
    \label{tab:model_size_comparison}
    \centering
    \begin{tabular}{llll}
        \toprule
        Framework & Ablation Selection                          & Number of critic parameters: \ac{mdp}  & Number of critic parameters: \ac{pomdp} \\
        \midrule
        \texttt{eQMARL} & $L=5$    & 265 (132 per agent, 1 central)                     & 817 (408 per agent, 1 central) \\
        \texttt{qfCTDE} & $L=5$    & 265                       & 817 \\
        \texttt{fCTDE} & $h=12$     & 889                       & 673 \\
        \texttt{sCTDE} & $h=12$     & 913 (444 per agent, 25 central)                    & 697 (336 per agent, 25 central) \\
        \bottomrule
    \end{tabular}
\end{table}
\section{Conclusion}\label{sec:conclusion}

In this paper we have proposed \ac{ours}, a novel quantum actor-critic framework for training decentralized policies using a split quantum critic with entangled input qubits and joint measurement.
Spreading the critic across the agents via a quantum channel eliminates sharing local observations, minimizes the classical communication overhead from sending model parameters or intermediate \acs{nn} activations, and reduces the centralized classical computational burden through optimization of a single quantum measurement observable parameter.
We have shown that $\Psi^{+}$ input entanglement improves agent cooperation and system generalizability across both \ac{mdp} and \ac{pomdp} environments.
For \ac{mdp}, we have shown that \ac{ours} converges to a cooperative strategy $10.8\%$ faster and with a higher score compared to \ac{sctde}.
Likewise, for \ac{pomdp}, we have shown that \ac{ours} converges to a cooperative strategy $17.8\%$ faster and with slightly higher score compared to \ac{qfctde}.
Further, we have also shown that \ac{ours} outperforms classical baselines and exhibits more stable performance than \ac{qfctde} in independent environments.
Lastly, we have shown that \ac{ours} requires $25$-times fewer centralized parameters compared to \ac{sctde}.
One limitation of this work is the computational complexity of quantum simulations on classical hardware, which is an ongoing topic of research for \ac{nisq} systems of many qubits. 
Indeed, many recent works on \ac{qmarl} \citep{Yun2022QuantumMultiAgentReinforcement,Yun2023QuantumMultiAgentActorCritic,Yun2022QuantumMultiAgentMeta,Chen2023AsynchronousTrainingQuantum,Park2023QuantumMultiAgentReinforcement,Kolle2024MultiAgentQuantumReinforcement} have similar hardware requirements to ours.
Further, many recent works on quantum networks \citep{VanMeter2022QuantumInternetArchitecture,Pettit2023PerspectivePathwayScalable,Lei2023QuantumOpticalMemory,Azuma2023QuantumRepeatersQuantum} propose methods for generating and storing entangled qubits, which can support the type of entanglement required in our system.

%%%%%%%%



% \bibliography{references/references_manual.bib}
\bibliography{main.bib}
\bibliographystyle{iclr2025_conference}

%%%%%%%%%%%%%%%%%%%%%%%%%%%%%%%%%%%%%%%%%%%%%%%%%%%%%%%%%%%%
% \newpage % Should this be a newpage?

\appendix

\renewcommand\thefigure{\thesection.\arabic{figure}} 
\renewcommand\thetable{\thesection.\arabic{table}} 
\renewcommand\theequation{\thesection.\arabic{equation}} 
\renewcommand\thealgocf{\thesection.\arabic{algocf}}

\section{Related Works}\label{app:related}

\Ac{qmarl} is a nascent field, with few works applying the quantum advantage to scenarios with multiple agents \citep{Yun2022QuantumMultiAgentReinforcement,Yun2023QuantumMultiAgentActorCritic,Yun2022QuantumMultiAgentMeta,Chen2023AsynchronousTrainingQuantum,Park2023QuantumMultiAgentReinforcement,Kolle2024MultiAgentQuantumReinforcement}. Further, the application of quantum to \ac{sl} is even newer, with \citet{Yun2023QuantumSplitNeural} being the only prior work.
In \citet{Yun2022QuantumMultiAgentReinforcement} and \citet{Yun2023QuantumMultiAgentActorCritic}, the authors 
propose a novel approach to \ac{qmarl} that integrates \ac{ctde} and quantum state encoding of environment states. Specifically, they deploy actor-critic \ac{qrl} with \acp{vqc} as the core for both actor and critic architectures. The \ac{qmarl} agents each deploy a local quantum actor network and learn collectively in \ac{ctde} fashion using a centralized critic and unified experience replay buffer. In particular, the localized actors each operate independently from the others in the network and yet critically must also share their environment state experience with the collective via the shared replay buffer.

In \citet{Yun2022QuantumMultiAgentMeta}, the authors propose the \ac{qm2arl} framework which uses a central meta Q-learning agent to train other local agents. In particular they define angle and pole training, a technique which tunes classical parameters to describe the Bloch sphere orientation of encoded quantum information. The first angle parameter training stage describes mapped Q-values as regions on the Bloch sphere, which are used to create the centralized meta agent. This meta agent is then subsequently used to train local agents by orienting pole parameters toward their specific environments.

In \citet{Chen2023AsynchronousTrainingQuantum}, the author proposes \ac{qa3c} as a framework for training decentralized \ac{qrl} agents, which leverages a \emph{global shared memory} and \emph{agent-specific memories} used in conjunction to train parallel agents. Akin to classical \ac{fl}, \ac{qa3c} deploys an actor-critic network within each decentralized agent. The gradients of these local networks are periodically sent to the global shared memory for aggregation into a global model. The global model parameters are then broadcast back to the agents to update their local models.

The work in \citet{Park2023QuantumMultiAgentReinforcement} proposes a \ac{qmarl} approach for autonomous mobility cooperation using actor-critic networks with \ac{ctde} in \ac{nisq} environments with a shared replay buffer. In particular, they consider deployment with \ac{nisq} era limitations, the most notable being a low number of qubits. They show that \ac{ctde}-based \ac{qmarl} can be deployed on near-term quantum hardware to coordinate robotic agents in smart factory environments. Critically, the coordination within their proposed system stems from the centralized critic network and shared replay buffer that has become synonymous with \ac{ctde} frameworks.

In \cite{Kolle2024MultiAgentQuantumReinforcement}, the authors propose a \ac{qmarl} approach using evolutionary optimization with a \ac{vqc} design based on quantum classification networks and agent policies implemented as independent \ac{vqc} models with shared local information. Variations of the parameters for these policies are trained in \emph{populations} to evaluate and compare performance. The authors assume fully observable environments in both cooperative and competitive settings, which is influenced via changes in the per-agent reward.

Finally, in \cite{Yun2023QuantumSplitNeural}, the authors propose a method for applying split learning to \ac{qml} for traditional \ac{ml} classification tasks. They deploy device-local \acp{qnn} to predict local labels and features which are transmitted classically to a central server and transformed into true set of labels and localized gradients for each device.

The resounding theme in \citet{Yun2022QuantumMultiAgentReinforcement,Yun2023QuantumMultiAgentActorCritic,Yun2022QuantumMultiAgentMeta,Chen2023AsynchronousTrainingQuantum,Park2023QuantumMultiAgentReinforcement,Kolle2024MultiAgentQuantumReinforcement,Yun2023QuantumSplitNeural} is the use of independent agents or branches that communicate and learn through centralized classical means.
No prior work, however, makes use of the quantum channel as a medium for system coupling or for multi-agent collaboration. 
Indeed, the quantum elements serve as drop-in replacements for classical \ac{nn} counterparts, and, importantly, the quantum channel between agents and the potential for sharing entangled qubit states go largely under-utilized. 
Simply put, entanglement and the quantum channel are potentially useful untapped cooperative resources intrinsic to \ac{qmarl} that have largely unknown benefits.

%%%%%%%

\section{More comprehensive preliminaries}\label{app:prelim}
\setcounter{figure}{0}
\setcounter{table}{0}
\setcounter{equation}{0}

\subsection{Quantum mutli-agent reinforcement learning}

We consider a \ac{rl} setting with multiple agents in environments with both full and partial information. The dynamics of a system with full information is described by a Markov game with an underlying \ac{mdp} with tuple $M_{\textrm{MDP}} = \langle \mathcal{N}, \mathcal{S}, \mathcal{A}, \mathcal{P}, \mathcal{R} \rangle$ where $\mathcal{N}$ is a set of $N$ agents, $\mathcal{S}=\{\mathcal{S}^{(n)}\}_{n\in\mathcal{N}}$ is the set of joint states across all agents, $\mathcal{A} = \{\mathcal{A}^{(n)}\}_{n \in \mathcal{N}}$ is the set of joint actions, $\mathcal{S}^{(n)}$ and $\mathcal{A}^{(n)}$ are the set of states and actions for agent $n$, $\mathcal{P}(s_{t+1} | s_t, a_t)$ is the state transition probability, and $\mathcal{R}(s_t, a_t) = \{r^{(n)}_t\}_{n \in \mathcal{N}} \in \mathbb{R}^{N}$ is the joint reward $\forall s_t \in \mathcal{S}, a_t \in \mathcal{A}$.
The dynamics of a system with partial information is described by a Markov game with an underlying \ac{pomdp} with tuple $M_{\textrm{POMDP}} = \langle \mathcal{N}, \mathcal{S}, \mathcal{A}, \mathcal{P}, \mathcal{R}, \Omega, \mathcal{O} \rangle$ where $\mathcal{N}, \mathcal{S}, \mathcal{A}, \mathcal{P}$ are the same as in $M_{\textrm{MDP}}$, however the full state of the environment $s_t$ at time $t$ is kept hidden from the players. Instead, at time $t$ the agents receive a local observation from the set of joint observations $\Omega=\{\Omega^{(n)}\}_{n\in\mathcal{N}}$, where $\Omega^{(n)}$ is the set of observations for agent $n$, with transition probability $\mathcal{O}(o_{t} | s_{t+1}, a_{t})$, $\forall o_{t}\in\Omega$, which is dependent on the hidden environment state after taking a joint action.
We treat $M_{\textrm{MDP}}$ as a special case of $M_{\textrm{POMDP}}$ where $o^{(n)}_t = s^{(n)}_t$, that is the observations represent the full environment state information. Hereinafter, all notations will use $o^{(n)}_{t}$ in place of the local environment state for brevity to encompass all cases.

\Ac{qmarl} is the application of quantum computing to \ac{marl}. A popular approach in \ac{marl} is through actor-critic architectures, which tune policies, called \emph{actors}, via an estimator for how good or bad the policy is at any given state of an environment, called a \emph{critic}. To do this, the critic needs access to the local agent environment observations to estimate the value for a particular environment state. Current state of the art approaches follow the \ac{ctde} framework which deploys the critic on a central server and the actors across decentralized agents. Because the critic and the agents are physically separated, \ac{ctde} requires the agents to transmit their local observations to the server for the critic to estimate the joint value, thereby publicizing potentially private local observations.
Quantum is often integrated as a drop-in replacement for classical \acp{nn}, called \acp{vqc}, within many \ac{marl} systems. These trainable quantum circuits tune the state of \emph{qubits}, the quantum analog of classical bits, using unitary \emph{gate} operations. 
%%%%%%%%%%%%%%%%%%%%%%%%%%%%

\subsection{Quantum computation}

%%%%%%%%%%%%%%%
\subsubsection{Qubit states}
A \emph{qubit} is the quantum mechanical analog to the classical bit. The state of a qubit is represented as a 2-dimensional unit vector in complex Hilbert space $\mathcal{H} \in \mathbb{C}^2$. 
The computational basis is the set of states  $\left\{ \ket{0}=\begin{bmatrix} 1 & 0 \end{bmatrix}^T, \ket{1}=\begin{bmatrix} 0 & 1 \end{bmatrix}^T \right\}$ which forms a complete and orthonormal basis in $\mathcal{H}$ (meaning $\braket{0}{1}=\braket{1}{0}=0$ and $\braket{0}{0}=\braket{1}{1}=1$). 
All qubit states can be expressed as a linear combination of any complete and orthonormal basis, such as the computational basis, which is called \emph{superposition}.
We adopt Dirac notation to describe arbitrary qubit states $\ket{\psi} = \begin{bmatrix} \alpha & \beta \end{bmatrix}^T = \alpha \ket{0} + \beta \ket{1} \in \mathcal{H}$ (called ``ket psi'') where $|\alpha|^2 + |\beta|^2 = 1$, their conjugate transpose $\bra{\psi} = \ket{\psi}^\dagger = \begin{bmatrix} \alpha^* & \beta^* \end{bmatrix} = \alpha^* \bra{0} + \beta^* \bra{1}$ (called ``bra psi''), the inner product $\braket{\psi_1}{\psi_2} = \alpha_1^* \alpha_2 + \beta_1^* \beta_2$, and the outer product $\ketbra{\psi_1}{\psi_2} = \begin{bmatrix} \alpha_1 \alpha_2^* & \alpha_1 \beta_2^* \\ \beta_1 \alpha_2^* & \beta_1 \beta_2^* \end{bmatrix}$.
Quantum systems with $D$ qubits can also be represented by extending the above notation using the Kronecker (tensor) product where $\mathcal{H} = \bigotimes_{d=0}^{D-1} \mathcal{H}_d = (\mathbb{C}^2)^{\otimes D}$ is the complex space of the system state $\ket{\psi} = \bigotimes_{d=0}^{D-1} \ket{\psi_d}$ for all $\ket{\psi_D} \in \mathcal{H}_D$.
States that can be represented as either a single ket vector, or a sum of basis states are called \emph{pure} states. For example, $\ket{0}$, $\ket{1}$, and $\frac{1}{\sqrt{2}}(\ket{0} + \ket{1})$ are all pure states in $\mathcal{H}$.

\subsubsection{Quantum gates}
A quantum \emph{gate} is an unitary operator (or matrix) $U$, such that $U U^\dagger = \mathbb{I}$, where $\mathbb{I}$ is the identity matrix, acting on the space $\mathcal{H}$ which maps between qubit states. Here, we use the single-qubit Pauli gates 
\begin{equation}
    X = \begin{bmatrix} 0&1\\1&0 \end{bmatrix},~
    Y = \begin{bmatrix} 0&-i\\i&0 \end{bmatrix},~
    Z = \begin{bmatrix} 1&0\\0&-1 \end{bmatrix},
\end{equation}
with their parameterized rotations 
\begin{equation}
    R_X(\theta) = e^{-i\frac{\theta}{2} X},~
    R_Y(\theta) = e^{-i\frac{\theta}{2} Y},~
    R_Z(\theta) = e^{-i\frac{\theta}{2} Z},
\end{equation}
where $\theta \in \mathbb{R} [0, 2\pi]$, the Hadamard gate 
\begin{equation}
    H = \frac{1}{\sqrt{2}}\begin{bmatrix} 1&1\\1&-1 \end{bmatrix},
\end{equation}
the 2-qubit controlled-$X$ ($CX$, also called $\textrm{CNOT}$) gate 
\begin{equation}
    {CX}_{1,2} = {\textrm{CNOT}}_{1,2} = \begin{bmatrix} \mathbb{I} & \mathbf{0} \\ \mathbf{0} & X \end{bmatrix},
\end{equation}
and the controlled-$Z$ ($CZ$) gate
\begin{equation}
    {CZ}_{1,2} = \begin{bmatrix} \mathbb{I} & \mathbf{0} \\ \mathbf{0} & Z \end{bmatrix},
\end{equation}
which are both controlled by qubit 1 and target qubit 2, where $\mathbf{0}$ is a $2 \times 2$ square matrix of zeros.
%%%%%%%%%%%%%%%

\subsubsection{Entanglement}

% What is entanglement?
Consider two arbitrary quantum systems $A$ and $B$, represented by Hilbert spaces $\mathcal{H}_{A}$ and $\mathcal{H}_{B}$. We can represent the Hilbert space of the combined system using the tensor product $\mathcal{H}_{A}\otimes\mathcal{H}_{B}$. If the quantum states of the two systems are $\ket{\psi}_{A}$ and $\ket{\psi}_{B}$, then the state of the combined system can be represented as $\ket{\psi}_{A}\otimes\ket{\psi}_{B}$. Quantum states that can be cleanly represented in this form, i.e., separated by tensor product, are said to be \emph{separable}. Not all quantum states, however, are separable. For example, if we fix a set of basis states $\left\{\ket{0}_{A},\ket{1}_{A}\right\}\in\mathcal{H}_{A}$ and $\left\{\ket{0}_{B},\ket{1}_{B}\right\}\in\mathcal{H}_{B}$, then a general state in the space of $\mathcal{H}_{A}\otimes\mathcal{H}_{B}$ can be represented as
% \begin{equation}
$
    \ket{\psi} = \sum_{a\in\{0,1\}} \sum_{b\in\{0,1\}} c_{a,b} \ket{a}_{A} \otimes \ket{b}_{B},
$
% \end{equation}
which is separable if there exists $c_{a,c} = c_{a} c_{b}$, $\forall a,b\in\{0,1\}$, producing the isolated states $\ket{\psi}_{A} = \sum_{a\in\{0,1\}} c_{a} \ket{a}_{A}$ and $\ket{\psi}_{B} = \sum_{b\in\{0,1\}} c_{b} \ket{b}_{B}$. If, however, there exists one $c_{a,c} \neq c_{a} c_{b}$, then the combined state is inseparable. In such cases, if a state is inseparable, it is said to be \emph{entangled}.

The four Bell states $\mathcal{B}=\left\{\ket{\Phi^{+}}_{AB}, \ket{\Phi^{-}}_{AB}, \ket{\Psi^{+}}_{AB}, \ket{\Psi^{-}}_{AB}\right\}$ form a complete basis for two-qubit systems $\mathcal{H}_{A}\otimes\mathcal{H}_{B}$, and have the form:
\begin{align}
    \ket{\Phi^{+}}_{AB} &= \frac{1}{\sqrt{2}}\left(\ket{0}_{A}\ket{0}_{B}+\ket{1}_{A}\ket{1}_{B}\right),\\
    \ket{\Phi^{-}}_{AB} &= \frac{1}{\sqrt{2}}\left(\ket{0}_{A}\ket{0}_{B}-\ket{1}_{A}\ket{1}_{B}\right),\\
    \ket{\Psi^{+}}_{AB} &= \frac{1}{\sqrt{2}}\left(\ket{0}_{A}\ket{1}_{B}+\ket{1}_{A}\ket{0}_{B}\right),\\
    \ket{\Psi^{-}}_{AB} &= \frac{1}{\sqrt{2}}\left(\ket{0}_{A}\ket{1}_{B}-\ket{1}_{A}\ket{0}_{B}\right).
\end{align}
Since it is impossible to separate the states of $\mathcal{B}$ into individual systems $\mathcal{H}_{A}$ and $\mathcal{H}_{B}$, the four Bell states are entangled. In particular, the Bell states are pure entangled states of the combined system $\mathcal{H}_{A}\otimes\mathcal{H}_{B}$, but cannot be separated into pure states of systems $\mathcal{H}_{A}$ and $\mathcal{H}_{B}$.
Additionally, the four Bell states can be generated by quantum circuits using a combination of Pauli operators with a constant $\ket{0}_{A}\ket{0}_{B}$ input state as follows:
\begin{align}
    \ket{\Phi^{+}}_{AB} &= \textrm{CNOT}_{1,2} \left(H\otimes\mathbb{I}\right) \ket{0}_{A}\ket{0}_{B}, \label{app:eqn:bell:phi+}\\
    \ket{\Phi^{-}}_{AB} &= \textrm{CNOT}_{1,2} \left(H\otimes\mathbb{I}\right) \left(X\otimes\mathbb{I}\right) \ket{0}_{A}\ket{0}_{B}, \label{app:eqn:bell:phi-}\\
    \ket{\Psi^{+}}_{AB} &= \textrm{CNOT}_{1,2} \left(H\otimes\mathbb{I}\right) \left(\mathbb{I}\otimes X\right) \ket{0}_{A}\ket{0}_{B}, \label{app:eqn:bell:psi+}\\
    \ket{\Psi^{-}}_{AB} &= \textrm{CNOT}_{1,2} \left(H\otimes\mathbb{I}\right) \left(X\otimes X\right) \ket{0}_{A}\ket{0}_{B} \label{app:eqn:bell:psi-}
\end{align}

\subsubsection{Projective measurements and observables}

A projective measurement is a Hermitian and unitary operator $O$, such that $O = O^\dagger$ and $O O^\dagger = \mathbb{I}$, called an \emph{observable}. The outcomes of a measurement are defined by an observable's spectral decomposition
\begin{equation}
    O = \sum_{m=0}^{M-1} \lambda_{m} P_{m},
\end{equation}
where $M = 2^n$ represents the number of measurement outcomes for $n$ qubits, and $m$ is a specific measurement outcome in terms of eigenvalues $\lambda_{m}$ and orthogonal projectors $P_{m}$ in the respective eigenspace. According to the \emph{Born rule} \cite{Born1926,Logiurato2012BornRuleNoncontextual,Masanes_2019}, the outcome of measuring an arbitrary state $\ket{\psi}$ will be one of the eigenvalues $\lambda_{m}$, and the state will be projected using the operator $P_{m} / \sqrt{p(m)}$ with probability:
\begin{equation}
    p(m) = \bra{\psi} P_{m} \ket{\psi} = \expval{P_{m}}_{\psi} \label{def:projective_measurement:eqn:expval_projection}
\end{equation}
The expected value of the observable with respect to the arbitrary state $\ket{\psi}$ is given by:
\begin{equation}
    \mathbb{E}_{\psi}[O] = \sum_{m=0}^{M-1} \lambda_{m} p(m) = \expval{O}_{\psi} \label{def:projective_measurement:eqn:expval_observable}
\end{equation}

%%%%%%% DO NOT NEED COMMUTING OBSERVABLES FOR JOINT CRITIC.

\subsubsection{Commuting Observables}

A set of observables $\{O_1, \dots, O_{K}\}$ share a common eigenbasis (i.e., a common set of eigenvectors with unique eigenvalues) if
\begin{equation}
    [O_i, O_j] = O_i O_j - O_j O_i = 0 ~~~~\forall i,j \in [1,K]
\end{equation}
i.e., their pair-wise commutator is zero. In such cases the observables in the set are said to be \emph{pair-wise commuting}, which in practice means that all observables in the set can be measured at the same time.

%%%%%%%%%%%%%%%%%%%%%
% \newpage
\section{Joint input entanglement circuits}\label{app:ent}
\setcounter{figure}{0}
\setcounter{table}{0}
\setcounter{equation}{0}

\begin{figure}[h]
    \vspace*{-1.5em}
    \centering
    \begin{subfigure}{0.74\linewidth}
        \centering
        \includegraphics[width=\linewidth]{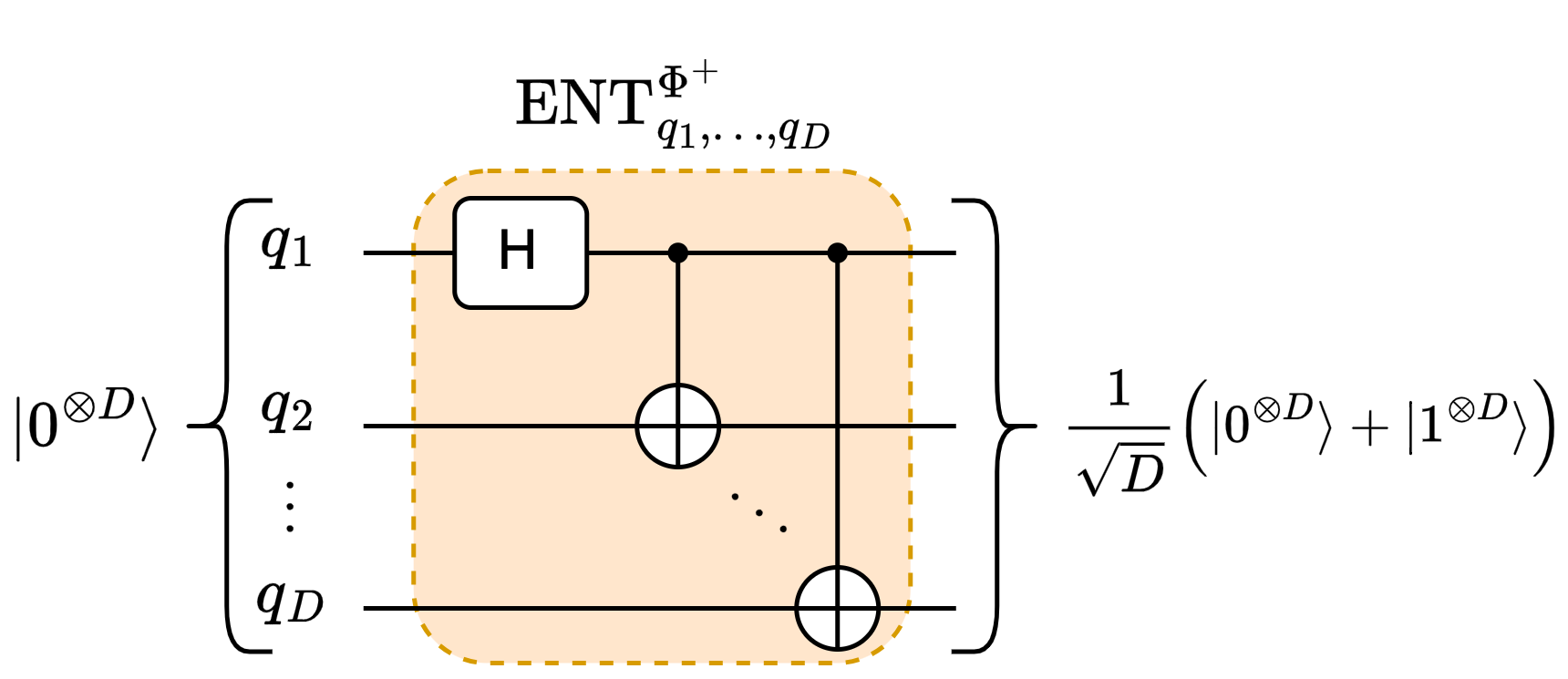}
        \vspace{0.1em}
        \caption{$\Phi^{+}$}
        \label{app:fig:joint_entanglement_circuit_phi+}
    \end{subfigure}\\
    \begin{subfigure}{0.74\linewidth}
        \centering
        \includegraphics[width=\linewidth]{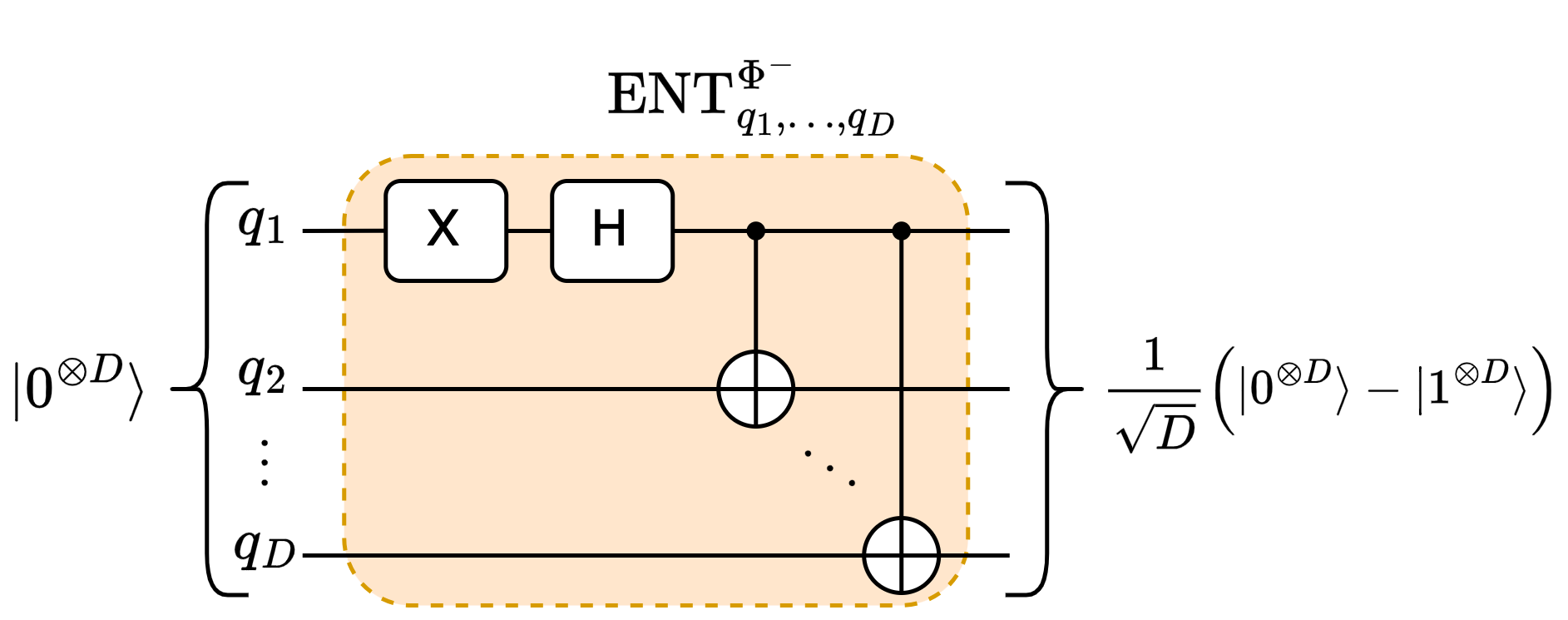}
        \vspace{0.1em}
        \caption{$\Phi^{-}$}
        \label{app:fig:joint_entanglement_circuit_phi-}
    \end{subfigure}\\
    \begin{subfigure}{0.74\linewidth}
        \centering
        \includegraphics[width=\linewidth]{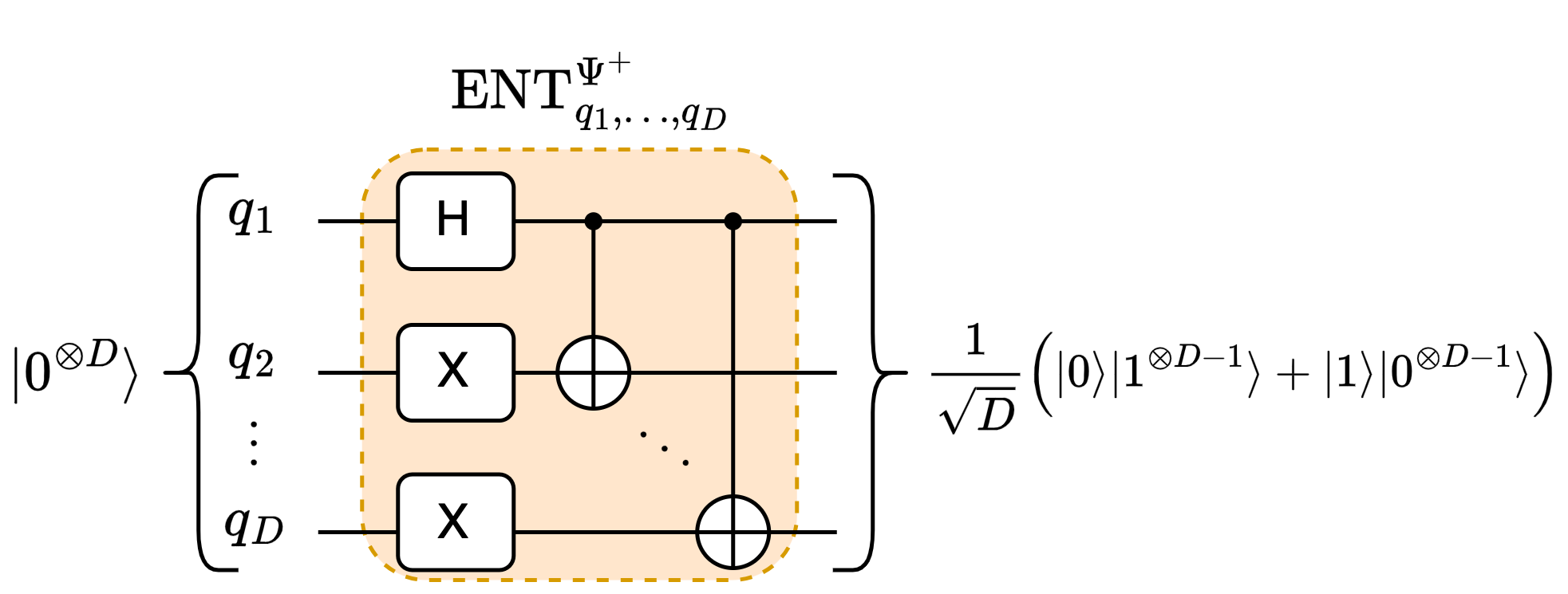}
        \vspace{0.1em}
        \caption{$\Psi^{+}$}
        \label{app:fig:joint_entanglement_circuit_psi+}
    \end{subfigure}\\
    \begin{subfigure}{0.74\linewidth}
        \centering
        \includegraphics[width=\linewidth]{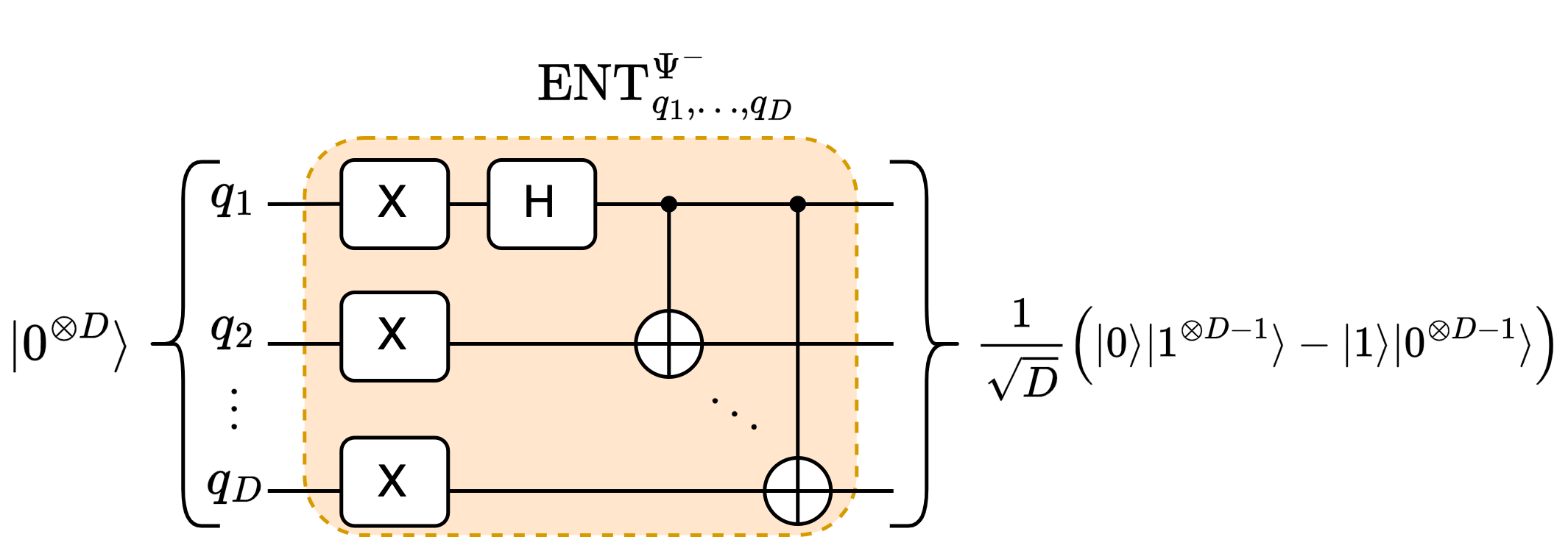}
        \vspace{0.1em}
        \caption{$\Psi^{-}$}
        \label{app:app:fig:joint_entanglement_circuit_psi-}
    \end{subfigure}
   \caption{Diagrams of joint entanglement operators based on the the four Bell states (a) $\Phi^{+}$, (b) $\Phi^{-}$, (c) $\Psi^{+}$, (d) $\Psi^{-}$.}
   \label{app:fig:joint_entanglement_circuit}
\end{figure}

We use a variation of Bell state entanglement to couple the input qubits of the agent \ac{vqc} branches. Specifically, we entangle based on the set of four Bell states $\mathcal{B} \in \{\Phi^{+}, \Phi^{-}, \Psi^{+}, \Psi^{-}\}$, as outlined in \cref{app:eqn:bell:phi+,app:eqn:bell:phi-,app:eqn:bell:psi+,app:eqn:bell:psi-}, using the circuits as shown in \cref{app:fig:joint_entanglement_circuit}. 
The circuits in \cref{app:fig:joint_entanglement_circuit} generate a quantum state across $D$ qubits, which has the combined Hilbert space $\mathcal{H}^{\otimes{D}}$.
In each of the entangled operators, we select one qubit, $q_{1}$, to serve as the master control, and all others, $q_2,\dots,q_D$, serve as targets. The control qubit functions identically to canonical Bell state entanglement. Here, we extend the gate operations that normally apply to the second qubit, in an $\mathcal{H}\otimes\mathcal{H}$ system, to all qubits in $\mathcal{H}^{\otimes{D-1}}$. The resulting state is an entangled pure state in $\mathcal{H}^{\otimes{D}}$.

%%%%%%%%%%%%%%%%%%%%%%%%%%%%%%%%%%%%%%%%%%%
% \newpage
\section{Full algorithm}\label{app:algorithm}
\setcounter{figure}{0}
\setcounter{table}{0}
\setcounter{equation}{0}
\setcounter{algocf}{0}

The following algorithm is an expanded version of \cref{alg:maa2c-summary}, as shown in \cref{alg:maa2c}. In \cref{alg:maa2c}, we include all operations necessary for training both the agents and the split critic. 
In \ac{ours}, there is a set of $N$ quantum agents $\mathcal{N}$ that are physically separated from each other (no cross-agent communication is assumed) and one central quantum server. Each agent $n\in\mathcal{N}$ employs an actor policy network $\pi_{\bm{\theta}_{\textrm{actor}}^{(n)}}(a | \bm{o}_{t}^{(n)})$ (which can either classical or quantum in nature) with parameters $\bm{\theta}_{\textrm{actor}}^{(n)}$, and a \ac{vqc} given by \cref{eqn:U_vqc} with unique parameters $\bm{\theta}_{\textrm{critic}}^{(n)}$ and $\bm{\lambda}^{(n)}$ that serves as one branch in the split critic network. 
In our experiments we simplify the setup by using policy sharing across the agents, as done in \cite{Yun2023QuantumMultiAgentActorCritic} and \cite{Chen2023AsynchronousTrainingQuantum}; in other words, $\bm{\theta}^{(n)}_{\textrm{actor}}=\bm{\theta}^{(k)}_{\textrm{actor}}~\forall n,k\in\mathcal{N}$. All agents interact with the environment independently and each has its own local data buffer $\mathcal{D}^{(n)}$ populated with local observations, actions, rewards, and next observations represented by the 4-tuple $\left( \bm{o}_{t}^{(n)},  a_{t}^{(n)}, r_{t}^{(n)}, \bm{o}_{t+1}^{(n)} \right)$ for any instant in time $t$. These local data are not shared amongst agents, with only the reward and action being communicated classically to the central quantum server (the action only being necessary for policy sharing).
Since we employ policy sharing, the final step in \ac{ours} with this in place is to also estimate the advantage value $A(\bm{o}_{\tau}, \bm{a}_{\tau}) = Q(\bm{o}_{\tau}, \bm{a}_{\tau}) - V(\bm{o}_{\tau})$ using the existing value and expected reward estimates, compute actor loss $\mathcal{L}_{\textrm{actor}}$, compute the gradient of the loss w.r.t~the shared actor parameters $\nabla_{\bm{\theta}_{\textrm{actor}}} \mathcal{L}_{\textrm{actor}}$, and update $\bm{\theta}_{\textrm{actor}}$. Note that here we use \texttt{Huber} loss for the critic, and for the actors we use entropy-regularized advantage loss. The loss functions are described in detail in \cref{app:loss}.

\begin{algorithm}[h]
  \caption{Full \ac{ours} training using \acrshort{maa2c} for a quantum entangled split critic.}\label{alg:maa2c}
  \footnotesize
  \KwRequire{Set of $N$ agents $\mathcal{N}$, 1 quantum entanglement source, and 1 quantum measurement server}
  Initialize $N$ agent actor networks with shared parameters $\bm{\theta}_{\textrm{actor}}$ and local replay buffer $\mathcal{D}^{(n)}=\{\}$, $\forall n \in \mathcal{N}$\;
  Initialize $N$ critic branches $U_{\textrm{vqc}}$ with $D$ qubits and parameters $\bm{\theta}_{\textrm{critic}}^{(n)}$, $\bm{\lambda}^{(n)}$, $\forall n \in \mathcal{N}$\;
  \For{episode=1, MaxEpisode}{
      \Comment{Localized environment interaction.}
      $t=0$\;\label{alg:maa2c:local_start}
      $done = False$\;
      \While{$done \neq True$ and $t < max\ steps$}{
          \For{each agent $n\in\mathcal{N}$}{
              Get local observation $\bm{o}_{t}^{(n)}$ from environment\;
              Compute $\pi_{\bm{\theta}_{\textrm{actor}}^{(n)}}(a | \bm{o}_{t}^{(n)})$ and sample $a_{t}^{(n)}$\;
              Apply action $a_{t}^{(n)}$ and get reward $r_{t}^{(n)}$ and next observation $\bm{o}_{t+1}^{(n)}$\;
              Update local replay buffer $\mathcal{D}^{(n)} = \mathcal{D}^{(n)} \cup \left\{\left( \bm{o}_{t}^{(n)},  a_{t}^{(n)}, r_{t}^{(n)}, \bm{o}_{t+1}^{(n)} \right)\right\}$\;
              If $\bm{o}_{t+1}^{(n)}$ is terminal then communicate $done=True$\;
          }
          $t=t+1$\;
      }\label{alg:maa2c:local_end}
      \Comment{\ac{ours} framework for training.}
      \For{$\tau \in [0,t-2]$}{\label{alg:maa2c:quantum_start}
          Central quantum server generates $2N$ sets of $D$ entangled qubits and transmits to agents via \textbf{quantum channel} (could be prepared a priori and stored in quantum memory)\;
          \For{each agent $n\in\mathcal{N}$}{
              Apply $U_{\textrm{vqc}}(\bm{\theta}^{(n)}_{\textrm{critic}},\bm{\lambda}^{(n)},\bm{o}_{\tau}^{(n)})$ and $U_{\textrm{vqc}}(\bm{\theta}^{(n)}_{\textrm{critic}},\bm{\lambda}^{(n)},\bm{o}_{\tau+1}^{(n)})$ from \cref{eqn:U_vqc} locally using assigned entangled input qubits\;
              Transmit via \textbf{quantum channel} the qubits after applying $U_{\textrm{vqc}}$ to central quantum server\;
              Transmit via \textbf{classical channel} the reward $r_{\tau}^{(n)}$ and action $a_{\tau}^{(n)}$ at the current time step to central quantum server (only reward if policy sharing is not used)\;
          }
          \Comment{Using notation $\bm{o}_{\tau}=\bigl(\bm{o}_{\tau}^{(n)}\bigr)_{n=1}^{N}$, $\bm{o}_{\tau+1}=\bigl(\bm{o}_{\tau+1}^{(n)}\bigr)_{n=1}^{N}$, $\bm{a}_{\tau}=\bigl(a_{\tau}^{(n)}\bigr)_{n=1}^{N}$.}
          Perform joint measurements on qubits across all agents to estimate $V(\bm{o}_{\tau})$ and $V(\bm{o}_{\tau+1})$ using \cref{eqn:value-estimate}\;
          Estimate $Q(\bm{o}_{\tau}, \bm{a}_{\tau}) = \sum_{n=1}^{N}{r_{\tau}^{(n)}} + \gamma V(\bm{o}_{\tau+1})$ using discount factor $\gamma$\;
          Estimate $A(\bm{o}_{\tau}, \bm{a}_{\tau}) = Q(\bm{o}_{\tau}, \bm{a}_{\tau}) - V(\bm{o}_{\tau})$ for policy sharing\;
          Compute $\nabla_{\bm{\theta}_{\textrm{actor}}} \mathcal{L}_{\textrm{actor}}$ and update $\bm{\theta}_{\textrm{actor}}$ for policy sharing\;
          Compute $\nabla_{x_{\textrm{split}}} \mathcal{L}_{\textrm{critic}}$ and transmit via \textbf{classical channel} to each agent to update $\bm{\theta}_{\textrm{critic}}^{(n)}$ locally using partial gradient from \cref{eqn:joint-critic-local-gradient}\;
      }\label{alg:maa2c:quantum_end}
  }
  \end{algorithm}

%%%%%%%%%%%%%%%%%%%%%%%%%%%%%%%%%%%%%%%%%%%

%%%%%%%%%%%%%%%%%%%%%%%%%%%%%%%%%%%%%%%%%%%
\section{Loss functions}\label{app:loss}
\setcounter{figure}{0}
\setcounter{table}{0}
\setcounter{equation}{0}

All of our actors and critics are trained using the same loss functions for each experiment. 
For the critics, we train using \texttt{Huber} loss
\begin{equation}
    \mathcal{L}_{\textrm{critic}} = \frac{1}{T-1} \sum_{\tau=0}^{T-1} \begin{cases}
        \frac{1}{2} (V(\bm{o}_{\tau}) - Q(\bm{o}_{\tau}, \bm{a}_{\tau}))^2 & \textrm{if $V(\bm{o}_{\tau}) - Q(\bm{o}_{\tau}, \bm{a}_{\tau}) \leq \delta$}, \\[0.5em]
        \delta |V(\bm{o}_{\tau}) - Q(\bm{o}_{\tau}, \bm{a}_{\tau})| - \frac{1}{2}\delta^2 & \textrm{otherwise},
    \end{cases}
\end{equation}
where $\delta$ controls the point in which the loss function turns from quadratic to linear. In this work we use $\delta=1$.
For the actors, we deploy policy sharing amongst the agents. As such, all agents us the same policy parameters, and thus the loss must aggregate the individual losses of each agent. We train using the entropy-regularized advantage function
\begin{equation}
    \mathcal{L}_{\textrm{actor}} = \frac{1}{n(T-1)} \sum_{n=1}^{N} \sum_{\tau=0}^{T-1} \Big[-A(o^{(n)}_{\tau}, a^{(n)}_{\tau}) \log_{e}{p(a^{(n)}_{\tau})} + \alpha H(a^{(n)}_{\tau}) \Big],
\end{equation}
where
\begin{equation}
    H(a^{(n)}_{\tau})=- p(a^{(n)}_{\tau}) \log_{e}{p(a^{(n)}_{\tau})},
\end{equation}
is the entropy of selecting an action, $\alpha$ controls the influence of entropy, $n\in\mathcal{N}$ is the agent index, and $p(a^{(n)}_{\tau})$ is the probability of chosen action at time step $\tau$.
%%%%%%%%%%%%%%%%%%%%%%%%%%%%%%%%%%%%%%%%%%%

%%%%%%%%%%%%%%%%%%%%%%%%%%
% \newpage
\section{Environment specifications}\label{app:env}
\setcounter{figure}{0}
\setcounter{table}{0}
\setcounter{equation}{0}
\setcounter{algocf}{0}

\subsection{\texttt{CoinGame}}\label{app:env:coin_game}

\begin{figure}[h]
    \vspace*{-2.5em}
     \centering
     \begin{minipage}[b]{0.4\linewidth}
         \begin{subfigure}{\linewidth}
             \centering
             \includegraphics[width=.8\linewidth]{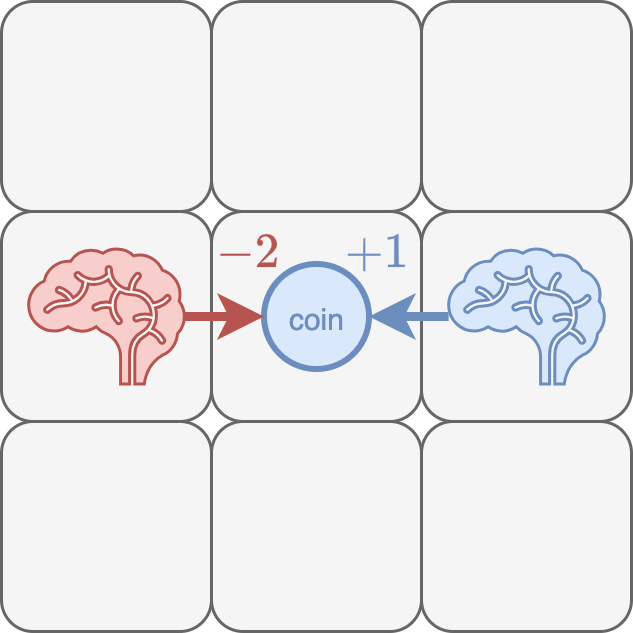}
             \caption{Diagram of \texttt{CoinGame-2} environment.}
             \label{fig:coin_game_diagram}
         \end{subfigure}
     \end{minipage}
     \hfill
     \begin{minipage}[b]{0.59\linewidth}
        \centering
         \begin{subfigure}{\linewidth}
             \centering
             \includegraphics[width=.75\linewidth]{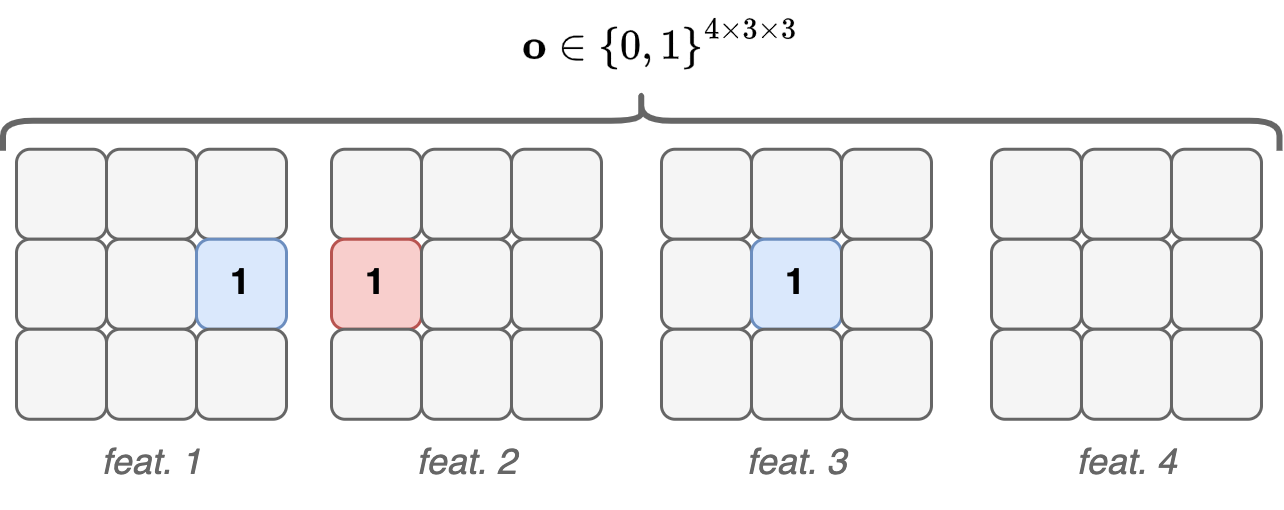}
             \caption{\acrshort{mdp} observation dynamics for blue player.}
             \label{fig:coin_game_obs_mdp}
         \end{subfigure}
         \vfill
         \begin{subfigure}{\linewidth}
             \centering
             \includegraphics[width=.58\linewidth]{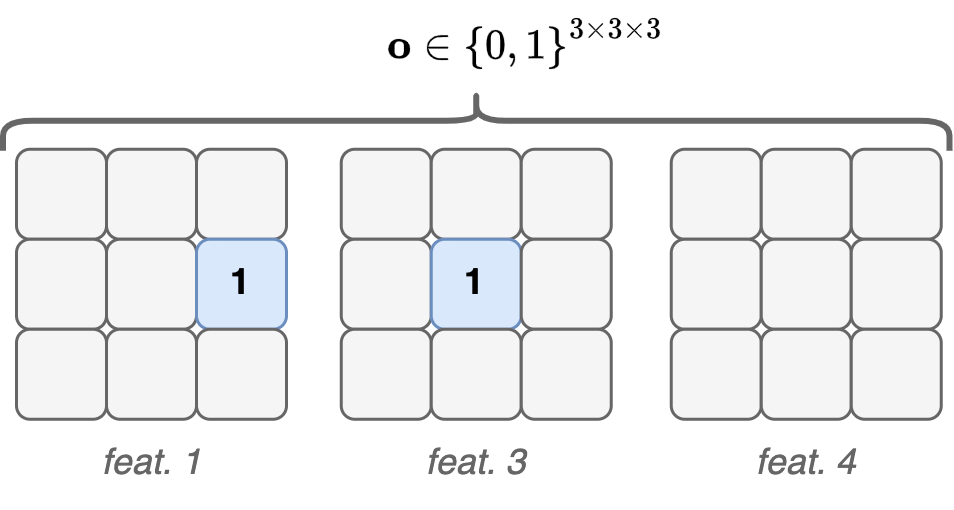}
             \caption{\acrshort{pomdp} observation dynamics for blue player.}
             \label{fig:coin_game_obs_pomdp}
         \end{subfigure}
     \end{minipage}
    \caption{Example of (a) \texttt{CoinGame-2} environment with two players, colored red and blue, and a single coin colored blue, with visualization of observation matrix for the blue player with (b) \acrshort{mdp} dynamics, and (c) \acrshort{pomdp} dynamics. Grid squares in (b) and (c) colored grey denote a $0$ value, and colored blue/red squares denote a $1$ value.}
    \label{fig:coin_game}
    \vspace*{-0.8em}
\end{figure}

In \ac{ours}, we train decentralized agents using the \texttt{CoinGame-2} environment first proposed in \cite{Lerer2018MaintainingCooperationComplex}, and as implemented in \cite{Phan2022EmergentCooperationMutual}. The \texttt{CoinGame-2} environment pits two agents of different colors (red and blue) on a $3\times3$ tile grid to collect coins with corresponding color. An example of \texttt{CoinGame-2} is shown in \cref{fig:coin_game_diagram}.
Agent observations are a sparse matrix $\mathbf{o}^{(n)} \in \{0,1\}^{4\times3\times3}$ $\forall n \in \mathcal{N}$ with 4 features each with a $3\times3$ grid world as shown in \cref{fig:coin_game_obs_mdp}. The features specifically are: 
\begin{enumerate*}[label=\arabic*)]
    \item A grid with a $1$ indicating the agent's location,
    \item A grid with a $1$ to indicate other agent locations,
    \item A grid with a $1$ for the location of coins that match the current agent's color, and
    \item A grid with a $1$ for all other coins (different colors).
\end{enumerate*}
Since these observations include all information about the game world, the game is considered fully observable and is described by an \ac{mdp}. We also experiment with a partially observable variant of this game which removes the second feature from agent observations (i.e., the location of other agents), which is a space matrix $\mathbf{o}^{(n)} \in \{0,1\}^{3\times3\times3}$ $\forall n \in \mathcal{N}$. In this partially observed setting, the game is described by a \ac{pomdp} since agents cannot see each other and thus the full state of the game board is unknown. An example of this observation space is shown in \cref{fig:coin_game_obs_pomdp}.
The agents can move along the grid by taking actions in the space $\mathcal{A}^{(n)}=\{\textrm{north},\textrm{south},\textrm{east},\textrm{west}\}$ $\forall n\in\mathcal{N}$. Each time an agent collects a coin of their corresponding color their episode reward is increased by $+1$, whereas a different color reduces their episode reward by $-2$. The goal for all agents is to maximize their discounted episode reward. 
The details of the environment are summarized in \cref{tab:env_spec}.

We evaluate agents using three metrics: \emph{score}, \emph{total coins collected}, and \emph{own coin rate}. The \emph{score} metric aggregates all agent undiscounted rewards over a single episode
\begin{equation}
    S = \sum_{n=1}^{N} \sum_{t=0}^{T-1} r^{(n)}_{t} \label{app:eqn:env:score}
\end{equation}
where $n\in\mathcal{N}$ is the agent index, $t\in[0,T-1]$ is the episode time index, $T$ is the episode time limit, and $r^{(n)}_{t}$ is the undiscounted agent reward at time $t$.
The \emph{total coins collected} metric gives insight into how active the agents were during the game
\begin{equation}
    TC = \sum_{n=1}^{N} \sum_{t=0}^{T-1} c^{(n)}_{t} \label{app:eqn:env:tc}
\end{equation}
where $c^{(n)}_{t}$ is the total number of coins collected by agent $n$ at time $t$.
Finally, the \emph{own coin rate} metric gives insight into how the agents achieve cooperation, specifically by being selective on which coins they procure
\begin{equation}
    OCR = \sum_{n=1}^{N} \sum_{t=0}^{T-1} {k^{(n)}_{t}}/{c^{(n)}_{t}} \label{app:eqn:env:ocr}
\end{equation}
where $k^{(n)}_{t}$ is the number of coins collected of the corresponding agent's color.

\begin{table}[t]
    \caption{Specifications for \texttt{CoinGame-2} environment with \emph{\ac{mdp}} and \emph{\ac{pomdp}} dynamics.}
    \label{tab:env_spec}
    \centering
    \begin{tabular}{ll}
        \toprule
        Parameter & Value \\
        \midrule
        \multirow{2}{*}{Observation for agent $n$ at time $t$} & \textbullet~~\ac{mdp}:~~$\bm{o}^{(n)}_{t} \in \{0,1\}^{4\times3\times3}$ (dimension is $36$) \\
                    & \textbullet~~\ac{pomdp}:~~$\bm{o}^{(n)}_{t} \in \{0,1\}^{3\times3\times3}$ (dimension is $27$) \\
        \cmidrule(lr){1-2}
        Number of players ($N$) & 2 \\
        Time limit ($T$) & 50 \\
        % \cmidrule(lr){1-2}
        Action for agent $n$ at time $t$ & $a^{(n)}_{t} \in \{\textrm{north},\textrm{south},\textrm{east},\textrm{west}\}$\\
        % \cmidrule(lr){1-2}
        Reward for agent $n$ at time $t$ & $r^{(n)}_{t} = \begin{cases}
            +1, & \textrm{if agent $n$ collects coin of same color}, \\
            -2, & \textrm{if agent $n$ collects coin of different color}, \\
            0, & \textrm{otherwise}
        \end{cases}$\\
        Discount factor ($\gamma$) & 0.99 \\
        Entropy coefficient ($\alpha$) & 0.001 \\
        \bottomrule
    \end{tabular}
\end{table}

\subsection{\texttt{CartPole}}\label{app:env:cartpole}

\begin{figure}[h]
    \vspace*{-0.8em}
    \centering
    \includegraphics[width=.6\linewidth]{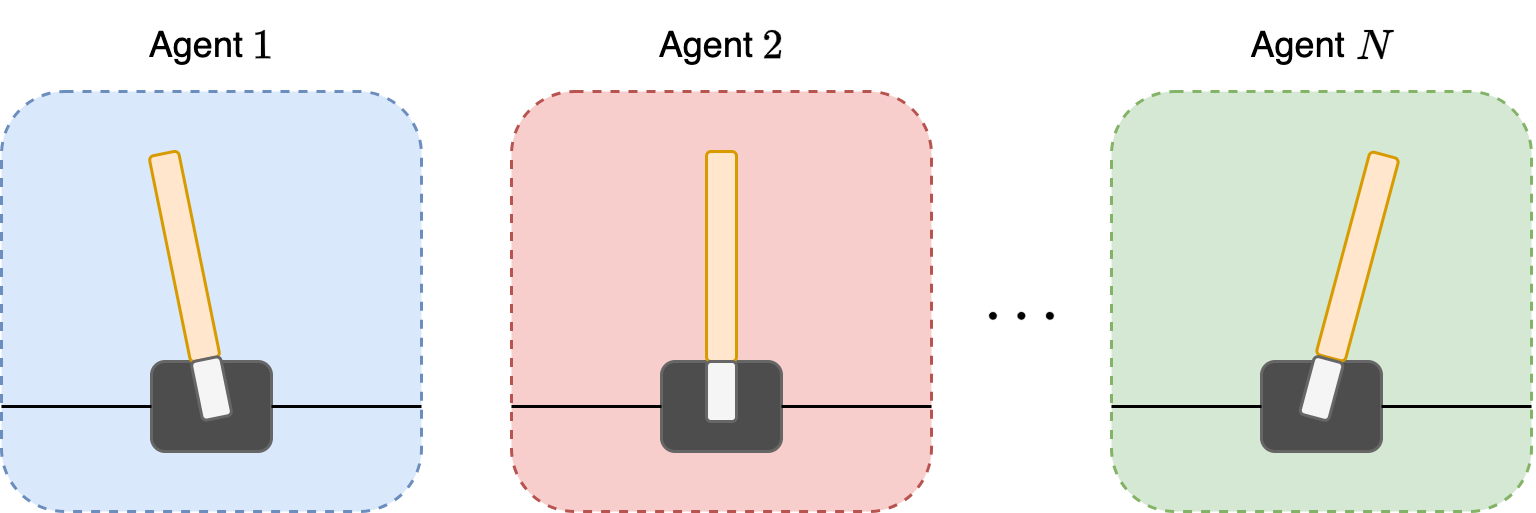}
    \caption{Example of an $N$-agent \texttt{CartPole} environment colored red, blue, and green.}
    \label{fig:multi_agent_cartpole}
    \vspace*{-0.8em}
\end{figure}

We also train \ac{ours} using a multi-agent variant of the \texttt{CartPole} environment as proposed in \citet{Barto1983NeuronlikeAdaptiveElements}. The multi-agent \texttt{CartPole} environment runs multiple independent instances of the single-agent variant in parallel. This environment setup is an interesting case study for multi-agent learning because the observations of each agent are completely independent from one another; that is, observations from sibling environments are not strictly necessary to develop a strategy for a specific environment instance. This allows us to examine the impacts of both explicit and implicit cooperation between independent agents. An example of \texttt{CartPole} with $N$ agents is shown in \cref{fig:multi_agent_cartpole}.
Agent observations are a matrix $\mathbf{o}^{(n)} \in \mathbb{R}^{4 \times 1}$, $\forall n \in \mathcal{N}$ with 4 real-valued features. The features are: 
\begin{enumerate*}[label=\arabic*)]
    \item Cart position with range $[-4.8, 4.8]$,
    \item Cart velocity with range $(-\infty, \infty)$,
    \item Pole angle in \emph{radians} with range $[-0.418, 0.418]$, and
    \item Pole angular velocity with range $(-\infty, \infty)$.
\end{enumerate*}
The pole is considered \emph{balanced} if the pole angle feature stays within the range $[-.2095, .2095]$ radians, and the cart position feature stays within the range $[-2.4, 2.4]$.
These observations include all information about the environment, and thus the environment under these conditions is considered \emph{fully observable} and described by an \ac{mdp}. We also consider a \emph{partially observed} variant of the environment which removes the second feature from agent observations (i.e., the cart velocity), which is a matrix $\mathbf{o}^{(n)} \in \mathbb{R}^{3 \times 1}$, $\forall n \in \mathcal{N}$. The environment is described by a \ac{pomdp} in this setting since agents are unaware of their cart's velocity, and thus the full state of the environment is unknown.
Notably, in this multi-agent variant of the environment the agent observations in both settings are independent from each other.
The agents interact with the environment by taking actions in the space $\mathcal{A}^{(n)} = \{\textrm{left},\textrm{right}\}$, $\forall n \in \mathcal{N}$, which correspond to pushing their cart to the left and right respectively. Similar to the observations, the agent actions are also independent and do not affect neighboring environments.
Each time step an agent is successful in keeping their pole balanced they receive a $+1$ episode reward. The episode terminates when an observation falls outside of the balanced range.
The goal for all agents is to maximize their expected total episode reward (i.e., the number of time steps they are able to keep the pole balanced). The details of the environment are summarized in \cref{tab:env_spec:cartpole}.

We evaluate the agents using the \emph{average reward} metric, which aggregates all agent rewards over a single episode
\begin{equation}
    AR = \frac{1}{N} \sum_{n=1}^{N} \sum_{t=0}^{T-1} r^{(n)}_{t}
\end{equation}
where $n\in\mathcal{N}$ is the agent index, $t\in[0,T-1]$ is the episode time index, $T$ is the episode time limit, and $r^{(n)}_{t}$ is the agent reward at time $t$.

\begin{table}[t]
    \caption{Specifications for multi-agent \texttt{CartPole} environment with \emph{\ac{mdp}} and \emph{\ac{pomdp}} dynamics.}
    \label{tab:env_spec:cartpole}
    \centering
    \begin{tabular}{ll}
        \toprule
        Parameter & Value \\
        \midrule
        \multirow{2}{*}{Observation for agent $n$ at time $t$} & \textbullet~~\ac{mdp}:~~$\bm{o}^{(n)}_{t} \in \mathbb{R}^{4\times1}$ \\
                    & \textbullet~~\ac{pomdp}:~~$\bm{o}^{(n)}_{t} \in \mathbb{R}^{3\times1}$ \\
        \cmidrule(lr){1-2}
        Number of players ($N$) & 2 \\
        Time limit ($T$) & 500 \\
        % \cmidrule(lr){1-2}
        Action for agent $n$ at time $t$ & $a^{(n)}_{t} \in \{\textrm{left},\textrm{right}\}$\\
        % \cmidrule(lr){1-2}
        Reward for agent $n$ at time $t$ & $r^{(n)}_{t} = \begin{cases}
            +1, & \textrm{if pole for agent $n$ is balanced}, \\
            0, & \textrm{otherwise}
        \end{cases}$\\
        \bottomrule
    \end{tabular}
\end{table}

\subsection{\texttt{MiniGrid}}\label{app:env:minigrid}

\begin{figure}[h]
    \vspace*{-0.8em}
    \centering
    \includegraphics[width=.8\linewidth]{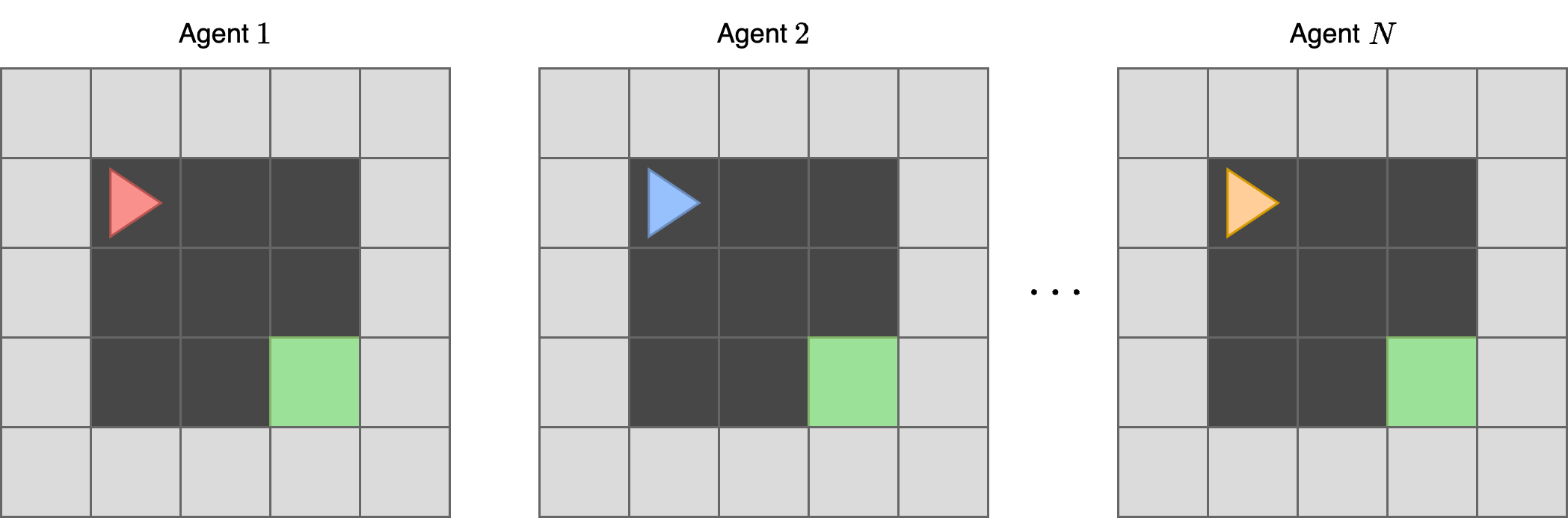}
    \caption{Example of an $N$-agent \texttt{MiniGrid} environment colored red, blue, and orange.}
    \label{fig:multi_agent_minigrid}
    \vspace*{-0.8em}
\end{figure}

We also train \ac{ours} using a multi-agent variant of the \texttt{MiniGrid} environment as proposed in \citet{MinigridMiniworld23}. The multi-agent \texttt{MiniGrid} environment runs multiple instances of the single-agent version in parallel. 
This environment configuration is an interesting case study for multi-agent learning because, similar to multi-agent \texttt{CartPole}, the local agent observations are independent of the others, and sharing observations is not necessary to solve the environment. This can, however, show how agent policies are affected with the seemingly added benefit of either directly shared or quantum-coupled local environment observations. An example of \texttt{MiniGrid} with $N$ agents is shown in \cref{fig:multi_agent_minigrid}, and the details of the environment are summarized in \cref{tab:env_spec:minigrid}.

In the \texttt{MiniGrid} environment, the task is to find an optimal grid traversal path from a starting position to a goal using the action set $\mathcal{A}^{(n)} = \{\texttt{turn left}, \texttt{turn right}, \texttt{move forward}\}$, $\forall n \in \mathcal{N}$. 
Agent observations are a matrix $\mathbf{o}^{(n)} \in \mathbb{Z}^{7 \times 7 \times 3}$, $\forall n \in \mathcal{N}$, where the agent has a $7 \times 7$ limited field of view of the maze grid, and each cell in the grid is encoded as the 3-tuple $\langle \texttt{object}, \texttt{color}, \texttt{state} \rangle$, where $\texttt{object} \in [0, 10]$ identifies the object at the cell (e.g., \texttt{empty}, \texttt{wall}, etc.), $\texttt{color} \in [0, 5]$ identifies the color of the cell (e.g., \texttt{red}, \texttt{green}, etc.), and $\texttt{state} \in [0, 2]$ identifies the state of the cell (e.g., \texttt{open}, \texttt{closed}, and \texttt{locked}).
We consider a $5 \times 5$ maze grid size in our experiments.
Notably, because agent observations are a \emph{limited field of view}, and not the full maze grid, we regard this environment as being described by \ac{pomdp} dynamics.
We use a reward shaping schedule of $-1$ for every step taken, $-2$ for standing still, $0$ for not reaching the goal, and $100 \times (1 - 0.9 \times ((t+1)/T)$ for reaching the goal, where $t \in [0, T-1]$ is episode the time index, and $T$ is the episode time limit. Additionally, we provide the reward bonus $1/\sqrt{c}$ for actions that explore less visited grid positions and action pairs, where $c$ is the count for a specific grid position and action pair.
In particular, there are four key factors that make this environment even more complex than the \texttt{CoinGame} and \texttt{CartPole} baselines. First, agents observe the environment using a limited field of view from their current position and rotational direction; hence, the agents must expend actions to both physically move and visually perceive the environment. Second, because the field of view is limited, the goal position is not always in view; meaning that the agent strategies must also learn to search for the goal position in addition to finding an optimal traversal route. Third, the rotational direction of the agent plays a major role in both grid-world visibility and traversal actions; whereby an agent must learn to optimize the total number of rotation actions (i.e., \texttt{turn left} and \texttt{turn right}) for visual exploration and navigation -- e.g., using a single \texttt{turn right} action (fewer steps) instead of three \texttt{turn left} actions (increased visibility). Fourth, agent actions are not limited when adjacent to grid-world \texttt{wall} positions; meaning that if an agent is facing \texttt{forward} to a \texttt{wall}, then the \texttt{move forward} action is still valid (even though clearly not optimal). Critically, the trade-off between increased visibility at the expense of rotation actions poses complex challenges here, and offers a unique opportunity for implicit observation sharing to improve navigation efficiency.

\begin{table}[t]
    \caption{Specifications for multi-agent \texttt{MiniGrid} environment with \emph{\ac{pomdp}} dynamics.}
    \label{tab:env_spec:minigrid}
    \centering
    \begin{tabular}{ll}
        \toprule
        Parameter & Value \\
        \midrule
        Grid size & $w=5$, $h=5$ \\
        Observation for agent $n$ at time $t$ & $\mathbf{o}^{(n)}_{t} \in \mathbb{Z}^{7 \times 7 \times 3}$ \\
        Number of players ($N$) & 2 \\
        Time limit ($T$) & 50 \\
        Grid position for agent $n$ at time $t$ & $g^{(n)}_{t} \in [0, w \times h - 1]$ \\
        Grid direction for agent $n$ at time $t$ & $d^{(n)}_{t} \in \{\texttt{up}, \texttt{down}, \texttt{left}, \texttt{right}\}$ \\
        % Step count for agent $n$ at time $t$  & $s^{(n)}_{t} \in [1, T]$ \\
        % \cmidrule(lr){1-2}
        Action for agent $n$ at time $t$ & $a^{(n)}_{t} \in \{\texttt{turn left}, \texttt{turn right}, \texttt{move forward}\}$\\
        % Count of position,action pair for agent $n$ at time $t$ & $c = \sum_{g \in 7 \times 7} \sum_{a \in \mathcal{A}} $ 
        % \cmidrule(lr){1-2}
        % \multirow{2}{*}{Reward for agent $n$ at time $t$}
        Reward for agent $n$ at time $t$ & $r^{(n)}_{t} = \frac{1}{\sqrt{c}} + \begin{cases}
            -2, & \textrm{if $g^{(n)}_{t} = g^{(n)}_{t-1}$}, \\
            100 (1 - 0.9 \frac{t+1}{T}) & \textrm{if goal is reached}, \\
            -1, & \textrm{otherwise}
        \end{cases}$ \\
        & where $c = \# \Bigl( \langle g^{(n)}_{t}, d^{(n)}_{t}, a^{(n)}_{t} \rangle \in \bigl\{\langle g^{(n)}_{i}, d^{(n)}_{i}, a^{(n)}_{i}\rangle \bigr\}_{i=0}^{t-1} \Bigr)$ \\
        \bottomrule
    \end{tabular}
\end{table}

%%%%%%%%%%%%%%%%%%%%%%%%%%%%%%%%%%%%%%%%%%%

%%%%%%%%%%%%%%%%%%%%%%%%%%%%%%%%%%%%%%%%%%%
\section{Quantum encoding transformations}\label{app:qenctran}
\setcounter{figure}{0}
\setcounter{table}{0}
\setcounter{equation}{0}

To encode environment observations into our quantum models we first apply a transformation on the observation matrix. This allows us to reduce its dimensions, thereby making it usable for the limited number of qubits available to \ac{nisq} systems, while also changing the range of matrix values to be suitable for input into one of the Pauli rotation gates.

\subsection{\texttt{CoinGame-2} Environment}

\paragraph{\ac{mdp} dynamics}
For the \texttt{CoinGame-2} environment with fully observed state dynamics we use the transformation
\begin{equation}
    f_{\textrm{MDP}}(\mathbf{o}_{i \times j \times k}) = \sum_{k} \mathbf{o}_{i \times j, k} 2^{-k}
\end{equation}
which sums over the last dimension of the observation matrix $\mathbf{o}_{i \times j \times k}$ with shape $i \times j \times k$. In the case of \texttt{CoinGame-2} with \ac{mdp} dynamics the observations have shape $4 \times 3 \times 3$. This transformation reduces the dimensions to $4 \times 3 \times 1$, which can be directly fed into the encoder architecture outlined in \cref{eqn:U_enc}.

\paragraph{\ac{pomdp} dynamics}
For the \texttt{CoinGame-2} environment with partially observed state dynamics our quantum models employ a small classical \acs{nn} at the input of the encoder for dimensionality reduction, as done in \citet{Chen2023AsynchronousTrainingQuantum}. In particular, we use the transformation
\begin{equation}
    f_{\textrm{POMDP}}(\mathbf{o}_{i \times j \times k}) = \mathbf{w}_{ijk \times 3d} \cdot \textrm{flatten}(\mathbf{o}_{i \times j \times k})^T + b
    \label{eqn:transform:coingame:pomdp}
\end{equation}
which flattens the observation matrix $\mathbf{o}_{i \times j \times k}$ with shape $i \times j \times k$ and passes it through a single fully-connected \ac{nn} layer with parameters $\mathbf{w}_{ijk \times 3d}$ and $b$, and $d$ is the number of qubits.
Note that, in \ac{pomdp}, the trainable quantum encoding parameters $\bm{\lambda}^{(n)}$ are no longer necessary. In this case, we set $\bm{\lambda}^{(n)}=\bm{1}$, where $\bm{1}$ is a matrix of ones.

\subsection{\texttt{CartPole} Environment}

\paragraph{Observation scaling}

For the \texttt{CartPole} environment we apply a constant observation scaling to both \ac{mdp} and \ac{pomdp} dynamics to normalize their values. In particular, we use the transformation
\begin{equation}
    f(\mathbf{o}_{i \times j}) = \mathbf{o}_{i \times j} / \mathbf{v}_{i}
\end{equation}
where
\begin{equation}
    \mathbf{v} = [2.4, 2.5, 0.21, 2.5]^\intercal
\end{equation}
is a constant scaling vector.

\paragraph{\ac{pomdp} dynamics}

For the \texttt{CartPole} environment with partially observed state dynamics we apply an additional transformation similar to the \texttt{CoinGame-2} \ac{pomdp} case to reduce input feature dimensions. In particular, we apply the transformation,
\begin{equation}
    f_{\textrm{POMDP}}(\mathbf{o}_{i \times j}) = \mathbf{w}_{ij \times 3d} \cdot \textrm{flatten}(\mathbf{o}_{i \times j})^T + b
\end{equation}
which flattens the observation matrix $\mathbf{o}_{i \times j}$ with shape $i \times j$ and passes it through a single fully-connected \ac{nn} layer with parameters $\mathbf{w}_{ij \times 3d}$ and $b$, and $d$ is the number of qubits.

\subsection{\texttt{MiniGrid} Environment}

For the \texttt{MiniGrid} environment we apply a similar transformation to the \texttt{CoinGame} and \texttt{CartPole} \ac{pomdp} environments, which is identical to \cref{eqn:transform:coingame:pomdp}. Specifically, we apply a fully-connected NN layer to reduce the dimensionality from the observation shape $7 \times 7 \times 3$ to $3d$, where $7 \times 7$ is the field of view of the agent as described in \cref{tab:env_spec:minigrid}, and $d$ is the number of qubits.

%%%%%%%%%%%%%%%%%%%%%%%%%%%%%%%%%%%%%%%%%%%

%%%%%%%%%%%%%%%%%%%%%%%%%%
% \newpage
% \clearpage
\section{Model hyperparameters}\label{app:hyp}
\setcounter{figure}{0}
\setcounter{table}{0}
\setcounter{equation}{0}
\setcounter{algocf}{0}

The hyperparameters for each of the models trained in our experiments, as discussed in \cref{sec:exp:setup}, are shown in \cref{app:tab:model_hyp:eqmarl,app:tab:model_hyp:fctde-sctde}. \Cref{app:tab:model_hyp:eqmarl} show the model parameters used in \ac{qfctde} and \ac{ours}. \Cref{app:tab:model_hyp:fctde-sctde} show the model parameters used in \ac{fctde} and \ac{sctde}. 

% \label{app:tab:model_hyp:eqmarl:minigrid}
% \label{app:tab:model_hyp:fctde-sctde:minigrid}

\begin{table}[h]
    \bufferspacefromheader
    \caption{Hyperparameters for \ac{qfctde} and \ac{ours}, actor and critic, used on all environments.}
    \label{app:tab:model_hyp:eqmarl}
    \centering
    \scriptsize
    % \resizebox{\linewidth}{!}{%
    \begin{tabular}{llll}
        \toprule
        Environment & Model & Parameter & Value \\
        \midrule
        \multirow{11}{*}{\shortstack[l]{\texttt{CoinGame-2},\\\texttt{CartPole}}} 
        & \multirow{11}{*}{\shortstack[l]{Actor,\\Critic}} 
        & \multirow{2}{*}{\ac{nn} encoder transform activation} & \textbullet~~\ac{mdp}:~~N/A \\
                                            & & & \textbullet~~\ac{pomdp}:~~linear \\
        \cmidrule(lr){3-4}
        & & \multirow{2}{*}{Flag $\bm{\lambda}^{(n)}$ as trainable} & \textbullet~~\ac{mdp}:~~True \\
                                            & & & \textbullet~~\ac{pomdp}:~~False \\
        \cmidrule(lr){3-4}
        & & Number of qubits per agent ($D$) & 4 \\
        & & (\ac{ours} only) Input entanglement type ($B$) for critic & $\Psi^{+}$ \\
        & & Number of layers ($L$) in $U_{\textrm{vqc}}$ & 5 \\
        & & Squash activation ($\phi$) & arctan \\
        & & Inverse temperature ($\beta$) & 1 \\
        & & Optimizer & Adam \\
        & & Learning rate & $[0.01, 0.1, 0.1]$ \\
        \cmidrule(lr){1-4}
        \multirow{11}{*}{\texttt{MiniGrid}} 
        & \multirow{9}{*}{Critic}  & \ac{nn} encoder transform activation & linear \\
        & & Flag $\bm{\lambda}^{(n)}$ as trainable & False \\
        & & Number of qubits per agent ($D$) & 4 \\
        & & (\ac{ours} only) Input entanglement type ($B$) for critic & $\Psi^{+}$ \\
        & & Number of layers ($L$) in $U_{\textrm{vqc}}$ & 5 \\
        & & Squash activation ($\phi$) & arctan \\
        & & Inverse temperature ($\beta$) & 1 \\
        & & Optimizer & Adam \\
        & & Learning rate & $[0.001, 0.001, 0.01, 0.1]$ \\
        \cmidrule(lr){2-4}
        & \multirow{3}{*}{Actor} & Optimizer & Adam \\
        & & Learning rate & $[0.0001]$ \\
        & & Number of hidden units & 100 \\
        \bottomrule
    \end{tabular}
    % }
\end{table}
\begin{table}[h]
    \bufferspacefromheader
    \caption{Hyperparameters for \ac{fctde} and \ac{sctde}, actor and critic, used on all environments.}
    \label{app:tab:model_hyp:fctde-sctde}
    \centering
    \scriptsize
    \begin{tabular}{llll}
        \toprule
        Environment & Model & Parameter & Value \\
        \midrule
        \multirow{3}{*}{\shortstack[l]{\texttt{CoinGame-2},\\\texttt{CartPole}}} 
        & \multirow{3}{*}{\shortstack[l]{Actor,\\Critic}} 
        & \ac{nn} hidden units ($h$) & $[12]$ \\
        & & Optimizer & Adam \\
        & & Learning rate & $0.001$ \\
        \cmidrule(lr){1-4}
        \multirow{3}{*}{\texttt{MiniGrid}} 
        & \multirow{3}{*}{\shortstack[l]{Actor,\\Critic}} 
        & \ac{nn} hidden units ($h$) & $[100]$ \\
        & & Optimizer & Adam \\
        & & Learning rate & $0.0001$ \\
        \bottomrule
    \end{tabular}
\end{table}

%%%%%%%%%%%%%%%%%%%%%%%%%%
% \newpage
% \clearpage
\section{Experiment results}\label{app:exp}
\setcounter{figure}{0}
\setcounter{table}{0}
\setcounter{equation}{0}
\setcounter{algocf}{0}

\subsection{Entanglement style comparison}

The empirical results for the entanglement comparison experiment, as discussed in \cref{sec:exp:results:ent}, are shown in \cref{tab:exp:ent:stat,tab:exp:ent:conv}.
\Cref{tab:exp:ent:stat} shows the score metric statistics mean, standard deviation, and $95\%$ confidence interval for each of the entanglement styles $\Psi^{+}$, $\Psi^{-}$, $\Phi^{+}$, $\Phi^{-}$, and $\mathtt{None}$.
\Cref{tab:exp:ent:conv} shows the convergence time, in epochs, to each of the score thresholds 20, 25, and also to the maximum score value (reported parenthetically in \textit{italics}) for each of the entanglement styles.
The best values in each column are highlighted in \textbf{bold}.

\begin{figure*}[h]
    \centering
    \begin{subfigure}{0.5\linewidth}
        \centering
        \includegraphics[width=\linewidth]{images/experiment_results/fig_maa2c_mdp_entanglement_compare/fig_maa2c_mdp_entanglement_compare-undiscounted_reward.pdf}
        \caption{Score - \ac{mdp}}
        \label{app:fig:fig_maa2c_mdp_entanglement_compare:undiscounted_reward}
    \end{subfigure}%
    \begin{subfigure}{0.5\linewidth}
        \centering
        \includegraphics[width=\linewidth]{images/experiment_results/fig_maa2c_pomdp_entanglement_compare/fig_maa2c_pomdp_entanglement_compare-undiscounted_reward.pdf}
        \caption{Score - \ac{pomdp}}
        \label{app:fig:fig_maa2c_pomdp_entanglement_compare:undiscounted_reward}
    \end{subfigure}
    \begin{subfigure}{0.5\linewidth}
        \centering
        \includegraphics[width=\linewidth]{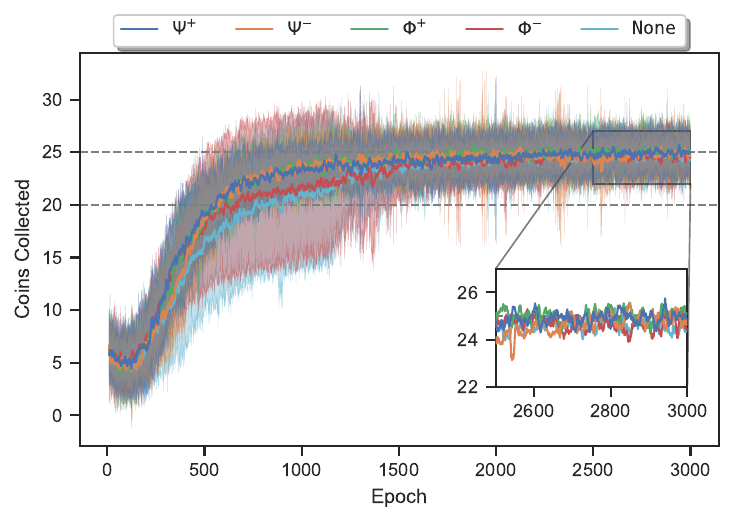}
        \caption{Total coins collected - \ac{mdp}}
        \label{app:fig:fig_maa2c_mdp_entanglement_compare:coins_collected}
    \end{subfigure}%
    \begin{subfigure}{0.5\linewidth}
        \centering
        \includegraphics[width=\linewidth]{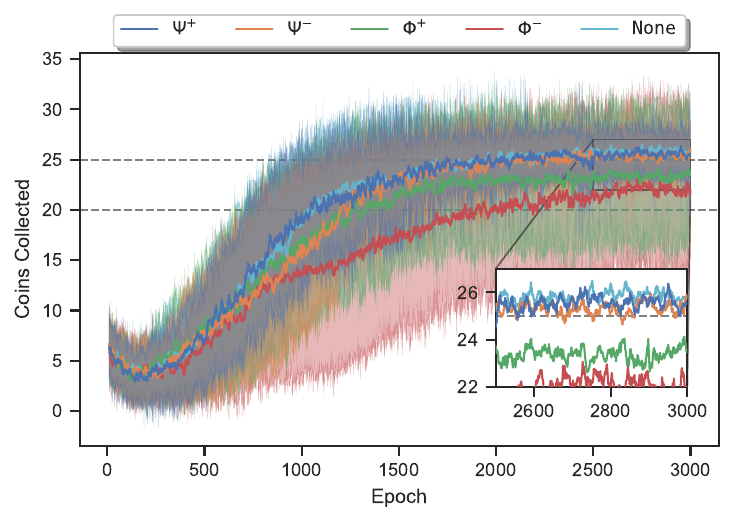}
        \caption{Total coins collected - \ac{pomdp}}
        \label{app:fig:fig_maa2c_pomdp_entanglement_compare:coins_collected}
    \end{subfigure}
    \begin{subfigure}{0.5\linewidth}
        \centering
        \includegraphics[width=\linewidth]{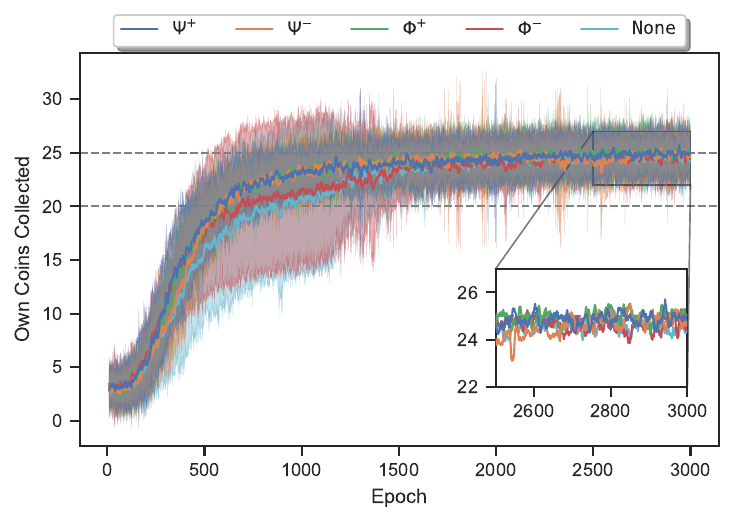}
        \caption{Own coins collected - \ac{mdp}}
        \label{app:fig:fig_maa2c_mdp_entanglement_compare:own_coins_collected}
    \end{subfigure}%
    \begin{subfigure}{0.5\linewidth}
        \centering
        \includegraphics[width=\linewidth]{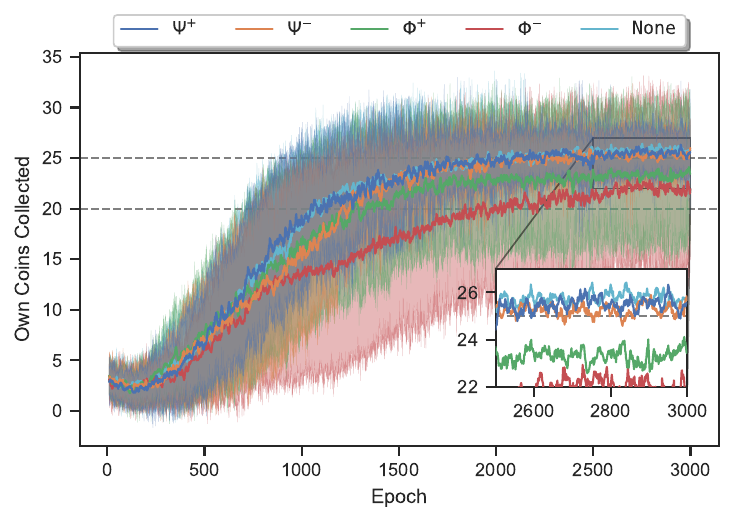}
        \caption{Own coins collected - \ac{pomdp}}
        \label{app:fig:fig_maa2c_pomdp_entanglement_compare:own_coins_collected}
    \end{subfigure}
    \begin{subfigure}{0.5\linewidth}
        \centering
        \includegraphics[width=\linewidth]{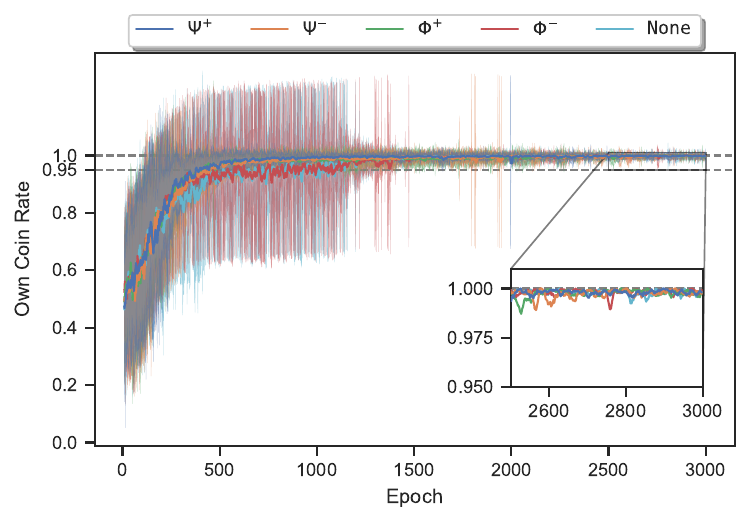}
        \caption{Own coin rate - \ac{mdp}}
        \label{app:fig:fig_maa2c_mdp_entanglement_compare:own_coin_rate}
    \end{subfigure}%
    \begin{subfigure}{0.5\linewidth}
        \centering
        \includegraphics[width=\linewidth]{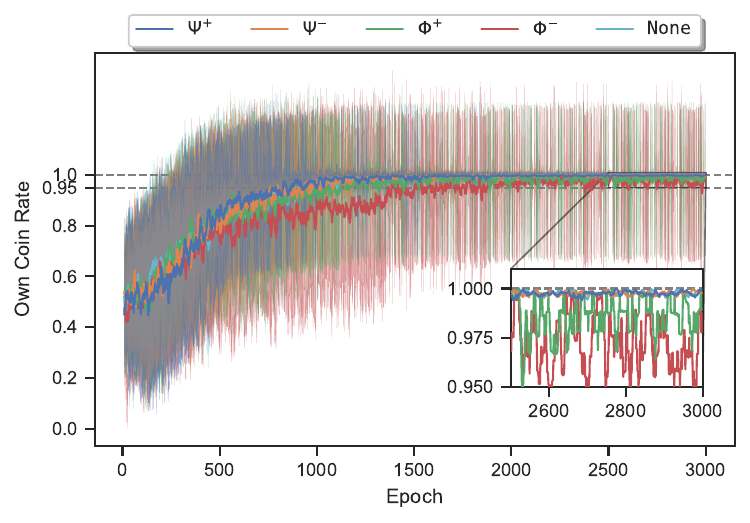}
        \caption{Own coin rate - \ac{pomdp}}
        \label{app:fig:fig_maa2c_pomdp_entanglement_compare:own_coin_rate}
    \end{subfigure}
   \caption{Comparison of \texttt{CoinGame-2} \ac{mdp} and \ac{pomdp} environment performance metrics (a,b) score, (c,d) total coins collected, (e,f) own coins collected, and (g,h) own coin rate for \ac{ours} with varying input quantum entanglement styles $\Psi^{+}$ (blue), $\Psi^{-}$ (orange), $\Phi^{+}$ (green), $\Phi^{-}$ (red), and $\texttt{None}$ (cyan) averaged over 10 runs, with $\pm 1$ std.\ dev.\ shown as shaded regions.}
   \label{app:fig:fig_maa2c_mdp_pomdp_entanglement_compare}
\end{figure*}

\Cref{app:fig:fig_maa2c_mdp_pomdp_entanglement_compare} shows the training results for the entanglement styles as discussed in \cref{sec:exp:results:ent}. In particular, we provide the full set of performance metrics of score, total coins collected, own coins collected, and own coin rate, as outlined in \cref{app:env}, \cref{app:eqn:env:score,app:eqn:env:tc,app:eqn:env:ocr}.
\Cref{app:fig:fig_maa2c_mdp_pomdp_entanglement_compare} shows the results for the entanglement styles $\Psi^{+}$, $\Psi^{-}$, $\Phi^{+}$, $\Phi^{-}$, and $\mathtt{None}$, as discussed in \cref{sec:exp:results:ent}. The left column, \cref{app:fig:fig_maa2c_mdp_entanglement_compare:undiscounted_reward,app:fig:fig_maa2c_mdp_entanglement_compare:coins_collected,app:fig:fig_maa2c_mdp_entanglement_compare:own_coins_collected,app:fig:fig_maa2c_mdp_entanglement_compare:own_coin_rate}, shows the performance for \ac{mdp} environment dynamics. Similarly, the right column, \cref{app:fig:fig_maa2c_pomdp_entanglement_compare:undiscounted_reward,app:fig:fig_maa2c_pomdp_entanglement_compare:coins_collected,app:fig:fig_maa2c_pomdp_entanglement_compare:own_coins_collected,app:fig:fig_maa2c_pomdp_entanglement_compare:own_coin_rate}, shows the performance for \ac{pomdp} environment dynamics.

% Tables to compare entanglement styles.
\begin{table}[h]
    \caption{Comparison of entanglement style score performance for \ac{mdp} and \ac{pomdp} \texttt{CoinGame-2} environment dynamics using mean, standard deviation, and $95\%$ confidence interval statistics. Best values are highlighted in bold.}
    \label{tab:exp:ent:stat}
    \centering
    \begin{tabular}{lllll}
        \toprule
                                    &                               & \multicolumn{3}{c}{Score} \\
                                                                     \cmidrule(lr){3-5}
        Dynamics                    & Entanglement                         & Mean          & SD            & $95\%$ CI \\
        \midrule
        \multirow{4}{*}{\ac{mdp}}   & $\Psi^{+}$ & \textbf{21.11} & 2.65 & (20.92, 21.29) \\
                                    & $\Psi^{-}$ & 20.85 & 2.70 & (20.61, 21.07) \\
                                    & $\Phi^{+}$ & 21.02 & \textbf{2.54} & (20.77, 21.30) \\
                                    & $\Phi^{-}$ & 20.43 & 3.85 & (20.20, 20.59) \\
                                    & $\mathtt{None}$ & 20.00 & 3.80 & (19.75, 20.20) \\
        \cmidrule(lr){1-5}
        \multirow{4}{*}{\ac{pomdp}} & $\Psi^{+}$ & 18.49 & \textbf{3.91} & (18.15, 18.74) \\
                                    & $\Psi^{-}$ & 17.77 & 4.05 & (17.40, 18.09) \\
                                    & $\Phi^{+}$ & 17.01 & 6.28 & (16.74, 17.25) \\
                                    & $\Phi^{-}$ & 14.73 & 7.63 & (14.45, 15.01) \\
                                    & $\mathtt{None}$ & \textbf{18.57} & 4.26 & (18.28, 18.82) \\
        \bottomrule
    \end{tabular}
\end{table}

\begin{table}[h]
    \caption{Comparison of entanglement style score convergence (in number of epochs) for \ac{mdp} and \ac{pomdp} \texttt{CoinGame-2} environment dynamics. Best values are highlighted in bold.}
    \label{tab:exp:ent:conv}
    \centering
    \begin{tabular}{lllll}
        \toprule
                                    &                               & \multicolumn{3}{c}{Epochs to Score Threshold} \\
                                                                     \cmidrule(lr){3-5}
        Dynamics                    & Entanglement                         & 20            & 25            & Max (\textit{value}) \\
        \midrule
        \multirow{4}{*}{\ac{mdp}}   & $\Psi^{+}$ & \textbf{568} & 2332 & 2942 (\textit{\textbf{25.67}}) \\
                                    & $\Psi^{-}$ & 595 & 1987 & 2849 (\textit{25.45}) \\
                                    & $\Phi^{+}$ & 612 & \textbf{1883} & 2851 (\textit{25.51}) \\
                                    & $\Phi^{-}$ & 691 & 2378 & 2984 (\textit{25.23}) \\
                                    & $\mathtt{None}$ & 839 & 2337 & \textbf{2495} (\textit{25.12}) \\
        \cmidrule(lr){1-5}
        \multirow{4}{*}{\ac{pomdp}} & $\Psi^{+}$ & \textbf{1049} & \textbf{1745} & 2950 (\textit{26.28}) \\
                                    & $\Psi^{-}$ & 1206 & 2114 & 2999 (\textit{25.95}) \\
                                    & $\Phi^{+}$ & 1269 & - & 2992 (\textit{24.1}) \\
                                    & $\Phi^{-}$ & 1838 & - & 2727 (\textit{22.8}) \\
                                    & $\mathtt{None}$ & 1069 & 1955 & \textbf{2841} (\textit{\textbf{26.39}}) \\
        \bottomrule
    \end{tabular}
\end{table}

\clearpage
\newpage
\subsection{\texttt{CoinGame} baselines comparison}

\begin{figure*}[h]
    \centering
    \begin{subfigure}{0.5\linewidth}
        \centering
        \includegraphics[width=\linewidth]{images/experiment_results/fig_maa2c_mdp/fig_maa2c_mdp-undiscounted_reward.pdf}
        \caption{Score - \ac{mdp}}
        \label{app:fig:fig_maa2c_mdp:undiscounted_reward}
    \end{subfigure}%
    \begin{subfigure}{0.5\linewidth}
        \centering
        \includegraphics[width=\linewidth]{images/experiment_results/fig_maa2c_pomdp/fig_maa2c_pomdp-undiscounted_reward.pdf}
        \caption{Score - \ac{pomdp}}
        \label{app:fig:fig_maa2c_pomdp:undiscounted_reward}
    \end{subfigure}
    \begin{subfigure}{0.5\linewidth}
        \centering
        \includegraphics[width=\linewidth]{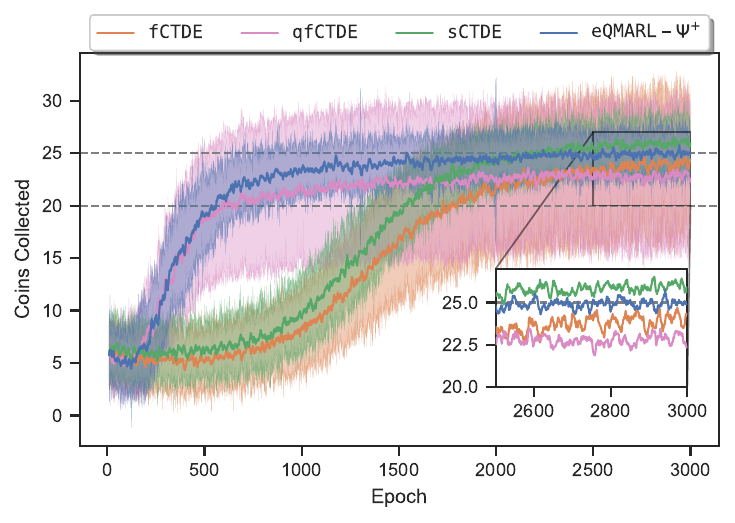}
        \caption{Total coins collected - \ac{mdp}}
        \label{app:fig:fig_maa2c_mdp:coins_collected}
    \end{subfigure}%
    \begin{subfigure}{0.5\linewidth}
        \centering
        \includegraphics[width=\linewidth]{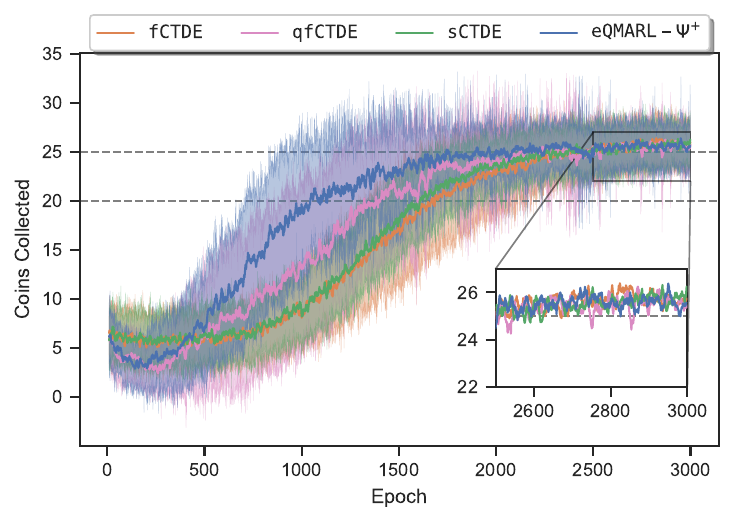}
        \caption{Total coins collected - \ac{pomdp}}
        \label{app:fig:fig_maa2c_pomdp:coins_collected}
    \end{subfigure}
    \begin{subfigure}{0.5\linewidth}
        \centering
        \includegraphics[width=\linewidth]{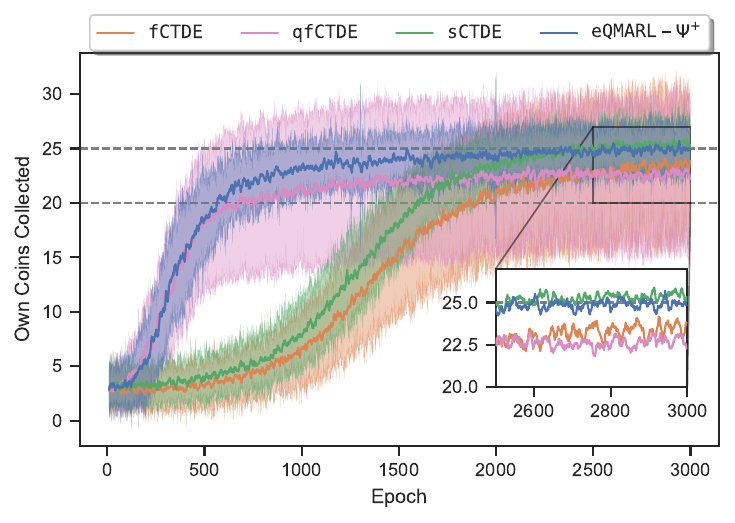}
        \caption{Own coins collected - \ac{mdp}}
        \label{app:fig:fig_maa2c_mdp:own_coins_collected}
    \end{subfigure}%
    \begin{subfigure}{0.5\linewidth}
        \centering
        \includegraphics[width=\linewidth]{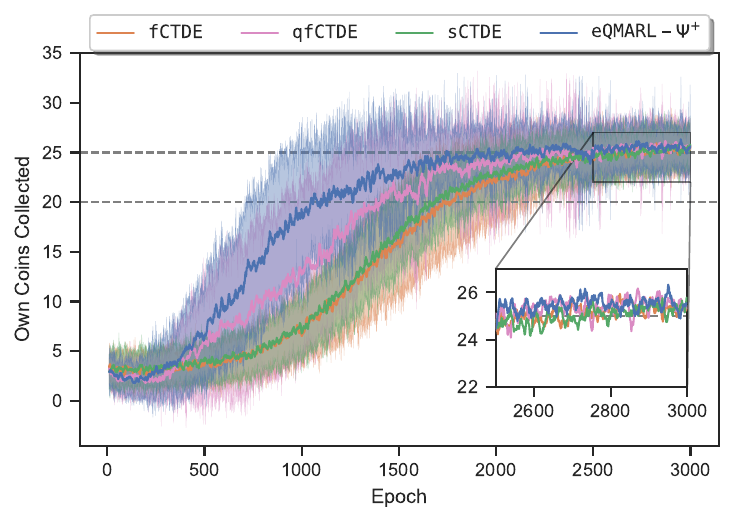}
        \caption{Own coins collected - \ac{pomdp}}
        \label{app:fig:fig_maa2c_pomdp:own_coins_collected}
    \end{subfigure}
    \begin{subfigure}{0.5\linewidth}
        \centering
        \includegraphics[width=\linewidth]{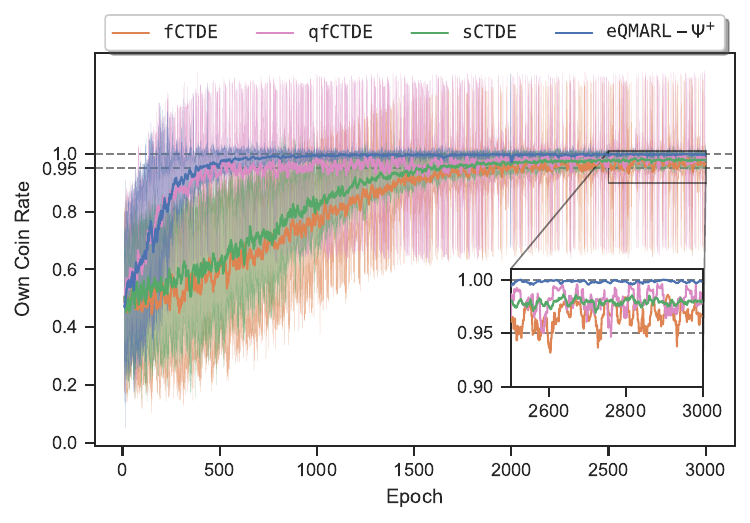}
        \caption{Own coin rate - \ac{mdp}}
        \label{app:fig:fig_maa2c_mdp:own_coin_rate}
    \end{subfigure}%
    \begin{subfigure}{0.5\linewidth}
        \centering
        \includegraphics[width=\linewidth]{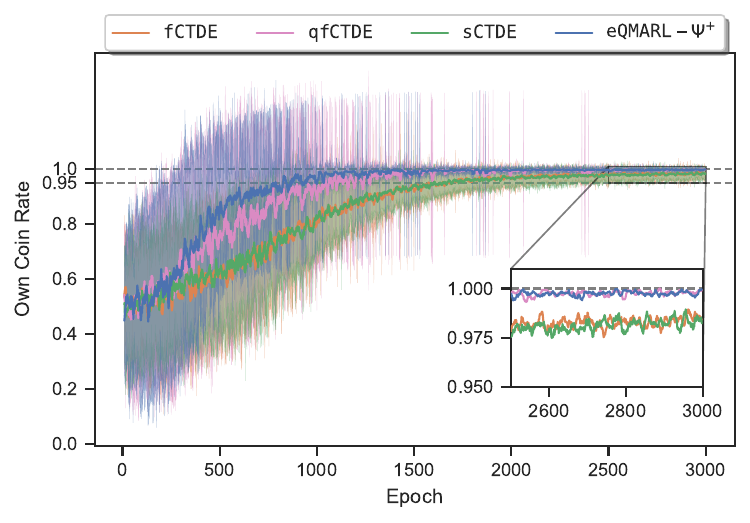}
        \caption{Own coin rate - \ac{pomdp}}
        \label{app:fig:fig_maa2c_pomdp:own_coin_rate}
    \end{subfigure}
   \caption{Comparison of \texttt{CoinGame-2} \ac{mdp} and \ac{pomdp} environment performance metrics (a,b) score, (c,d) total coins collected, (e,f) own coins collected, and (g,h) own coin rate for \texttt{fCTDE} (orange), \texttt{qfCTDE} (magenta), \texttt{sCTDE} (green), and \texttt{eQMARL-$\Psi^{+}$} (blue) averaged over 10 runs, with $\pm 1$ std.\ dev.\ shown as shaded regions.}
   \label{app:fig:fig_maa2c_mdp_pomdp}
\end{figure*}

The empirical results for \texttt{eQMARL-$\Psi^{+}$}, \texttt{qfCTDE}, \texttt{fCTDE}, and \texttt{sCTDE}, as discussed in \cref{sec:exp:results:coingame}, are shown in \cref{tab:exp:ent:conv,tab:exp:models,app:fig:fig_maa2c_mdp_pomdp}. 
The performance for \ac{mdp} dynamics is shown in \cref{app:fig:fig_maa2c_mdp:undiscounted_reward,app:fig:fig_maa2c_mdp:coins_collected,app:fig:fig_maa2c_mdp:own_coins_collected,app:fig:fig_maa2c_mdp:own_coin_rate}, and for \ac{pomdp} dynamics is shown in \cref{app:fig:fig_maa2c_pomdp:undiscounted_reward,app:fig:fig_maa2c_pomdp:coins_collected,app:fig:fig_maa2c_pomdp:own_coins_collected,app:fig:fig_maa2c_pomdp:own_coin_rate}.
Importantly, \cref{app:fig:fig_maa2c_mdp_pomdp} shed light on when, and how, a cooperative strategy is achieved by each framework. Further, through \cref{app:fig:fig_maa2c_mdp_pomdp} we also observe the relationship between the metrics outlined in \cref{app:env:coin_game}. This connection is important, as a single metric in isolation only paints part of the performance picture. A full comparison can be achieved by considering the metrics as as group, and, particularly, the relationship between agent score, i.e., the sum of rewards, and own coin rate, i.e., the priority given to coins of matching color.

% Tables to compare baselines.
\begin{table}[h]
    \caption{Comparison of model score and own coin rate performance for \ac{mdp} and \ac{pomdp} \texttt{CoinGame-2} environment dynamics using mean, standard deviation, and $95\%$ confidence interval statistics. Best values are highlighted in bold.}
    \label{tab:exp:models:stat}
    \centering
    \begin{tabular}{llllllll}
        \toprule
                                    &                               & \multicolumn{3}{c}{Score}                                      & \multicolumn{3}{c}{Own Coin Rate} \\
                                                                     \cmidrule(lr){3-5} \cmidrule(lr){6-8}
        Dynamics                    & Framework                     & Mean          & SD            & $95\%$ CI           & Mean          & SD            & $95\%$ CI \\
        \midrule
        \multirow{4}{*}{\ac{mdp}}   & \texttt{eQMARL-$\Psi^{+}$}    & \textbf{21.11} & \textbf{2.65} & \textbf{(20.91, 21.37)} & \textbf{0.9640} & \textbf{0.0347} & \textbf{(0.9606, 0.9667)} \\ 
                                    & \texttt{qfCTDE}               & 19.41 & 6.23 & (19.22, 19.60) & 0.9398 & 0.1020 & (0.9367, 0.9423) \\ 
                                    & \texttt{sCTDE}                & 14.18 & 2.69 & (13.87, 14.53) & 0.8504 & 0.0928 & (0.8436, 0.8558) \\ 
                                    & \texttt{fCTDE}                & 12.36 & 4.41 & (12.01, 12.66) & 0.8202 & 0.1379 & (0.8153, 0.8255) \\ 
        \cmidrule(lr){1-8}
        \multirow{4}{*}{\ac{pomdp}} & \texttt{eQMARL-$\Psi^{+}$}    & \textbf{18.49} & 3.91 & \textbf{(18.24, 18.75)} & \textbf{0.9226} & \textbf{0.0831} & \textbf{(0.9173, 0.9281)} \\
                                    & \texttt{qfCTDE}               & 16.79 & 4.66 & (16.43, 17.19) & 0.9040 & 0.1135 & (0.8991, 0.9094) \\
                                    & \texttt{sCTDE}                & 13.70 & \textbf{2.79} & (13.33, 14.07) & 0.8466 & 0.0985 & (0.8407, 0.8525) \\
                                    & \texttt{fCTDE}                & 13.46 & 3.24 & (13.08, 13.75) & 0.8443 & 0.1026 & (0.8389, 0.8495) \\
        \bottomrule
    \end{tabular}
\end{table}

\begin{table}[h]
    \caption{Comparison of model score and own coin rate convergence (in number of epochs) for \ac{mdp} and \ac{pomdp} \texttt{CoinGame-2} environment dynamics. Best values are highlighted in bold.}
    \label{tab:exp:models}
    \centering
    \begin{tabular}{llllllll}
        \toprule
                                    &                               & \multicolumn{3}{c}{Epochs to Score Threshold}                                      & \multicolumn{3}{c}{Epochs to Own Coin Rate Threshold} \\
                                                                     \cmidrule(lr){3-5} \cmidrule(lr){6-8}
        Dynamics                    & Framework                     & 20            & 25            & Max (\textit{value})           & 0.95          & 1.0           & Max (\textit{value}) \\
        \midrule
        \multirow{4}{*}{\ac{mdp}}   & \texttt{eQMARL-$\Psi^{+}$}    & \textbf{568}  & \textbf{2332} & 2942 (\textit{\textbf{25.67}}) & \textbf{376}  & \textbf{2136} & \textbf{2136} (\textit{\textbf{1.0}}) \\
                                    & \texttt{qfCTDE}               & 678           & -             & \textbf{2378} (\textit{23.38}) & 397           & -             & 2832 (\textit{0.9972}) \\
                                    & \texttt{sCTDE}                & 1640          & 2615          & 2631 (\textit{25.3})           & 1511          & -             & 2637 (\textit{0.9864}) \\
                                    & \texttt{fCTDE}                & 1917          & -             & 2925 (\textit{23.67})          & 1700          & -             & 2909 (\textit{0.9857}) \\
        \cmidrule(lr){1-8}
        \multirow{4}{*}{\ac{pomdp}} & \texttt{eQMARL-$\Psi^{+}$}    & \textbf{1049} & \textbf{1745} & 2950 (\textit{\textbf{26.28}}) & \textbf{773}  & -             & \textbf{2533} (\textit{0.9997}) \\
                                    & \texttt{qfCTDE}               & 1382          & 2124          & 2871 (\textit{26.09})          & 1038          & \textbf{2887} & 2887 (\textit{\textbf{1.0}}) \\
                                    & \texttt{sCTDE}                & 1738          & 2750          & 2999 (\textit{25.33})          & 1588          & -             & 2956 (\textit{0.9894}) \\
                                    & \texttt{fCTDE}                & 1798          & 2658          & \textbf{2824} (\textit{25.49}) & 1574          & -             & 2963 (\textit{0.9894}) \\
        \bottomrule
    \end{tabular}
\end{table}

%%%%%%%%%%%%%%%%%%%%%%%%%%%%%%%%%%%%%%%%

\clearpage
\newpage
\subsection{\texttt{CartPole} baselines comparison}

\begin{figure*}[h]
    \vspace*{-1em}
    \centering
    \begin{subfigure}{0.5\linewidth}
        \centering
        \includegraphics[width=\linewidth]{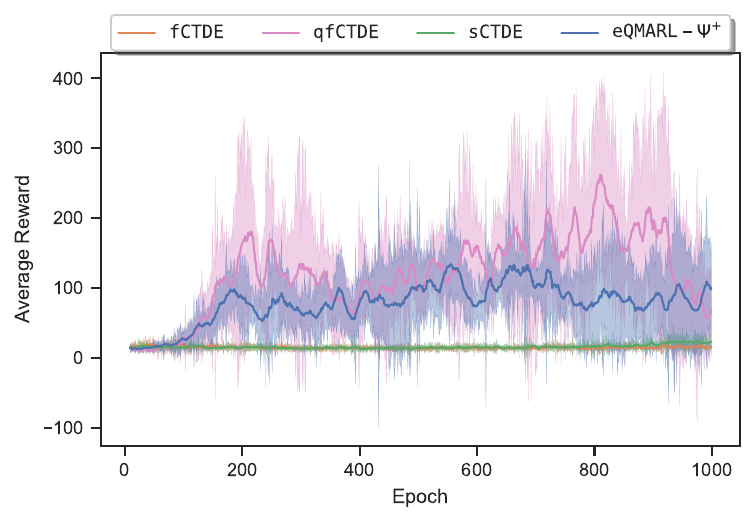}
        
        \vspace*{-0.1em}
        \caption{Average Reward - \ac{mdp}}
        \vspace*{-0.25em}
        \label{fig:fig_cartpole_maa2c_mdp:reward_mean}
    \end{subfigure}%
    % \hfill%
    \begin{subfigure}{0.5\linewidth}
        \centering
        \includegraphics[width=\linewidth]{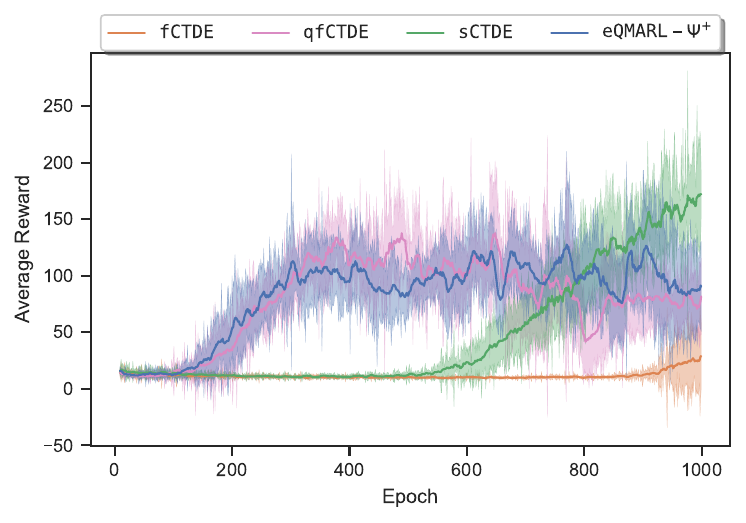}
        
        \vspace*{-0.1em}
        \caption{Average Reward - \ac{pomdp}}
        \vspace*{-0.25em}
        \label{fig:fig_cartpole_maa2c_pomdp:reward_mean}
    \end{subfigure}
   \caption{Comparison of \texttt{CartPole} \ac{mdp} and \ac{pomdp} environment average reward performance for \texttt{fCTDE} (orange), \texttt{qfCTDE} (magenta), \texttt{sCTDE} (green), and \texttt{eQMARL-$\Psi^{+}$} (blue) averaged over 5 runs of 1000 epochs, with $\pm 1$ std.\ dev.\ shown as shaded regions.}
   \label{app:fig:fig_cartpole_maa2c_mdp_pomdp}
   \vspace*{-0.8em}
\end{figure*}

The empirical results for \texttt{eQMARL-$\Psi^{+}$}, \texttt{qfCTDE}, \texttt{fCTDE}, and \texttt{sCTDE}, as discussed in \cref{sec:exp:results:cartpole}, are shown in \cref{tab:exp:cartpole:baseline:stat,tab:exp:cartpole:baseline:conv,app:fig:fig_cartpole_maa2c_mdp_pomdp}. The performance for \ac{mdp} dynamics is shown in \cref{fig:fig_cartpole_maa2c_mdp:reward_mean}, and for \ac{pomdp} dynamics is shown in \cref{fig:fig_cartpole_maa2c_pomdp:reward_mean}. Importantly, from this we see that the classical models do not perform well overall in either setting, and \texttt{qfCTDE} experiences high variance in the \ac{mdp} case. Even though \texttt{sCTDE} has a higher reward at the end of training in the \ac{pomdp} case, it converges considerably more slowly, experiencing high variance at the end, and requires over 400 more epochs achieve a mean value less than half of \ac{ours}. In contrast, \ac{ours} is more stable than \texttt{qfCTDE}, and more rapidly converges to a higher mean reward than \texttt{fCTDE} and \texttt{sCTDE} across both settings.

% Tables to compare baselines.
\begin{table}[h]
    \vspace*{-0.8em}
    \caption{Comparison of model average reward performance for \ac{mdp} and \ac{pomdp} \texttt{CartPole} environment dynamics using mean, standard deviation, and $95\%$ confidence interval statistics.}
    \label{tab:exp:cartpole:baseline:stat}
    \centering
    \begin{tabular}{lllll}
        \toprule
                                    &                               & \multicolumn{3}{c}{Reward} \\
                                                                     \cmidrule(lr){3-5}
        Dynamics                    & Framework                         & Mean          & SD            & $95\%$ CI \\
        \midrule
        \multirow{4}{*}{\ac{mdp}}   & \texttt{eQMARL-$\Psi^{+}$} & 79.11 & 50.62 & (77.26, 81.01) \\
                                    & \texttt{qfCTDE} & 121.35 & 110.13 & (117.95, 124.59) \\
                                    & \texttt{sCTDE} & 16.07 & 22.15 & (15.90, 16.21) \\
                                    & \texttt{fCTDE} & 15.14 & 17.43 & (15.06, 15.22) \\
        \cmidrule(lr){1-5}
        \multirow{4}{*}{\ac{pomdp}} & \texttt{eQMARL-$\Psi^{+}$} & 82.28 & 44.24 & (80.80, 83.91) \\
                                    & \texttt{qfCTDE} & 79.03 & 44.06 & (76.27, 81.02) \\
                                    & \texttt{sCTDE} & 47.59 & 29.71 & (44.71, 50.86) \\
                                    & \texttt{fCTDE} & 11.62 & 32.02 & (11.45, 11.82) \\
        \bottomrule
    \end{tabular}
\end{table}

\begin{table}[h]
    \vspace*{-1.0em}
    \caption{Comparison of model average reward convergence (in number of epochs) for \ac{mdp} and \ac{pomdp} \texttt{CartPole} environment dynamics.}
    \label{tab:exp:cartpole:baseline:conv}
    \centering
    \begin{tabular}{llll}
        \toprule
                                    &                               & \multicolumn{2}{c}{Epochs to Average Reward Threshold} \\
                                                                     \cmidrule(lr){3-4}
        Dynamics                    & Framework                         & Mean (\textit{value})        & Max (\textit{value}) \\
        \midrule
        \multirow{4}{*}{\ac{mdp}}   & \texttt{eQMARL-$\Psi^{+}$} & 166 (\textit{79.11}) & 555 (\textit{134.16}) \\
                                    & \texttt{qfCTDE} & 189 (121.35) & 810 (\textit{262.43}) \\
                                    & \texttt{sCTDE} & 23 (\textit{16.07}) & 978 (\textit{24.64}) \\
                                    & \texttt{fCTDE} & 9 (\textit{15.14}) & 44 (\textit{19.43}) \\
        \cmidrule(lr){1-4}
        \multirow{4}{*}{\ac{pomdp}} & \texttt{eQMARL-$\Psi^{+}$} & 251 (\textit{82.28}) & 770 (\textit{127.60}) \\
                                    & \texttt{qfCTDE} & 276 (\textit{79.03}) & 648 (\textit{137.66}) \\
                                    & \texttt{sCTDE} & 669 (\textit{47.59}) & 998 (\textit{172.16}) \\
                                    & \texttt{fCTDE} & 9 (\textit{11.62}) & 999 (\textit{28.83}) \\
        \bottomrule
    \end{tabular}
\end{table}

%%%%%%%%%%%%%%%%%%%%%%%%%%%%%%%%%%

\clearpage
\newpage
\subsection{\texttt{MiniGrid} baselines comparison}\label{app:exp:minigrid}

\begin{figure*}[h]
    \centering
    \includegraphics[width=0.5\linewidth]{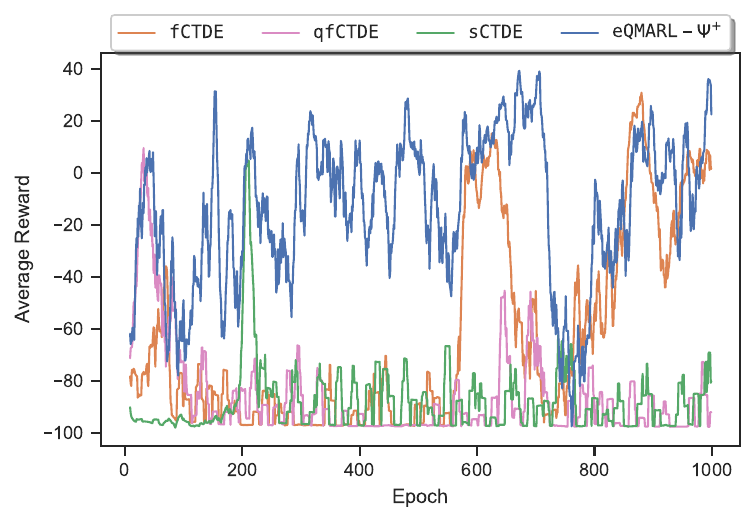}
    \caption{Comparison of \texttt{MiniGrid} \ac{pomdp} environment average reward performance for \texttt{fCTDE} (orange), \texttt{qfCTDE} (magenta), \texttt{sCTDE} (green), and \texttt{eQMARL-$\Psi^{+}$} (blue) for 2 agents over 1000 epochs, with a maximum step limit of 50.}
    \label{app:fig:fig_minigrid_maa2c}
    \vspace*{-0.8em}
\end{figure*}

The empirical results for \texttt{eQMARL-$\Psi^{+}$}, \texttt{qfCTDE}, \texttt{fCTDE}, and \texttt{sCTDE}, as discussed in \cref{sec:exp:results:minigrid}, are shown in \cref{app:tab:exp:minigrid,app:fig:fig_minigrid_maa2c}. Importantly, from \cref{app:fig:fig_minigrid_maa2c} we see that the \texttt{qfCTDE}, \texttt{fCTDE}, and \texttt{sCTDE} baselines have an average reward that is clustered near $-100$ for the majority of training. This implies that the baselines learn to exhaust many steps by simply spinning in place, since the maximum step size is 50 and the agents receive a $-2$ reward for staying in the same position as a previous time step. In contrast, we see that the average reward of \texttt{eQMARL-$\Psi^{+}$} is spread out higher over the training regime. From \cref{app:tab:exp:minigrid} we specifically see that \texttt{eQMARL-$\Psi^{+}$} achieves an average overall reward of $-13.32$, which is 4.5-times higher than the baselines. Indeed, this negative reward means that \texttt{eQMARL-$\Psi^{+}$} also expends actions turning in place, but the fact the reward is so close to zero implies these events occur at a vastly reduced frequency than the baselines.

In addition, from \cref{app:tab:exp:minigrid} we see that both \texttt{eQMARL-$\Psi^{+}$} and \texttt{qfCTDE} reduce the overall critic size by a factor of 8 compared to the classical baselines. This reduction in size means that \texttt{eQMARL-$\Psi^{+}$} and \texttt{qfCTDE} are more computationally efficient than the classical baselines. Further, we see that \texttt{eQMARL-$\Psi^{+}$} is even more efficient because it only requires a single centralized trainable parameter, which is a significant 200-times reduction in size compared to \texttt{sCTDE}.

In testing, \texttt{eQMARL-$\Psi^{+}$} was able to traverse to the goal in as little as 9 steps, whereas \texttt{fCTDE} required 17 steps, and both \texttt{qfCTDE} and \texttt{sCTDE} were unable to find the goal within the 50 step limit. This is a marked 50\% improvement in the exploration and navigation speed of \texttt{eQMARL-$\Psi^{+}$} over \texttt{fCTDE}, with the bonus of no observation sharing, and an 8-times smaller overall critic size. Hence, we have shown that \texttt{eQMARL-$\Psi^{+}$} can indeed be applied to more complex environments, such as grid-world navigation with limited visibility, and provide learning benefits over baselines without the need for observation sharing.

\begin{table}[h]
    \vspace*{-0.8em}
    \caption{Comparison of model average reward performance for \ac{pomdp} \texttt{MiniGrid} environment dynamics using mean and $95\%$ confidence interval statistics, and comparison of model size in number of trainable critic parameters for each framework.}
    \label{app:tab:exp:minigrid}
    \centering
    % \scriptsize
    \begin{tabular}{lllll}
        \toprule
                                            & \multicolumn{2}{c}{Reward} \\
        \cmidrule(lr){2-3}
        Framework                           & Mean & $95\%$ CI & Number of Trainable Critic Parameters \\
        \midrule
        \texttt{fCTDE} & -63.04 & (-65.16, -61.06) & 29,601 \\
        \texttt{qfCTDE} & -85.86 & (-87.03, -84.72) & 3,697 \\
        \texttt{sCTDE} & -88.02 & (-88.69, -87.10) & 29,801 (14,800 per agent, 201 central) \\
        \texttt{eQMARL-$\Psi^{+}$} & -13.32 & (-14.68, -11.91) & 3,697 (1,848 per agent, 1 central) \\
        \bottomrule
    \end{tabular}
\end{table}

% Model	Reward Mean	Reward 95% CI	Number of Trainable Critic Parameters
% fCTDE	-63.04	(-65.16, -61.06)	29,601
% qfCTDE	-85.86	(-87.03, -84.72)	3,697
% sCTDE	-88.02	(-88.69, -87.10)	29,801
% eQMARL-	-13.32	(-14.68, -11.91)	3,697

%%%%%%%%%%%%%%%%%%%%%%%%%%%%%%%%%%

\clearpage
\newpage
\subsection{Ablation Study}\label{app:exp:ablation}

\begin{figure*}[h]
    \begin{subfigure}{0.5\linewidth}
        \centering
        \includegraphics[width=\linewidth]{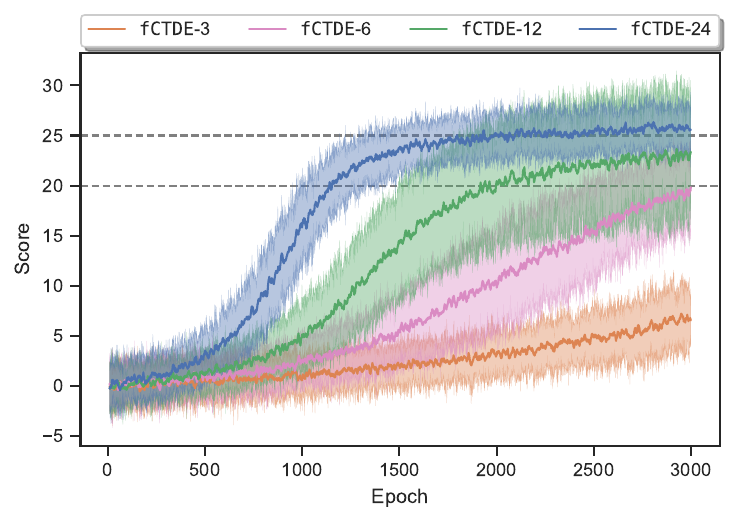}
        \caption{fCTDE - MDP - Score}
        \vspace*{-0.25em}
    \end{subfigure}%
    \begin{subfigure}{0.5\linewidth}
        \centering
        \includegraphics[width=\linewidth]{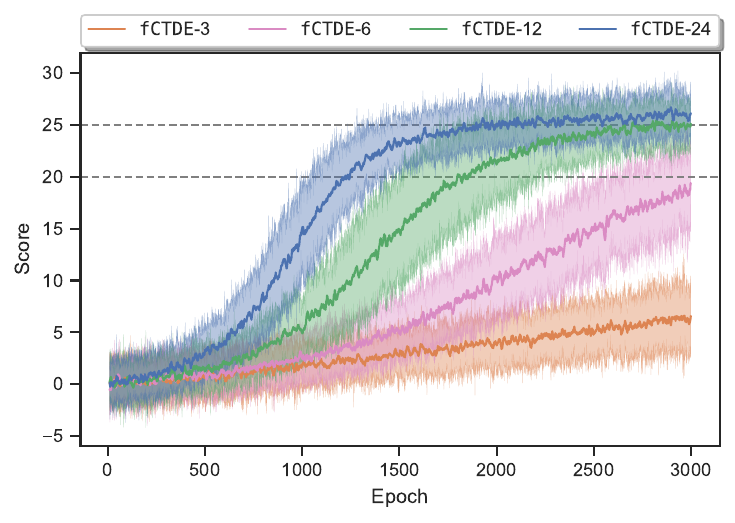}
        \caption{fCTDE - POMDP - Score}
        \vspace*{-0.25em}
    \end{subfigure}\\
    \begin{subfigure}{0.5\linewidth}
        \centering
        \includegraphics[width=\linewidth]{images/experiment_results/fig_coingame2_maa2c_mdp_ablation_sctde/fig_coingame2_maa2c_mdp_ablation_sctde-undiscounted_reward.pdf}
        \caption{sCTDE - MDP - Score}
        \vspace*{-0.25em}
    \end{subfigure}%
    \begin{subfigure}{0.5\linewidth}
        \centering
        \includegraphics[width=\linewidth]{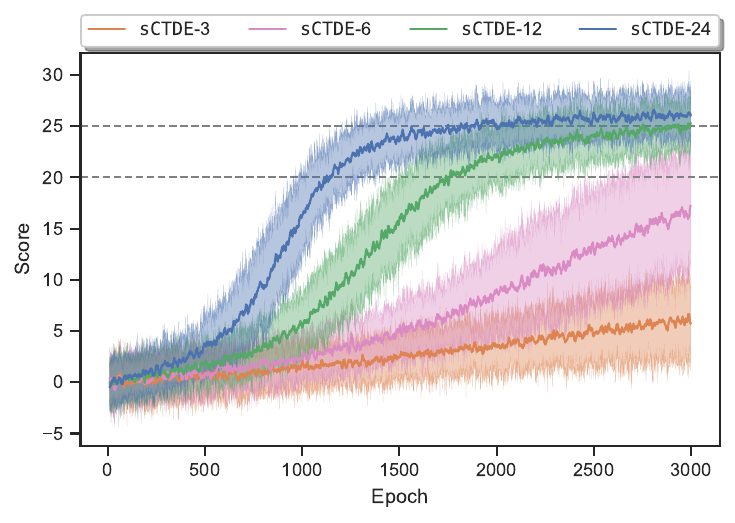}
        \caption{sCTDE - POMDP - Score}
        \vspace*{-0.25em}
    \end{subfigure}\\
    \begin{subfigure}{0.5\linewidth}
        \centering
        \includegraphics[width=\linewidth]{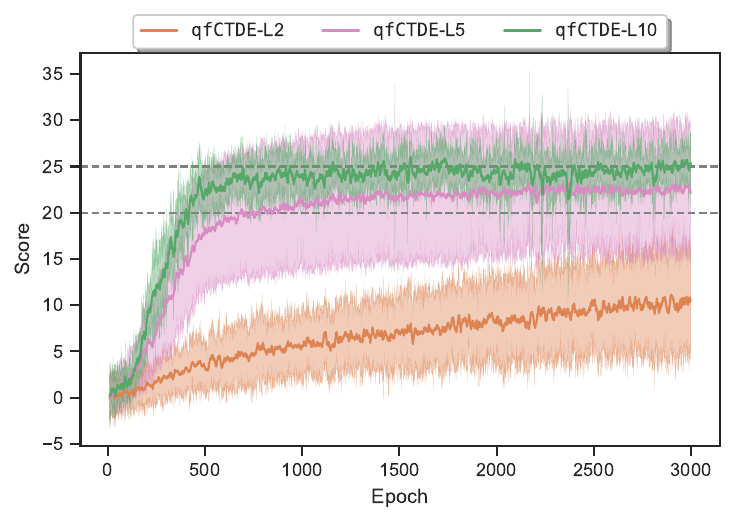}
        \caption{qfCTDE - MDP - Score}
        \vspace*{-0.25em}
    \end{subfigure}%
    \begin{subfigure}{0.5\linewidth}
        \centering
        \includegraphics[width=\linewidth]{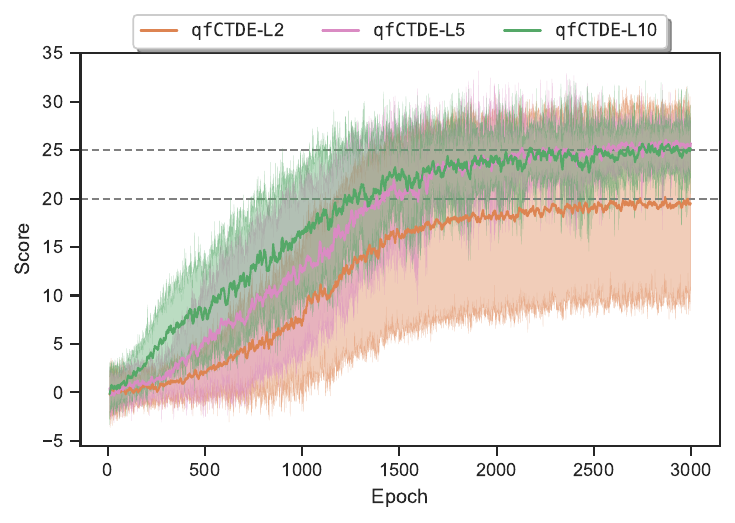}
        \caption{qfCTDE - POMDP - Score}
        \vspace*{-0.25em}
    \end{subfigure}\\
    \begin{subfigure}{0.5\linewidth}
        \centering
        \includegraphics[width=\linewidth]{images/experiment_results/fig_coingame2_maa2c_mdp_ablation_eqmarl_psi+/fig_coingame2_maa2c_mdp_ablation_eqmarl_psi+-undiscounted_reward.pdf}
        \caption{eQMARL-$\Psi^{+}$ - MDP - Score}
        \vspace*{-0.25em}
    \end{subfigure}%
    \begin{subfigure}{0.5\linewidth}
        \centering
        \includegraphics[width=\linewidth]{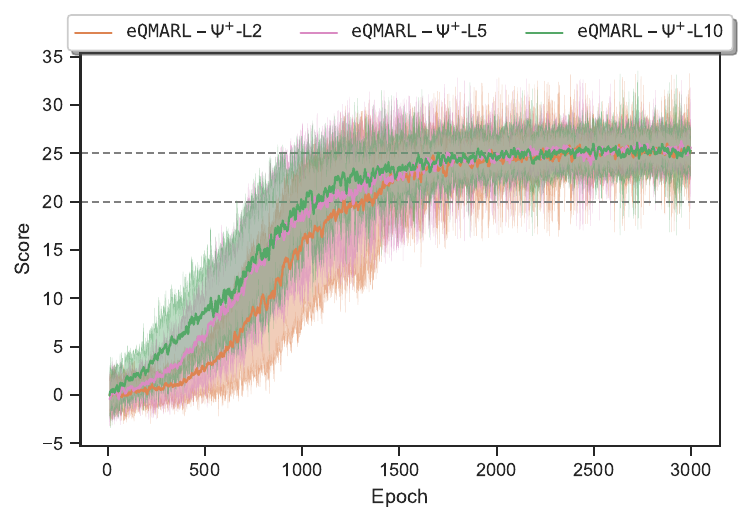}
        \caption{eQMARL-$\Psi^{+}$ - POMDP - Score}
        \vspace*{-0.25em}
    \end{subfigure}
    \caption{Score performance for ablation study using \texttt{CoinGame-2} for (a,b) fCTDE, and (c,d) sCTDE, and (e,f) qfCTDE, and (g,h) eQMARL-$\Psi^{+}$ with hidden layer units $h \in \{3,6,12,24\}$ and VQC layers $L \in \{2,5,10\}$, averaged over 10 runs of 3000 epochs, with $\pm 1$ std.\ dev.\ shown as shaded regions.}
    \label{app:fig:fig_maa2c_mdp_pomdp_ablation:score}
\end{figure*}

\begin{figure*}[h]
    \begin{subfigure}{0.5\linewidth}
        \centering
        \includegraphics[width=\linewidth]{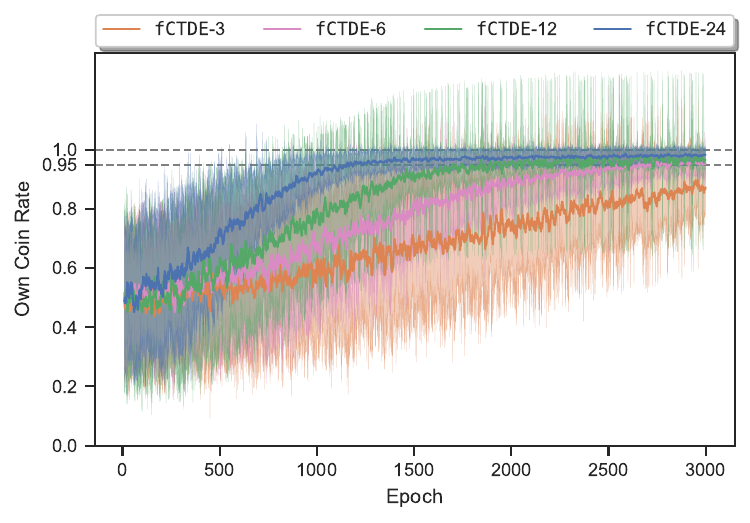}
        \caption{fCTDE - MDP - Own Coin Rate}
        \vspace*{-0.25em}
    \end{subfigure}%
    \begin{subfigure}{0.5\linewidth}
        \centering
        \includegraphics[width=\linewidth]{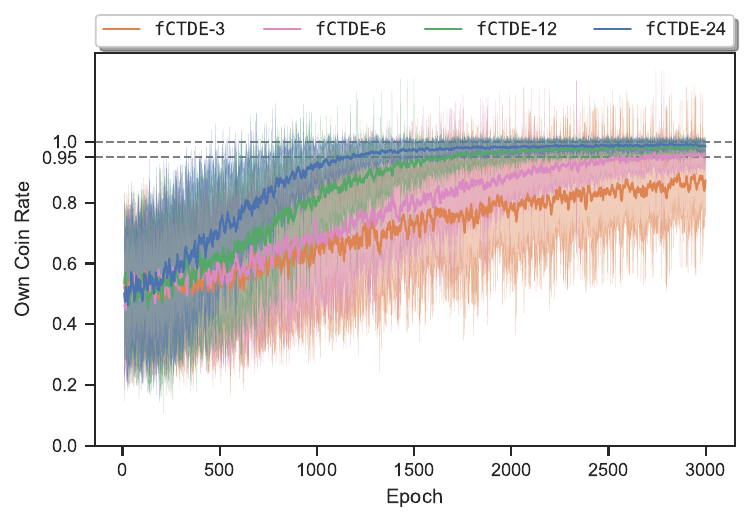}
        \caption{fCTDE - POMDP - Own Coin Rate}
        \vspace*{-0.25em}
    \end{subfigure}\\
    \begin{subfigure}{0.5\linewidth}
        \centering
        \includegraphics[width=\linewidth]{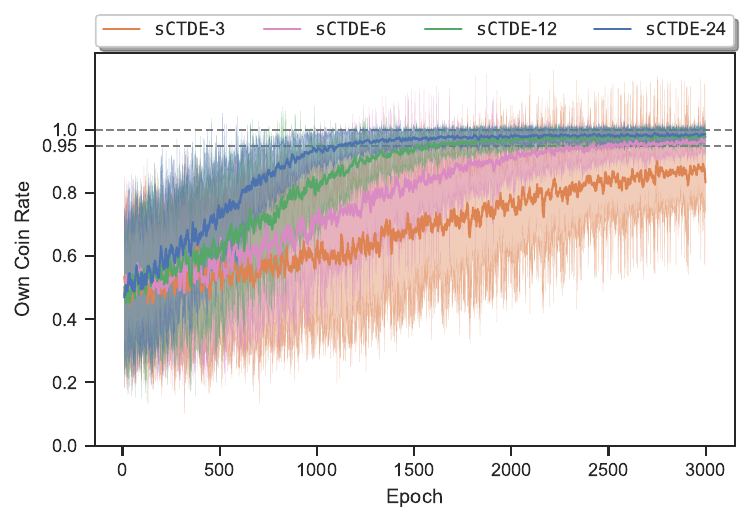}
        \caption{sCTDE - MDP - Own Coin Rate}
        \vspace*{-0.25em}
    \end{subfigure}%
    \begin{subfigure}{0.5\linewidth}
        \centering
        \includegraphics[width=\linewidth]{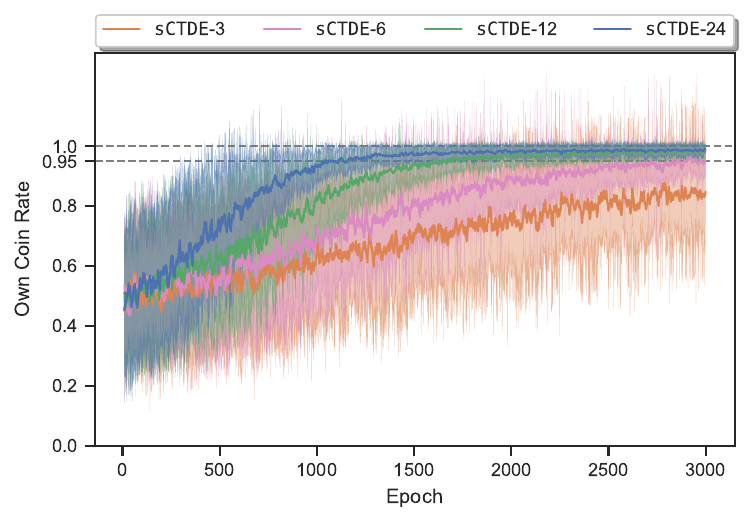}
        \caption{sCTDE - POMDP - Own Coin Rate}
        \vspace*{-0.25em}
    \end{subfigure}\\
    \begin{subfigure}{0.5\linewidth}
        \centering
        \includegraphics[width=\linewidth]{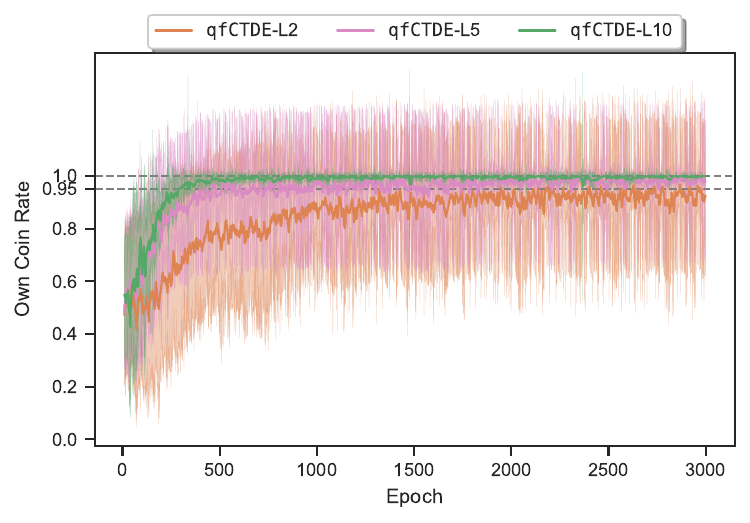}
        \caption{qfCTDE - MDP - Own Coin Rate}
        \vspace*{-0.25em}
    \end{subfigure}%
    \begin{subfigure}{0.5\linewidth}
        \centering
        \includegraphics[width=\linewidth]{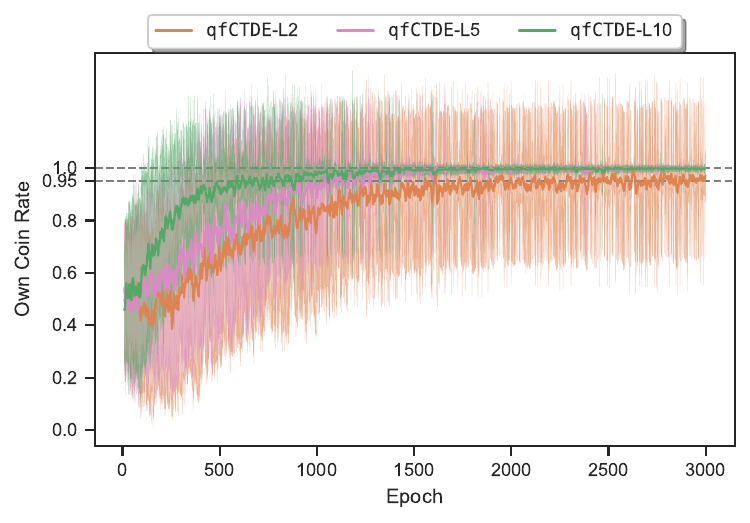}
        \caption{qfCTDE - POMDP - Own Coin Rate}
        \vspace*{-0.25em}
    \end{subfigure}\\
    \begin{subfigure}{0.5\linewidth}
        \centering
        \includegraphics[width=\linewidth]{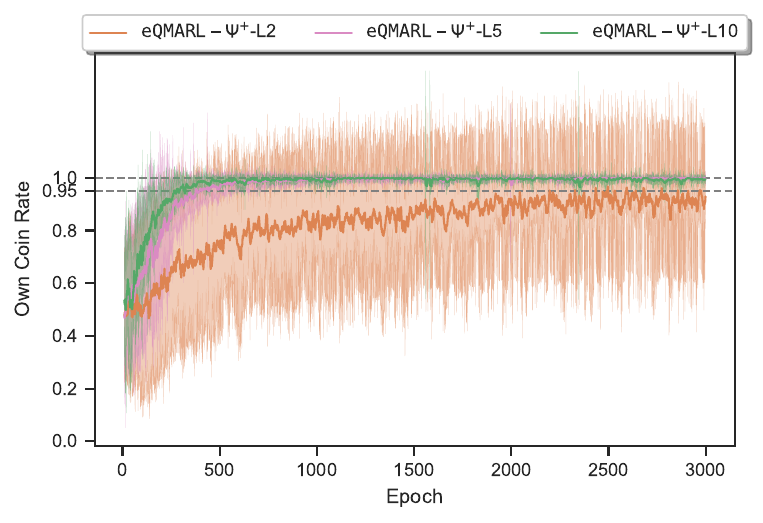}
        \caption{eQMARL-$\Psi^{+}$ - MDP - Own Coin Rate}
        \vspace*{-0.25em}
    \end{subfigure}%
    \begin{subfigure}{0.5\linewidth}
        \centering
        \includegraphics[width=\linewidth]{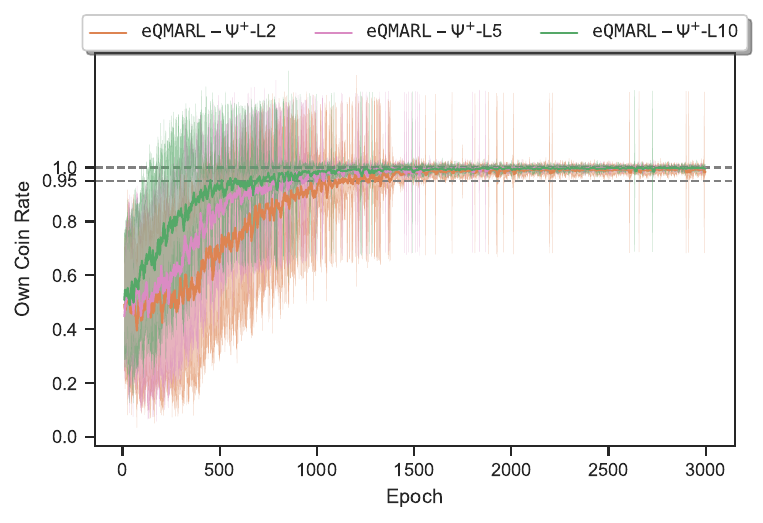}
        \caption{eQMARL-$\Psi^{+}$ - POMDP - Own Coin Rate}
        \vspace*{-0.25em}
    \end{subfigure}
    \caption{Own Coin Rate performance for ablation study using \texttt{CoinGame-2} for (a,b) fCTDE, and (c,d) sCTDE, and (e,f) qfCTDE, and (g,h) eQMARL-$\Psi^{+}$ with hidden layer units $h \in \{3,6,12,24\}$ and VQC layers $L \in \{2,5,10\}$, averaged over 10 runs of 3000 epochs, with $\pm 1$ std.\ dev.\ shown as shaded regions.}
    \label{app:fig:fig_maa2c_mdp_pomdp_ablation:own_coin_rate}
\end{figure*}

The empirical results for the ablation study as discussed in \cref{sec:exp:results:ablation} are shown in \cref{tab:exp:coingame:ablation:stat} and \cref{app:fig:fig_maa2c_mdp_pomdp_ablation:score,app:fig:fig_maa2c_mdp_pomdp_ablation:own_coin_rate}.
Looking at the statistics in \cref{tab:exp:coingame:ablation:stat} and the convergence in \cref{app:fig:fig_maa2c_mdp_pomdp_ablation:score,app:fig:fig_maa2c_mdp_pomdp_ablation:own_coin_rate}, we can see that 
our selection of $h=12$ hidden units for the baselines and $L=5$ for the quantum models is fair because of the significant performance drops and increased variation incurred by reducing, and the limited gains by increasing, the number of units and layers. This choice for architecture results in the most comparable performance across all baselines.
From the results of this ablation study, we can more concretely represent a comparison of the final sizes of the actor and critic models, in number of trainable parameters, for \texttt{eQMARL}, \texttt{qfCTDE}, \texttt{fCTDE}, and \texttt{sCTDE}. The final selected model sizes, in number of trainable parameters, are shown in \cref{app:tab:model_size_comparison}.

\begin{table}[h]
    \vspace*{-0.8em}
    \caption{Ablation study with classical model hidden layer units $h \in \{3,6,12,24\}$ and quantum VQC layers $L \in \{2,5,10\}$. Compares model size in number of trainable critic parameters with score and own coin rate performance for \ac{mdp} and \ac{pomdp} \texttt{CoinGame-2} environment dynamics using mean, standard deviation, and $95\%$ confidence interval statistics.}
    \label{tab:exp:coingame:ablation:stat}
    \centering
    \scriptsize
    \begin{tabular}{lllllllll}
        \toprule
                                    &                               & & \multicolumn{3}{c}{Score} & \multicolumn{3}{c}{Own Coin Rate} \\
                                                                     \cmidrule(lr){4-6} \cmidrule(lr){7-9}
        Dynamics                    & Framework                             & Params & Mean & SD & $95\%$ CI & Mean & SD & $95\%$ CI \\
        \midrule
        \multirow{14}{*}{\ac{mdp}}   
                                    & \texttt{fCTDE-3}                                & 223         &   2.42 & 2.35 & (2.35, 2.49)     & 0.6720 & 0.2024 & (0.6685, 0.6769) \\
                                    & \texttt{fCTDE-6}                                & 445         &   7.41 & 3.46 & (7.19, 7.65)     & 0.7658 & 0.1414 & (0.7610, 0.7712) \\
                                    & \texttt{fCTDE-12}                               & 889         &   12.36 & 4.41 & (12.09, 12.67)  & 0.8202 & 0.1379 & (0.8139, 0.8262) \\
                                    & \texttt{fCTDE-24}                               & 1777        &   17.63 & 2.58 & (17.25, 17.91)  & 0.8823 & 0.0751 & (0.8770, 0.8875) \\
                                    \cmidrule(lr){2-9}
                                    & \texttt{sCTDE-3}                                & 229         &   3.24 & 3.09 & (3.16, 3.33)     & 0.6852 & 0.1991 & (0.6821, 0.6897) \\
                                    & \texttt{sCTDE-6}                                & 457         &   8.54 & 3.67 & (8.29, 8.78)     & 0.7857 & 0.1327 & (0.7804, 0.7924) \\
                                    & \texttt{sCTDE-12}                               & 913         &   14.18 & 2.69 & (13.90, 14.60)  & 0.8504 & 0.0928 & (0.8454, 0.8553) \\
                                    & \texttt{sCTDE-24}                               & 1825        &   18.18 & 2.41 & (17.84, 18.53)  & 0.8936 & 0.0673 & (0.8896, 0.8979) \\
                                    \cmidrule(lr){2-9}
                                    & \texttt{qfCTDE-L2}                              & 121         &   6.58 & 3.92 & (6.47, 6.66)     & 0.8482 & 0.1921 & (0.8435, 0.8518) \\
                                    & \texttt{qfCTDE-L5}                              & 265         &   19.41 & 6.23 & (19.23, 19.59)  & 0.9398 & 0.1020 & (0.9366, 0.9426) \\
                                    & \texttt{qfCTDE-L10}                             & 505         &   22.08 & 2.22 & (21.91, 22.26)  & 0.9691 & 0.0247 & (0.9665, 0.9723) \\
                                    \cmidrule(lr){2-9}
                                    & \texttt{eQMARL-$\Psi^{+}$-L2}                   & 121         &   5.38 & 3.74 & (5.30, 5.46)     & 0.8271 & 0.2213 & (0.8234, 0.8300) \\
                                    & \texttt{eQMARL-$\Psi^{+}$-L5}                   & 265         &   21.11 & 2.65 & (20.92, 21.35)  & 0.9640 & 0.0347 & (0.9601, 0.9667) \\
                                    & \texttt{eQMARL-$\Psi^{+}$-L10}                  & 505         &   22.45 & 2.23 & (22.28, 22.62)  & 0.9719 & 0.0219 & (0.9685, 0.9745) \\
        \cmidrule(lr){1-9}
        \multirow{14}{*}{\ac{pomdp}} 
                                    & \texttt{fCTDE-3}                                & 169         &   2.98 & 2.47 & (2.91, 3.05)      & 0.7082 & 0.1890 & (0.7039, 0.7123) \\
                                    & \texttt{fCTDE-6}                                & 337         &   7.15 & 3.06 & (6.95, 7.37)      & 0.7711 & 0.1388 & (0.7658, 0.7781) \\
                                    & \texttt{fCTDE-12}                               & 673         &   13.46 & 3.24 & (13.09, 13.76)   & 0.8443 & 0.1026 & (0.8396, 0.8506) \\
                                    & \texttt{fCTDE-24}                               & 1345        &   17.38 & 2.65 & (17.06, 17.73)   & 0.8889 & 0.0752 & (0.8840, 0.8945) \\
                                    \cmidrule(lr){2-9}
                                    & \texttt{sCTDE-3}                                & 175         &   2.68 & 2.60 & (2.61, 2.74)      & 0.6834 & 0.1942 & (0.6792, 0.6866) \\
                                    & \texttt{sCTDE-6}                                & 349         &   6.35 & 3.53 & (6.18, 6.54)      & 0.7677 & 0.1488 & (0.7633, 0.7725) \\
                                    & \texttt{sCTDE-12}                               & 697         &   13.70 & 2.79 & (13.44, 13.99)   & 0.8466 & 0.0985 & (0.8411, 0.8515) \\
                                    & \texttt{sCTDE-24}                               & 1393        &   17.97 & 2.60 & (17.67, 18.25)   & 0.8948 & 0.0723 & (0.8898, 0.9004) \\
                                    \cmidrule(lr){2-9}
                                    & \texttt{qfCTDE-L2}                              & 745         &   12.34 & 7.56 & (12.09, 12.60)   & 0.8335 & 0.2058 & (0.8277, 0.8386) \\
                                    & \texttt{qfCTDE-L5}                              & 817         &   16.79 & 4.66 & (16.45, 17.04)   & 0.9040 & 0.1135 & (0.8994, 0.9091) \\
                                    & \texttt{qfCTDE-L10}                             & 937         &   18.14 & 4.28 & (17.83, 18.31)   & 0.9476 & 0.0660 & (0.9443, 0.9508) \\
                                    \cmidrule(lr){2-9}
                                    & \texttt{eQMARL-$\Psi^{+}$-L2}                   & 745         &   17.14 & 3.98 & (16.77, 17.47)   & 0.8834 & 0.1106 & (0.8769, 0.8896) \\
                                    & \texttt{eQMARL-$\Psi^{+}$-L5}                   & 817         &   18.49 & 3.91 & (18.23, 18.80)   & 0.9226 & 0.0831 & (0.9172, 0.9272) \\
                                    & \texttt{eQMARL-$\Psi^{+}$-L10}                  & 937         &   19.09 & 3.44 & (18.86, 19.46)   & 0.9485 & 0.0603 & (0.9458, 0.9523) \\
        \bottomrule
    \end{tabular}
\end{table}

\begin{table}[h]
    \vspace*{-0.8em}
    \caption{Comparison of the best model size in number of trainable parameters for each framework used on \texttt{CoinGame-2} environment with \emph{\ac{mdp}} and \emph{\ac{pomdp}} dynamics.}
    \label{app:tab:model_size_comparison}
    \centering
    \scriptsize
    \begin{tabular}{lllll}
        \toprule
                                            & &           & \multicolumn{2}{c}{Number of Trainable Parameters} \\
        \cmidrule(lr){4-5}
        Framework                           & Ablation Selection & Model     & \emph{\ac{mdp} dynamics}  & \emph{\ac{pomdp} dynamics} \\
        \midrule
        \multirow{2}{*}{\texttt{eQMARL}}    & $L=5$ & Actor     & 136                       & 412 \\
                                            & $L=5$ & Critic    & 265 (132 per agent, 1 central)                     & 817 (408 per agent, 1 central) \\ \cmidrule(lr){1-5}
                                            % 144 quantum, 672 classical
        \multirow{2}{*}{\texttt{qfCTDE}}    & $L=5$ & Actor     & 136                       & 412 \\
                                            & $L=5$ & Critic    & 265                       & 817 \\ \cmidrule(lr){1-5}
        \multirow{2}{*}{\texttt{fCTDE}}     & $h=12$ & Actor     & 496                       & 388 \\
                                            & $h=12$ & Critic    & 889                       & 673 \\ \cmidrule(lr){1-5}
        \multirow{2}{*}{\texttt{sCTDE}}     & $h=12$ & Actor     & 496                       & 388 \\
                                            & $h=12$ & Critic    & 913 (444 per agent, 25 central)                    & 697 (336 per agent, 25 central) \\
        \bottomrule
    \end{tabular}
\end{table}

%%%%%%%%%%%%%%%%%%%%%%%%%%%%%%%%%%%%%%%%%%%%%%%%%%%%%%%%%%%%

\end{document}